\documentstyle[epsfig]{article}
\def\degr{\hbox{$^\circ$}}

\def\keywords{\@keywords}

\def\@keywords{\list{}{%
  \leftmargin 0.0pt \listparindent 0em
  \itemindent 0pt \labelwidth 0pt \labelsep 0pt \rightmargin 0pt
  \parsep 0pt plus 1pt}\item[]
  \large{\it Keywords}~:}
\begin{document}
\large
\baselineskip 17pt
\pagenumbering{roman}
\setcounter{page}{1}
\title {\bf A stellar photo-polarimeter}
\author {{\bf A. V. Raveendran$^*$, G. Srinivasulu, S. Muneer, 
M. V. Mekkaden$^*$,}\\
{\bf N. Jayavel$^*$, M. R. Somashekar, K. Sagayanathan, S. Ramamoorthy,}\\
{\bf M. J. Rosario\thanks{Retired from service}\, and K. Jayakumar$^*$}\\ \\
{\Large {\bf {Indian Institute of Astrophysics}}}\\ {\large {\bf { Bangalore 560 034, India}}}}
\date{}
\maketitle
\begin{abstract}
A new astronomical photo-polarimeter that
can measure linear polarization of point sources
simultaneously in three spectral bands was designed and built in the Institute.
The polarimeter has a Calcite beam-displacement prism as the analyzer.
The ordinary and extra-ordinary emerging beams in each spectral band are
quasi-simultaneously detected by the same photomultiplier by using a high speed
rotating chopper. The effective chopping frequency can be set to as high as
200~Hz.
 A rotating superachromatic Pancharatnam halfwave
plate is used to modulate the light incident on the analyzer. The spectral
bands are isolated using appropriate dichroic and glass filters.

A detailed analysis shows that the reduction of 50\% in the efficiency of
the polarimeter because of the fact that the intensities of the two beams
are measured alternately is partly compensated by the reduced time to be
spent on the observation of the sky background. The use of a beam-displacement
prism as the analyzer completely removes the polarization of
background skylight, which
is a major source of error during moonlit nights, especially, in
the case of faint stars.

The field trials that were carried out by observing several
polarized and unpolarized stars show a very high mechanical stability for the
polarimeter. The position angle of polarization produced by the Glan-Taylor
prism in the light path is found to be slightly wavelength-dependent,
indicating that the fixed super-achromatic halfwave plate in the beam does not
fully compensate for the variation in the position 
angle of the effective optical axis of the rotating plate. However, the
total amplitude of variation in the $U-I$ spectral region is only 0.\degr 92.
The polarization efficiency is also found to be wavelength-dependent with a
total amplitude of 0.271\% in the $U-I$ region; its mean value is 99.211\%.
The instrumental polarization is found to be very low. It is
nearly constant in the $V-I$ spectral region ($\sim$ 0.04\%), and
apparently, it increases slightly towards the ultraviolet. The observations
of polarized stars show that the agreement between the measured polarization
values and those available in the literature to be excellent.

\end{abstract}


\newpage
\tableofcontents
\newpage
\listoftables
\newpage
\listoffigures
\newpage
\pagenumbering{arabic}
\setcounter{page}{1}
\section{Introduction}

The linear polarization observed in stars, in general, is very small. A
polarization of 0.01 (1 per cent) is considered to be quite large
by the usual standards. It is essential that a high precision is achieved
in polarimetry if the measurements of stellar polarization and
attempts to detect any likely variation in it are to be meaningful.
In principle, it is possible to attain a very high precision in
polarimetry when compared to photometry since the former essentially 
uses the
technique of differential photometry: we do not express the results
in terms of the polarized flux, but in terms of the degree of polarization
where the mean brightness of the star itself serves as the reference
light level.
This procedure eliminates to a large extent the adverse effects of
atmospheric scintillation, seeing and transparency variation,
the factors which usually limit the photometric accuracy. 
While achieving an external photometric consistency of 1 per cent is
quite difficult
even for bright objects, with suitable instrumentation  and proper
care it is possible to obtain an external
consistency of 0.01 per cent in polarimetry, if sufficient photons
are available.

There are several physical mechanisms that produce polarization
of light from astronomical objects (Serkowski 1974b; Scarrott 1991).
One of the most common processes that give rise to linear polarization
of starlight involves circumstellar dust grains; they produce
a net polarization in starlight
integrated over the stellar disc either by
scattering in an asymmetric envelope or by selective dichroic extinction.
Measurements of polarization in multibands
are required to study the nature of grains and their formation in the
circumstellar environments. The observed polarization, in general, will
contain a contribution from the interstellar dust grains, which could be
quite significant in some cases.
With reasonable assumptions about the interstellar component,
it is possible to derive the intrinsic
polarization exhibited by starlight if the
wavelength dependence of the observed polarization is known.
Polarimetric observations of Vega-like stars indicate that most of them
show intrinsic polarization probably arising from scattering by dust grains
confined to circumstellar discs (Bhatt \& Manoj 2000).
In order 
to improve the sample of such stars, `it is essential that the polarimeter
should be able to measure polarizations which are extremely low
($<$~0.1 per cent) because the material that constitutes the
circumstellar discs is expected to be very tenuous.
The observations done at 
Vainu Bappu Observatory during the past several years indicate that in
some of the RV Tauri stars circumstellar grains either
condense, or get aligned cyclically
during each pulsational cycle, possibly triggered by the periodic
passage of atmospheric shocks (Raveendran 1999a).
In order to look for a possible evolution
in the grain size distribution, in case it is the dust condensation
that occurs, multiband observations are absolutely
essential. Similarly, the identification of the polarigenic mechanism
in T Tauri objects, which again constitute another group of objects
extensively studied at the Institute (Mekkaden 1998, 1999),
also requires a knowledge of the wavelength dependence of polarization.
Some of the other polarimetrically interesting objects
that are studied in the Institute using multi-spectral band data are
Herbig Ae/Be stars (Ashok et al. 1999),
Luminous Blue variables (Parthasarathy, Jain \& Bhatt 2000),
Young Stellar Objects (Manoj, Maheswar, \& Bhatt 2002) and R CrB stars
(Kameswara Rao \& Raveendran 1993).

There was a need for an efficient photo-polarimeter for observations with the
1-m Carl Zeiss Telescope at Vainu Bappu Observatory, Kavalur. The star--sky
chopping polarimeter, which was built by Jain \& Srinivasulu (1991) almost a 
quarter century
ago, was the only available instrument for polarimetric studies at the
Observatory for several ongoing programmes.
It is a very inefficient instrument. It uses a rotating HNP'B sheet as the
analyzer and observations can be done only in one spectral band at a time.
Further, the time spent on sky observations is always the same as that spent
on star observations, irrespective of the relative brightnesses of the star and
the sky background, and thus underutilizing the available telescope time.

A project to build a new
photo-polarimeter for observations of point sources in $UBVRI$ bands with the
1-m telescope
was initiated in the Indian Institute of Astrophysics
quite sometime back. Unfortunately, due to unforeseen reasons there
were delays at various stages of the execution of the project.

In this write-up we give 
the details of the three band, two beam photo-polarimeter for point sources
that we designed and built in the Institute. 
In section~\ref{s:worpri} we give the working principle of
the instrument. The optical layout, and the functions
and details of important components are described in section~\ref{s:instr}.
Several factors that have gone into the selection of the optical
elements and the design of the various mechanical parts
are  briefly described in that section. The responses of the
dichroic mirrors and the glass filters supplied by the
manufacturer, and the mean wavelengths
of the $UBVRI$ passbands of the polarimeter computed from the
responses of the various components including the
detector are also presented in the same section.
A brief description of the polarized light in terms of the
Stokes parameters and the effects of polarizer and retarders
on these parameters when introduced in the light path
are presented in the next section.

In section~\ref{s:linpol} we describe the various schemes of
determination of linear
polarization included in the reduction
program, and discuss how the available time should be optimally
distributed between the
observations of object and background brightness in order to minimize
the error in the linear polarization.
A brief description 
of the electronics system that controls the polarimeter operations 
and data acquisition are given in section~\ref{s:conele}.
The data reduction and display program is briefly
described in section~\ref{s:datacq}.

The polarimeter was mounted onto the 1-m Carl Zeiss telescope at Kavalur
and observations of several polarized and unpolarized stars were made
during 14~April--30~May~2014 to evaluate its performance. 
Due to prevalent poor sky conditions, the instrument could be
used effectively only on a few nights during this period. The instrument was
found to have a very high degree of mechanical stability, but
a comparatively low polarization efficiency of 94.72\%.
In order to reduce the scattered light inside the instrument,
we blackened the few internal mechanical parts that were
left out earlier and improved light shields, 
and again made observations during February--April~2015;
in section~\ref{s:obsres} we present a detailed analysis
of these observations and the results obtained.

\section  {Principle of operation} \label{s:worpri}
\begin{figure}[htb]
\centerline {\psfig{figure=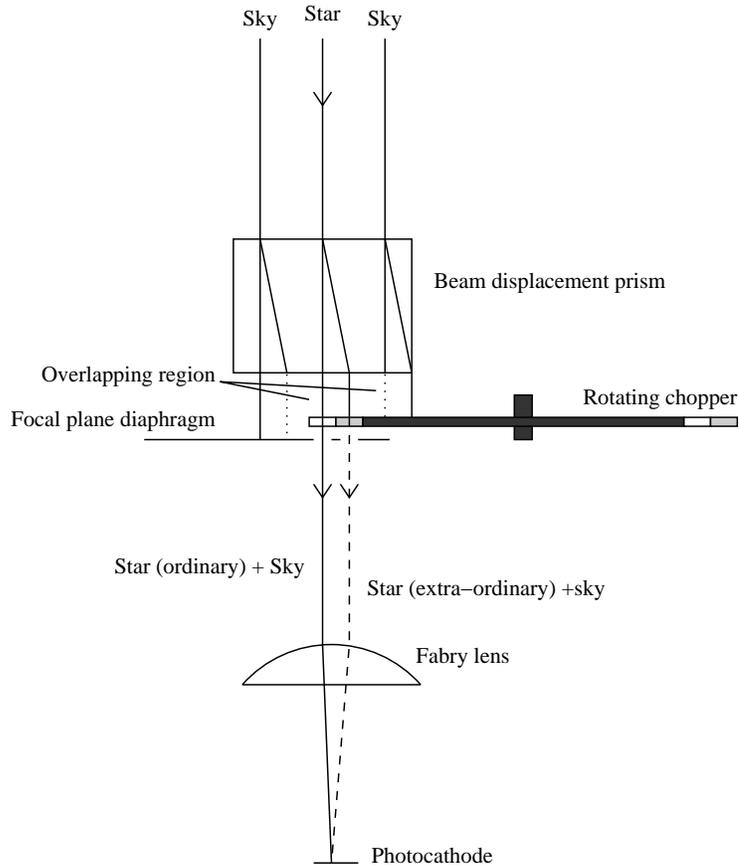,width=10cm,clip=}}
\caption {Working principle of the polarimeter.} \label{f:worpri}
\end{figure}

The working principle of the polarimeter is illustrated in 
Figure~\ref{f:worpri}.
An astronomical polarimeter based on this principle was first built
by Piirola (1973).
A beam displacement prism divides the incident 
light into two beams with mutually perpendicular planes of polarization.
The ordinary-beam, with the plane of
polarization perpendicular to the plane containing the optical axis
and the incident beam, obeys the Snell's law of refraction at the prism
surface and travels along the initial ray direction.
The extraordinary-beam with vibrations lying in the
above plane does not obey Snell's law, and hence, travels in
the prism with a speed that
varies with the direction.  As a result of these, the incident light beam
will emerge from the prism as two spatially separated beams. 
The background sky, which acts as an extended object, illuminates
the entire top surface of the
beam displacement prism, and hence, produces two broad emergent
beams whose centres are spatially separated by the same amount as
above. There will be a considerable overlap between these two beams
about the geometrical axis of the prism, and wherever the
beams overlap in the focal plane of the telescope that portion remains
unpolarized by the prism and observations give the background
sky brightness directly.
The situation is different with respect to the starlight;
the star being a point source, there will be no overlap between the
two emergent beams from the prism. The net result of
these two effects is that we observe two images of the star
with mutually
perpendicular planes of polarization at the focal plane superposed on
the unpolarized background sky. Two identical apertures are used to
isolate these images, and a rotating chopper is used to alternately
block one of the images allowing the other to be detected by the
same photomultiplier tube.

The main advantages of such an arrangement are:
(i) The contribution of background sky polarization is completely eliminated
from the data, thereby, facilitating the observations of fainter stars during
moonlit nights without compromising on the accuracy that is achievable
during dark nights. This is possible because the background sky
is not modulated, it just appears
as a constant term that can be removed from the data accurately.
(ii) Since the same photomultiplier tube is used to
detect both the beams, the effect of any time-dependent
variations in its sensitivity
as a result of variations in the associated electronics, like, high voltage
supply, is negligible.
(iii) The quasi-simultaneous detection of both the beams using a fast
rotating chopper essentially eliminates the effects of variations
in sky-transparency and reduces the errors due to
atmospheric scintillation for bright stars significantly.
Scintillation noise is independent of the brightness of the star
and dominates the photon noise for bright objects; with a 1-m
telescope the low frequency ($<$ 50~Hz) scintillation noise is
expected to be larger than the photon noise for stars brighter
than B~$=$~7.0~mag at an airmass 1.0 (Young 1967).
The averaging of the data by the process of long integration will reduce
the scintillation noise
only to a certain extent because of its log-normal
distribution. The frequency spectrum of scintillation
noise is flat up to about 50~Hz; the noise amplitude decreases rapidly
above this frequency and it becomes negligibly small above 250~Hz.
With the fast chopping of the beams and automatic removal of the background
sky polarization it is possible to make polarization
measurements that are essentially photon-noise limited.

The two factors that determine the overall efficiency of a polarimeter
attached to a telescope are:
(i) the faintest magnitude that could be reached with a specified
accuracy in a given time, and (ii) the maximum amount of information
on wavelength dependence that could be obtained during the same time.
It is evident that
the former depends on the efficiency in the
utilization of photons collected
by the telescope and the latter on the number of spectral bands
that are simultaneously available for observation.
In the beam displacement prism-based polarimeters, where the image
separation is small, the intensities
of the two beams produced by the analyzer are measured alternately
using the same detector, and hence, only 50 per cent of the light
collected by the telescope is effectively utilized.
In polarimeters using Wollaston prism
as the analyzer, the two beams, which are well-separated
without any overlap, can be detected simultaneously by two
independent photomultiplier tubes, fully utilizing
the light collected (Magalhaes, Benedetti \& Roland~1984;
Deshpande et al. 1985). For simultaneous multi-spectral band
observations that make use of the incident light fully,
the beams emerging from the analyzer will have to be well-separated so as
to accommodate a large number of photomultiplier tubes,
making the resulting instrument both heavy and large in size
(Serkowski, Mathewson \& Ford 1975),
and hence, not suitable for the 1-m telescope.
Usually, in Wollaston and Foster
prism-based polarimeters, when multispectral bands are available
simultaneously, different spectral bands are distributed
among the two emergent beams and at a time only 50 per cent of the light
collected in each spectral region is made use of (Kikuchi 1988;
Hough, Peacock \& Bailey 1991).

Separate observations are
needed to remove the background sky polarization
when there is no overlap between the emergent beams, and a significant
fraction of the time spent on object integration will have to be
spent for such observations if the objects are faint.
In beam displacement prism based polarimeters, time has to be spent
only to determine the brightness of the background sky, and
for the maximum signal-to-noise ratio it is only a negligible
fraction of the object integration
time even for relatively faint objects; the available time for observation
can be almost entirely utilized
to observe the object. Therefore, the non-utilization of 50 per cent
light in beam displacement prism based polarimeters does not
effectively reduce
their efficiency, especially, while observing faint objects where the
photons lost actually matter. The provisions for multi-spectral
band observations can be easily incorporated in the design, making
beam displacement prism based polarimeters to have an overall
efficiency significantly higher than that of the usual
Wollaston or Foster prism based polarimeters
(Magalhaes, Benedetti \& Roland~1984; Deshpande et al. 1985;
Hough, Peacock \& Bailey 1991).
Another advantage of
these polarimeters is that they can be easily converted into
conventional photometers,
just by pulling the beam displacement prism out of the light path.

\section{The instrument}\label{s:instr}

We adopted a design based on the beam displacement prism for the
polarimeter because with such an instrument
polarimetry can be done without compromising
much on the precision even under moderate variations in background
sky brightness and atmospheric transparency.
The brightness in the blue spectral band,
when the sky is dark, is about 15~mag, typically,
if a focal plane diaphragm of 20 $arcsec$
diameter is used for observation.
The background brightness continuously increases for a few hours
after moon-rise and decreases for a few hours before moon-set. Even
when the moon is high up in the sky the background brightness can 
change appreciably during the observation if the object integration
lasts several minutes. Large changes in the background brightness can also
occur if the sky is partially cloudy and moonlit.
When Wollaston prism-based polarimeters are used
an exact removal of the background polarization, which
could be several tens of percentage, is extremely difficult.
Sometimes, it may
be even impossible to remove the background polarized flux accurately,
if large changes occur in it. These instruments are effective in observing
faint objects during dark periods of nights. But, with a
beam displacement prism-based polarimeter
reliable results can be obtained during fairly bright moonlit periods
even when the sky is partially cloudy. In fact,
most of the spectroscopic nights can be utilized for polarimetric
observations with such an instrument.
In the present polarimeter
observations can be made simultaneously in three spectral bands.
More sophisticated versions of the beam displacement prism
are available elsewhere (Magalhaes \& Velloso 1988;
Scaltriti et al. 1989; Schwarz \& Piirola 1999).

\begin{figure}[htb]
\centerline{\psfig{figure=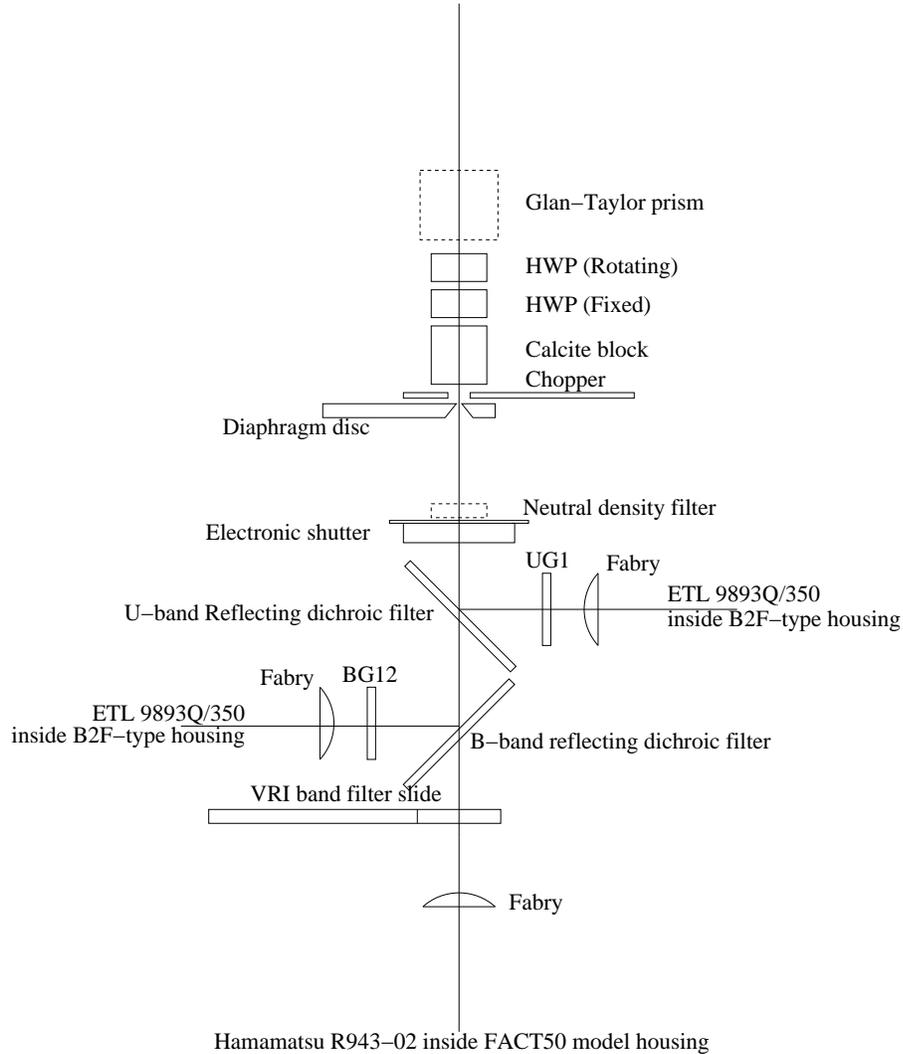,width=12cm}}
\caption {A schematic layout of the polarimeter indicating its
main components.} \label{f:schelay}
\end{figure}

The layout of the polarimeter indicating the positions of the main
components is shown schematically in Figure~\ref{f:schelay}.

\begin{figure}[htb]
\centerline{\psfig{figure=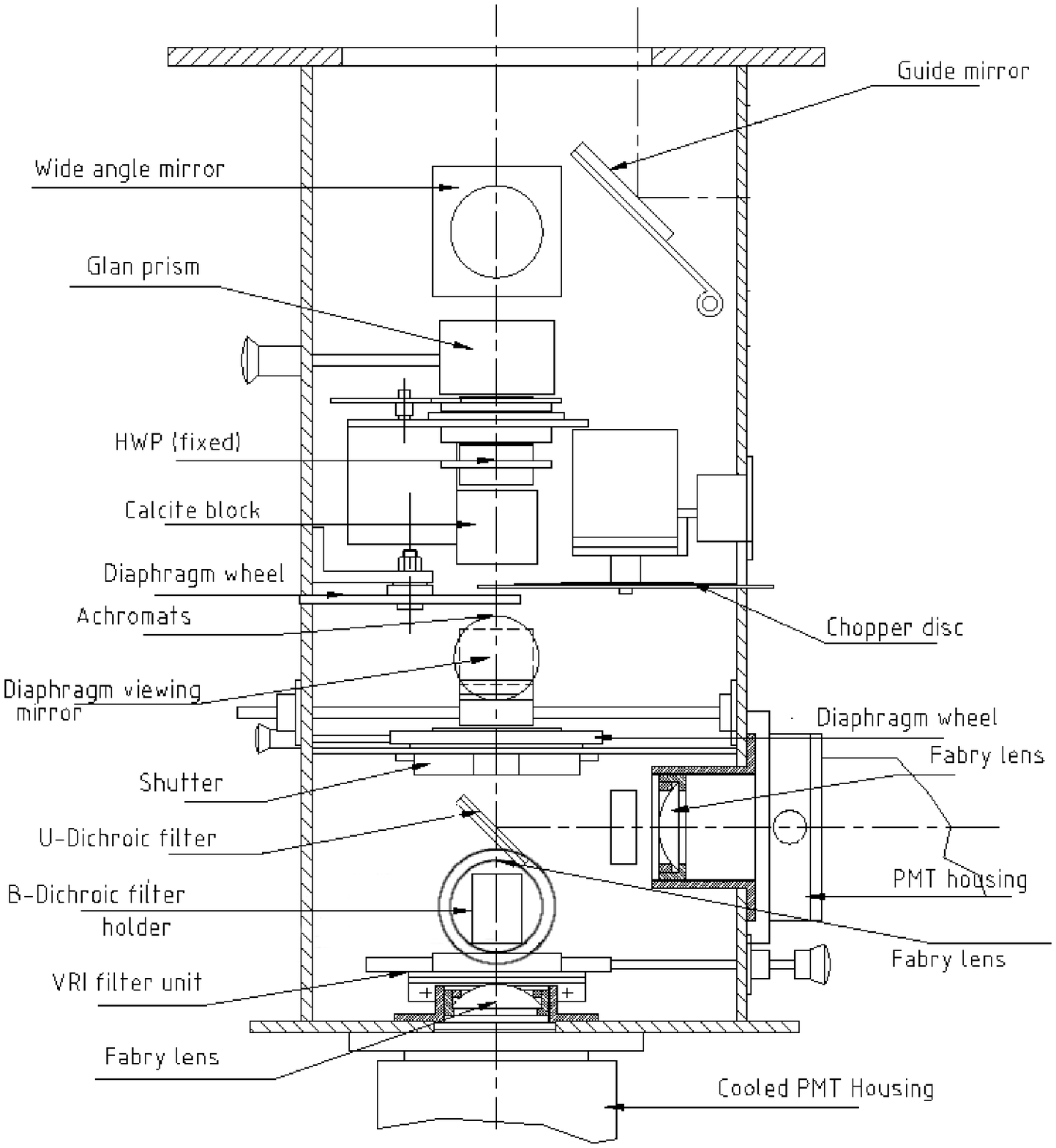,width=12cm}}
\caption {Schematic front-view of the polarimeter.} \label{f:frontview}
\end{figure}

\begin{figure}[htb]
\centerline{\psfig{figure=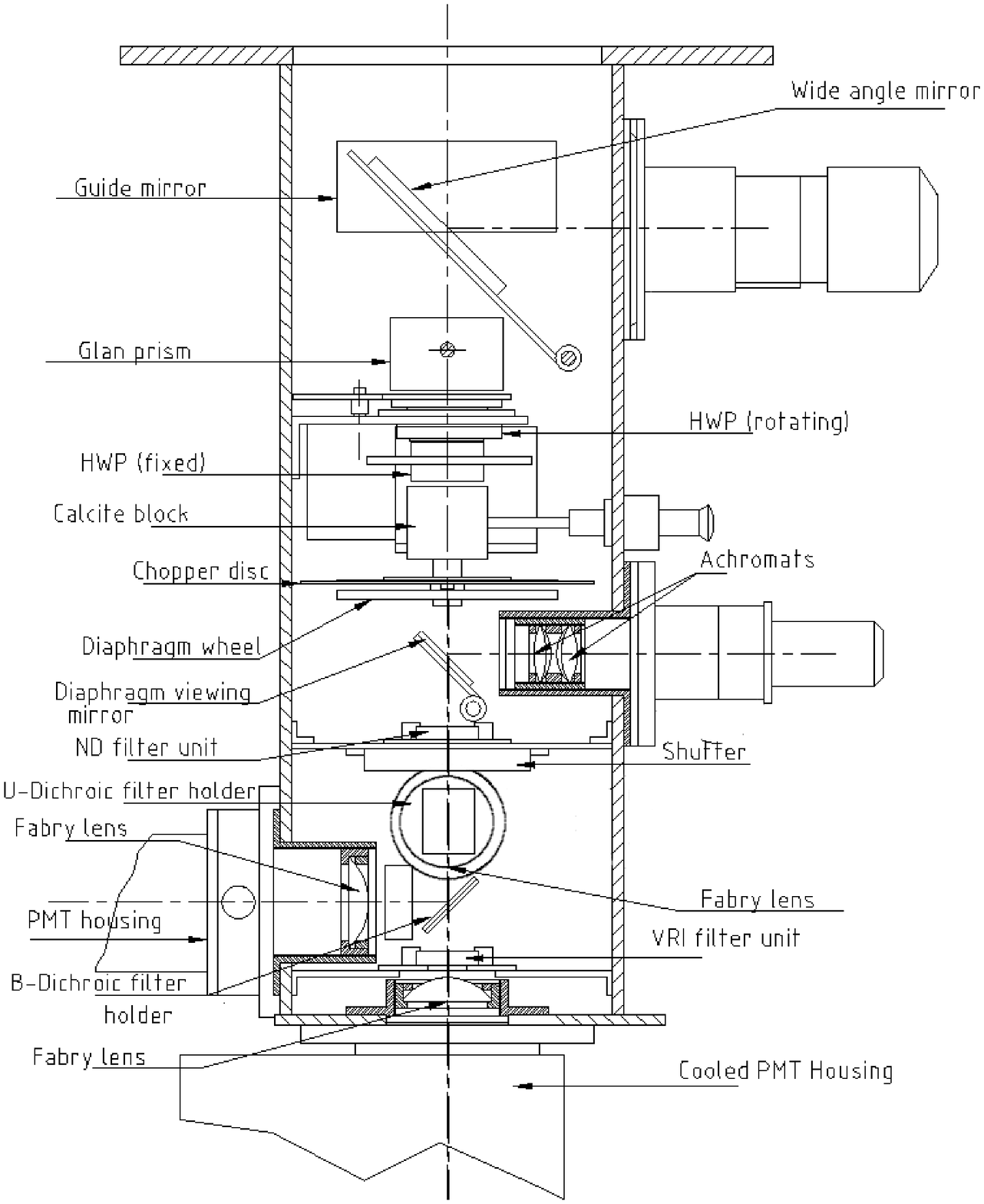,width=12cm}}
\caption {Schematic side-view of the polarimeter.} \label{f:sideview}
\end{figure}

The front and side views of the polarimeter are schematically
shown in Figures~\ref{f:frontview} and \ref{f:sideview},
 indicating the important components.
In the following subsections we describe
the various components and the details which have gone into either their
design, or their selection.
The instrument was designed basically for use at the
1-m Carl Zeiss telescope of Vainu Bappu Observatory, as
mentioned earlier;
in principle, it can be used with any telescope having an f/13 beam
with suitable modifications in the sizes of the twin diaphragms. 

\subsection {Wide angle viewing}

The first element in the polarimeter is the wide angle viewing arrangement
for object field identification. The star field is brought into the field
of view of the eyepiece
by flipping a plane mirror to a 45-degree position with respect to the
telescope axis. 
The costs of the optical elements, like,
half-wave plate and beam displacement prism, go up drastically with increase
in their clear apertures. The Glan prism, which is used to
calibrate the instrument for 100 per cent polarization,
has a constant aperture-to-thickness ratio, and hence, larger
the aperture, larger is its thickness.
Since no re-imaging is done the diameter of the light beam
produced by the telescope increases linearly with distance on either side
of the image-isolating, focal plane diaphragms.
In order to keep their clear apertures as small as possible, all
the optical elements should be kept as close to the diaphragm as possible.
The minimum distance between the centre of the wide angle mirror and the
diaphragm, which could be accomplished, is 195~$mm$. At this distance
the diameter of the F/13 beam is
15~$mm$. Assuming a field-stop of 50~$mm$ for a 2-$in$ barrel eyepiece (the
maximum field-stop possible for such an eyepiece is 46~$mm$) the minimum
width of the mirror should be 65~$mm$ and its length should be 92~$mm$. Since
the mirror is introduced in a converging beam rather than a parallel beam,
the centre of the mirror should be offset by about 1.5~$mm$,
along the length of the mirror towards the side closer to
the telescope back-end. We have used a 70~$mm$ $\times$ 100~$mm$ mirror.
Because it is slightly oversized the mirror need not be offset to accommodate
all the light collected by the telescope over the field.
The flat mirror blank is made of borosilicate crown and an overcoating 
of silicon monoxide is given to protect the reflective aluminium
coating for longer life.

For the object identification a field of view of about 10 $arcmin$ 
diameter is needed;
this requires eyepieces of fairly long focal lengths with large field stops.
Several makes
of eyepieces with focal lengths around 50~$mm$ are available in the market.
Gordon (1988), who made a comparative study of the performance of these
eyepieces, has reported that the 4-element super Plossl eyepiece of
56~$mm$ focal length made by Meade has the best overall performance.
The exit pupil,
the image of the primary formed by the eyepiece, has a size of 4.3~$mm$. 
The eye-relief, which is the distance between the last physical element
of the eyepiece and the exit pupil, is excellent ($>$~38~$mm$). The eyepiece,
like other makes of similar focal lengths, is not parfocal with other
eyepieces, and hence, refocusing is needed if the eyepiece has to be replaced
with another one of a different focal length. It has the largest
apparent field of 52 degree among the eyepieces investigated
by Gordon, and since a larger
apparent field implies a higher magnification it also has the highest
magnification.
The apparent field of an eyepiece
is the angle subtended at the eye by the circular
patch that appears to the eye when viewed
through it. The outer boundary is defined by the field stop,
a metal ring that gives a sharp boundary by restricting the observer
from seeing too far off axis, where the quality of star images becomes
poor. The diameter of the field stop is approximately 46~$mm$. With an
image scale of 15.5 $arcsec$ per $mm$ at the focal plane of the
1-m telescope
the true field that can be viewed is
about 12 $arcmin$, which is fairly good for object identification.
 
The axis of the field-viewing eyepiece is 100~$mm$ from the top of
the mounting flange.
With the position angle device, which has a thickness of 50~$mm$,
there is sufficient clearance between the observer's head and the back-end
of the telescope, making the wide angle viewing convenient.

\subsection {Offset guiding}

\begin{figure}[htb]
\centerline{\psfig{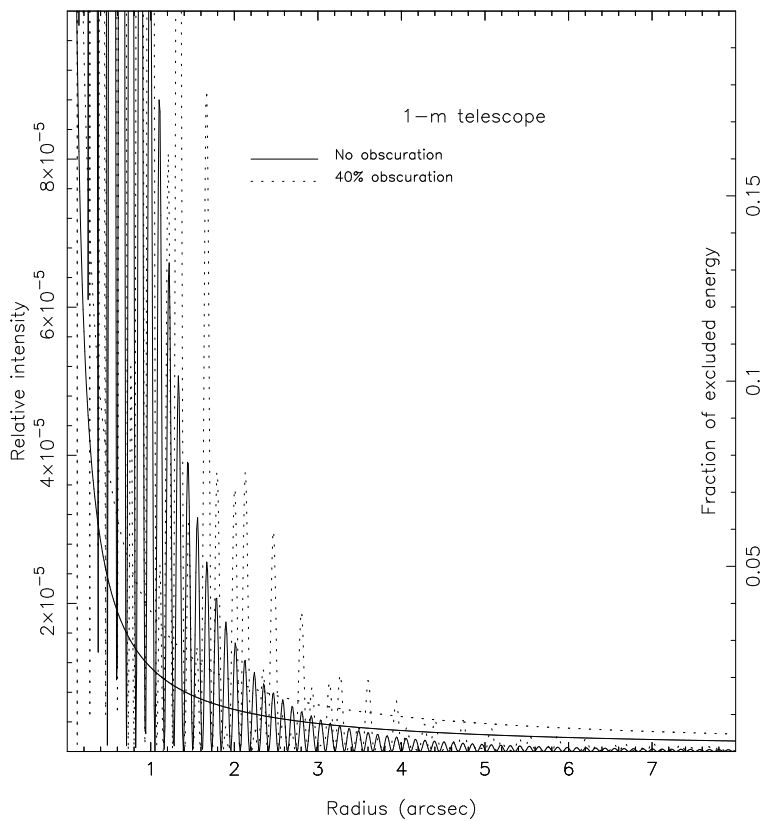}}
\caption {Intensity distribution and the excluded energy due to diffraction.}
\label{f:exclint}
\end{figure}

Because of the diffraction phenomenon occurring at its
aperture the starlight collected by a telescope will be distributed
spatially in the image plane
with the intensity falling off asymptotically as $r^{-3}$, where
$r$ is the radial distance from the central axis. The reflector telescopes
usually have a central obscuration, and for a typical Ritchie-Chretien
telescope it is about 40 per cent of the full aperture.
Figure~\ref{f:exclint}
 shows the intensity distribution due to diffraction effects
at the focal plane of a 1-m telescope with and without the
central obscuration.
It also shows the energy excluded as a function of the diaphragm radius
for both the cases. The main effect
of the central obscuration is the redistribution of intensity in the
outer rings, and thereby, spreading out farther
the light collected by the telescope.
It is clear from the figure that
even with an aperture of 16-$arcsec$ diameter at the focal plane of
a 1-m Cassegrain telescope the excluded energy is more than 0.5 per cent.
If the aperture is not exactly at
the focal plane the excluded energy will be significantly larger than
the above (Young 1970).

The actual measurements of light lost in the focal plane
diaphragm by Kron \& Gordon (1958) show that the excluded
energy can be several times larger
than that indicated in Figure~\ref{f:exclint}, where only the effects due to
the classical diffraction phenomenon are included.
The increase in excluded energy in practice arises
from the extended rings of the star image primarily caused by the
surface roughness of the telescope mirrors or the presence of dust
on them (Kormendy 1973; Young et al. 1991).

If the star trails inside the diaphragm due to a
poor tracking of the telescope, the included
light will vary because the light thrown out on one side of the diaphragm
will not be the same as that brought in on the other side of the diaphragm,
thus, resulting in a variation in the output from the photomultiplier tube.
The two images formed by the beam displacement prism passes through
different regions of the various optical elements, like, filters.
The transmittance of the glass filters and the reflectivity of the
dichroic mirrors, in general, will not be the same across their surfaces;
a spatial drift in the images will cause a variation in the
differential throughput of the beams. The
accuracy in the polarimetric measurements primarily depends on the accuracy
in the determination of the ratios of the intensities of the two beams
over a full rotational cycle of the signal modulator.
If the star images trail inside the diaphragms there will be additional
modulations of the ratio of intensities of the two beams, contributing
to the errors in the measurement.
For precise polarimetry it is essential that the star image should not
drift inside the diaphragm even by a minute fraction of its
size while observation, emphasizing the importance of the availability
of a proper guiding facility.

Accurate guiding is possible if the image scale
available for guiding is similar to that used for observation, making
offset guiding the best option.
Another plane mirror, which is also inclined at an angle of
45 degree to the telescope axis, is provided to reflect the starlight
to the right side of the polarimeter for the purpose of guiding. Using 
a separate mirror for a guide star ensures that
the alignment of the wide angle
viewing arrangement with respect to the diaphragm is not disturbed, and
thereby, saving the trouble of re-aligning it every time either
the instrument is mounted onto the telescope, or a new
object is acquired for observation, and improving slightly the
efficiency in the use of telescope time.
The centre of the mirror is at a distance of 210~$mm$ from the
focal plane diaphragm.
The width of the plane mirror used is 70~$mm$, which gives a projected
cross-section of 50~$mm$, and its length is 120~$mm$.
The unvignetted field of the 1-m telescope is 40 $arcmin$.
The central clearance of the mounting flange of the polarimeter
has a diameter of
168~$mm$, corresponding to a slightly smaller unvignetted field of
38 $arcmin$. The positioning of the rectangular mirror with respect
to the telescope axis and the circular clearance of the mounting
flange defines a field of 18 $arcmin$ $\times$ 8 $arcmin$ for choosing
a guide star. 

A suitable X-Y stage which can hold an intensified CCD is yet to be
finalized.

\subsection {Glan-Taylor prism}

In order to measure the degree of polarization of the incident
light accurately it is essential that
the instrument should not cause any depolarization;
it should have an efficiency of 100 per cent as a polarizer, because
a good analyzer has to be a good polarizer.
The polarization
efficiency of the instrument should be frequently
measured to look for any malfunctioning, and
the availability of a provision for doing so at the
telescope is highly desirable. A fully plane polarized
beam that is necessary for this purpose can be obtained
by inserting the Glan-Taylor prism in the path of the
starlight. The prism is made of calcite, and its entrance and exit faces
are normal to the direction of light path. Its angle has been cut
such that the O-ray is internally reflected and absorbed by the  black
mounting material within the prism housing. 
The two halves of the prism are separated by an airspace for greater
ultraviolet
transmission. The usable wavelength range of the prism is 300 to 2700~$nm$.
The extinction ratio is less than $10^{-5}$ for the undeviated ray.
The polarization effect is maintained in a field of view of 13 to
7.5 degree, but it is symmetrical about the normal to the incident surface
only at one wavelength. The symmetric field of view decreases with increase
in wavelength; at 1000~$nm$ the full field of view is about 4.5 degree,
and hence, the prism is suitable for an F/13 beam where
the convergence angle is 4.4 degree.
The clear aperture of the prism used is 24.5~$mm$ and its length is 25~$mm$.
The deviation of the
extraordinary beam on emergence from its initial direction is less than
2 $arcmin$. The lateral shift caused in the image will be less than
15~$\mu m$, and a re-centring of the star image inside the
diaphragm may not be
necessary  after the introduction of the prism in the light path.
Because of its fairly large length, when the prism is inserted in the
light path a re-focusing of the image will be needed. The image plane will
be shifted down by about 8.2-mm from its normal position when the prism
is introduced.  Among the polarizing prisms that are usually employed,
Glan-Taylor prisms have the lowest length to aperture ratio ($\sim$ 0.85).

\subsection {Signal modulation}

The intensity of light at different position angles,
which is needed for the
determination of the degree of polarization, may be measured
either by rotating the analyzer,
or by rotating the entire instrument about the telescope axis. Both the
procedures are cumbersome and involve a large amount of overhead time at
the expense of the actual observational time. The star image will have to be
centred inside the diaphragm at each position of
the instrument or the analyzer during the rotation
as it is extremely difficult to align the axis of the instrument
or the analyzer with the respective
rotational axes. When the analyzer is rotated the plane of polarization of 
light incident on the photocathode changes. Since the sensitivities
of a cathode for different
planes of polarization are not same, the rotation of the analyzer would 
either cause spurious polarization, or increase the error in the measured
polarization unless a depolarizer efficient over the entire wavelength region
of observation is inserted after the analyzer. These problems can be
avoided if a rotating half-wave plate is introduced in the light path
for modulating the starlight. If 
the half-wave plate is rotated by an angle $\psi$, 
most of the disturbing instrumental effects, in particular,
those caused by image motion on the photocathode will have
modulations with
$\psi$ and 2$\psi$ angles, while the linear polarization will have a
modulation of 4$\psi$. This effectively reduces the risk of spurious
polarization caused by the image motion. However, there will be an increase
in the error of measurement depending on the amplitudes
of modulation with angles $\psi$ and 2$\psi$.
The retarder rotated should be extremely plane
parallel and the rotational axis should be exactly aligned with the
normal to its light-incident face. 

The modulating half-wave plate should be placed in front of
the optical components
which isolate spectral regions for observations because the latter usually
polarize light; it is ideal to have the half-wave plate as the first
element in the optical path inside the polarimeter.

\begin{figure}[htb]
\centerline{\psfig{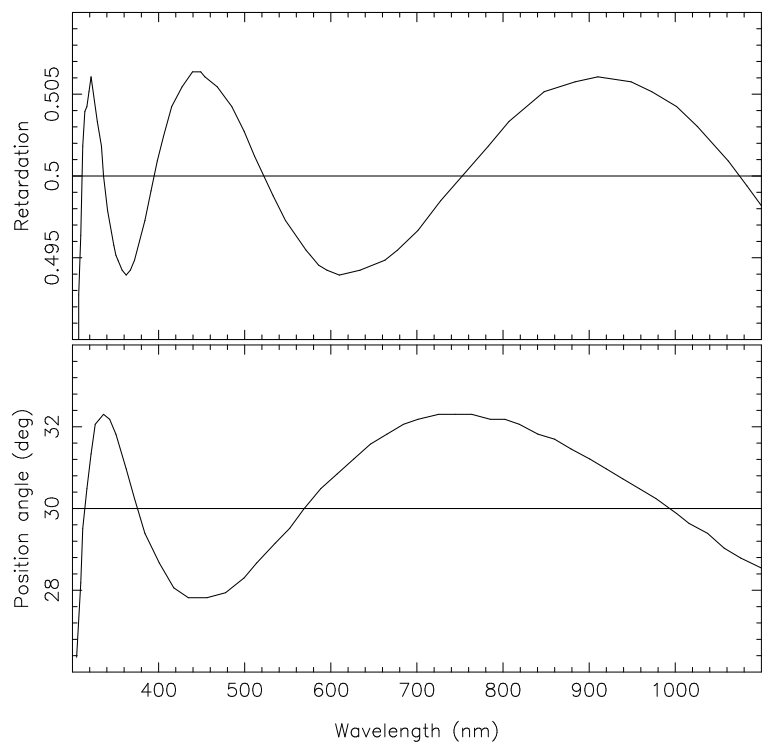}}
\caption {Calculated position angle of the effective optical
axis and retardation for a superachromatic half-wave plate.}
\label{f:hwpret}
\end{figure}

A half-wave retarder made of a single plate will act as the same only
at a particular wavelength and on either side of this the retardance
continuously changes. If we have to cover a wide spectral range, the retardance
should change as little as possible over the entire range.
Achromatic retarders
are produced by combining two plates of different birefringent materials.
For good achromatism the surfaces should be flat to $\lambda$/10 and plane
parallel to about 1 $arcsec$. The pair of magnesium fluoride and quartz
provides a good combination and the retardance of an achromatic 
half-wave plate made out of these materials
does not deviate more than 45 degree from 180 degree
over the spectral range 300--1000~$nm$. These materials are transparent
over a wide spectral range and are hard, and hence, easy to polish with
a high precision. A very high degree of achromatism,
super-achromatism, over the above spectral range can be obtained by using
a Pancharatnam retarder which consists of three achromatic retarders,
each made by a combination of quartz and magnesium fluoride and operating
in a particular wavelength region. Such a retarder was used for the
first time in an astronomical polarimeter by Frecker \& Serkowski (1976).
The achromatic retarders, which form the superachromatic combination,
are cemented  to each other.
The calculated path difference for such a superachromatic
half-wave plate is plotted in the top panel of Figure~\ref{f:hwpret}. 
In the spectral
range 310--1100~$nm$ the path difference (R) lies within $\pm$1.3 per cent
of $\lambda$/2, and during the manufacture it is possible to achieve the
theoretical retardation to an accuracy of about $\pm$3 per cent. The error
introduced by this in the measurement of linear polarization
will be almost non-existent (see section~\ref{ss:rotmod}).

The main drawback of the Pancharatnam retarders is the wavelength
dependence of position angle of their effective optical axis as seen in
the bottom panel of Figure~\ref{f:hwpret}.
  Over the spectral region 310 to 1100~$nm$
the orientation of the expected effective
optical axis on the surface rotates by about
$\pm$2.5 degree; the actual range could be slightly larger than the theoretical
value. This would cause difficulties in the accurate determination of the
position angle of polarization when broad spectral bands, as in the
present case, are used for observations because corrections that are to be 
applied to the position angles would depend on the energy distribution
of the observed objects. Such an inconvenience can, fortunately, be avoided
by introducing another stationary identical Pancharatnam plate 
immediately in the light path. The introduction of such a half-wave plate
ensures that the signal modulation is only a function of the relative 
positions of the effective optical axis of the two half-wave plates.
The stationary half-wave plate is mounted with its
effective optical axis approximately parallel to the principal
plane of the analyzer, which in this case is the calcite block mounted
just below this. This is to reduce the contributions to the
modulation of intensity
of transmitted light from terms that depend on the sine of position
angle of optical axis.

For a birefringent material the reflection coefficient even for normal
incidence depends on the position angle of the plane of vibration, and
hence unpolarized light may become polarized after passing through a retarder.
To reduce the polarization by refraction, the six plates of the
superachromatic half-wave plates are cemented with cover plates made from
fused silica Suprasil, which is an isotropic material, and the outer
surfaces are single-reflection coated. The two identical
half-wave plates of 19-$mm$ clear aperture
used in the polarimeter are acquired
from Bernhard Halle Nachfl. GmbH, Berlin, Germany.

The half-wave plate is rotated using a stepper motor of Model No. MO61-FC02,
procured from Superior Electric. It makes one rotation in 400 half-steps, and
is coupled to the halfwave plate through a 1:1 anti-backlash spur gear system.

\subsection {Calcite block}

The role of the analyzer is performed by the calcite block introduced
in the light path immediately after the stationary half-wave plate.
The optical components, dichroic filters and diffraction grating, that are
commonly used to isolate spectral regions in multichannel
instruments change the state
of polarization of the incident light, and hence,
the analyzer should be placed in front of all such components. 

\begin{figure}[htb]
\centerline{\psfig{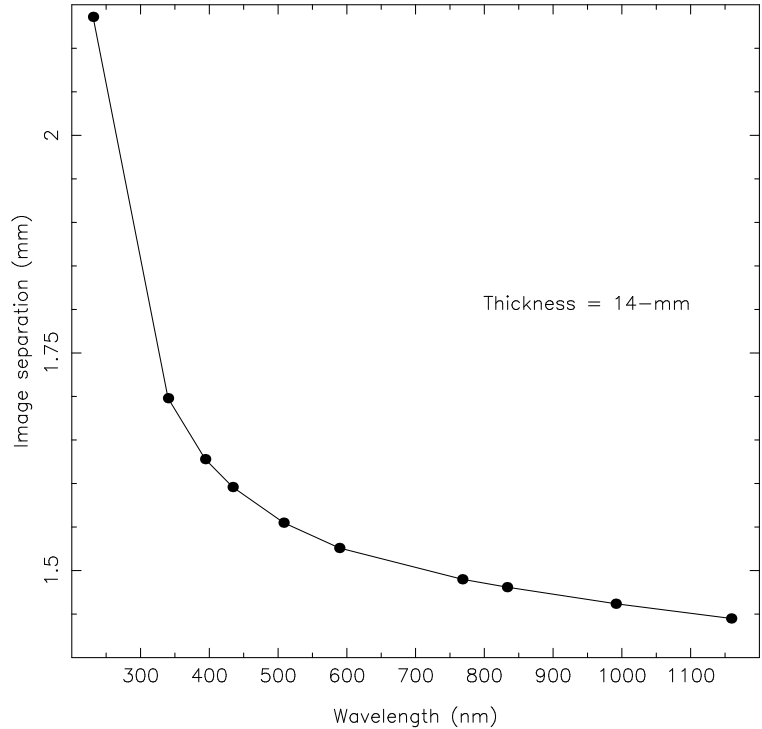}}
\caption {Separation between the ordinary and extra-ordinary images
produced by a single calcite plate.}
\label{f:imgsep}
\end{figure}

As mentioned already the light beam splits into two at the incident
surface of a beam-displacement prism. If the optical axis is parallel
to the incident ray both beams travel with the same velocity inside
the prism and if it is perpendicular they travel along the same
direction with an ever increasing phase difference between them;
in both cases the
emergent beams produce a single image. When the optical axis
makes any other angle with the incident ray the two beams travel in
different directions and produce two spatially separated beams on emerging
from the crystal. The maximum separation between the emergent
beams occurs when the optical axis
is inclined at an angle of 45 degree with the incident beam. Since the
light beam is incident normal to its face, a beam displacement
prism is cut with its optical axis making an angle of
45 degree with the incident face. The Huygens wave front of the ordinary
vibrations is spherical while that of the extraordinary vibrations is
an ellipsoid of revolution about the optical axis, with the major axis
inclined at an angle of 45 degree to the incident beam, and this causes
the splitting of the incident beam. If  $\theta$ is the deviation of
the extraordinary beam from the ordinary beam then
$$ \tan\theta = \frac{n_e^2 - n_o^2}{n_e^2 + n_o^2} \, , $$ where
$n_e$ and $n_o$ are the refractive indices of the extraordinary and
ordinary rays. If $t$ is
the thickness of the prism, the separation $d$ of the two beams on emerging 
from it will be
$$ d = t\, \tan\theta. $$
It is clear from the above two relations that the separation
between the images depends on the difference
in the refractive indices of the ordinary and extraordinary rays. Of the
commonly available crystals, quartz, magnesium fluoride, sapphire,
calcite, KDP and lithium niobate,
calcite has  the largest difference between $n_e$ and $n_o$, and hence,
for a given thickness a calcite plate
gives the largest separation between the beams.
Calcite is a negative crystal; at $\lambda$ = 589~$nm$,
$n_o$ = 1.65835, $n_e - n_o$ = $-$0.17195, and
the deviation of the extraordinary beam from the ordinary beam,
$\theta$ = 6.22 degree.
Since the difference in the refractive indices of the two beams is a function
of wavelength, the spatial separation of the two beams also will be 
a function of wavelength. The separation as a function of wavelength
expected for a calcite
plate of 14~$mm$ thickness is plotted in Figure~\ref{f:imgsep}; the 
refractive indices needed for the computation are taken from Levi (1980).

The introduction of the calcite plate in the converging beam will
shift the focal plane away from it. If $t$ is the thickness and $n$ the
refractive index, the shift produced in the focal plane is given by
$$ \Delta F \approx  \frac{t}{n} \, (n-1). $$
With $t$ = 28~$mm$, at $\lambda$ = 589~$nm$, for the ordinary beam
$\Delta F$ $\approx$ 11~$mm$.
Because of the variations in the refractive
index with wavelength even the ordinary image will show
chromatic aberration.
The extraordinary image will show astigmatism additionally because
the corresponding beam propagates obliquely through the plate
at an angle of about 6 degree with respect to the normal to the surface.
The predominant aberrations shown by the image will be very similar in
magnitude to those that would be produced by a beam of light
passing through a glass plate at an angle same as above (Brand 1971).
Consequently, the focal planes, planes containing the circle of
least confusion, for
the two beams will not be the same. Further, as a result of the 
Fresnel reflection
losses from the surface being different for the two orthogonal planes
of polarization the two beams will have slightly different intensities
even for unpolarized objects.
Both these inconvenient features of a single calcite plate can be avoided
if two cemented plane parallel plates with their principal planes
crossed at right angles are used (Serkowski 1974a);
the light which is ordinary in the first
plate becomes extraordinary, and vice versa in the second.
Hence, both the images of a star produced by a calcite block
with two crossed plates will be affected similarly
and the images will have circles of least confusion in
the same plane. This will enable one to obtain the same degree
of focus for both the images inside the corresponding apertures.
\begin{table}[htb]
\begin{center}
\caption{Parameters of the images formed by the
calcite block.}\label{t:parimg}
\medskip
\begin{tabular}{ll}
\hline
&\\
Parameter & Value ($mm$) \\
&\\
\hline
&\\
Single calcite plate: & \\
Separation of images at 320~$nm$  & 1.780 \\
Separation of images at 990~$nm$  & 1.462 \\
Length of the image strip & 0.318\\
Mean offset of the image centre from the initial direction & 1.621\\
&\\
Two crossed calcite plates: & \\
Centre-to-centre distance of the two images & 2.292\\
Offset of the centre of the images from the initial direction$^*$ & 1.146\\
&\\
\hline
\multicolumn{2}{l}{\footnotesize{* The calcite block is rotated such that the 
offset is towards the observer}}
\end{tabular}
\end{center}
\end{table}

We have used a 20-$mm$ clear aperture calcite block made of two
cross-mounted plates of each 14-$mm$ thickness; this is also
acquired from Bernhard Halle Nachfl. GmbH. The transmission of 
such a block over the spectral region 300--1000~$nm$ is close to
90 per cent (McCarthy 1967).
The computed parameters of the images are given in Table~\ref{t:parimg}.
The calcite block is mounted such that the line joining the centres
of the images is parallel to the front side of the polarimeter. It 
can be removed from the light path, if needed,
to use the instrument as a three channel
photometer with or without the star-sky chopping facility.  

\subsection {Diaphragms and image chopper}

The maximum back focal length from the last mounting plate with the
position angle device of the 1-m Carl Zeiss telescope at
Kavalur is 324$~mm$. The diaphragms are placed at a distance 
of 295~$mm$ from the top of the mounting flange of the polarimeter, which is
well within the allowed range.
The details of the diaphragms that are available are given in
 Table~\ref{t:detaper}.
The disc on which these diaphragms are mounted is made to protrude
slightly outside the instrument-body to facilitate its rotation
manually
for choosing the required set of twin diaphragms for observation.
The centres of the twin diaphragms are aligned along the radial
direction of the disc.
They should always be accurately positioned
at the same locations with respect to the axis of the instrument
to avoid any change either in the angle of incidence on the
photocathode or in the region of illumination on its surface by the
two images.
A possible light leak through the slot made on the side plate of
the instrument is arrested by using a cap to the protruding portion.
It is seen from from Figure~\ref{f:exclint}
that the excluded energy is nearly flat
beyond a radius of about 7 $arcsec$, and hence, with the 1.3-$mm$ diaphragm,
which corresponds to an angular diameter of 20 $arcsec$,
the excluded light is less than 0.5 per cent.

The 1.3-$mm$ single diaphragm is included for the purpose
of using the instrument as a conventional photometer
without star-sky chopping or as a polarimeter in single image mode.
The 4.5-$mm$ single diaphragm is included for the purpose of
field-viewing to make centring of the images inside the smaller
diaphragms for observation easy.
It is highly desirable if the metal focal plane diaphragms that are used
currently are replaced with
those made of a non-metallic material.
Serkowski (1974a) has reported that when
the star image was on the edge of a metallic diaphragm such that
only about half of the light entered the polarimeter a linear polarization
over 0.2 per cent with the position angle parallel to the edge of the diaphragm
was found. Hence, if the star image drifts during the observation it could 
increase the errors in the polarization measurements
significantly, if metallic apertures are used. 

The twin diaphragms need not be of equal sizes if the instrument is
used only in the double image mode polarimetry because the sky
intensity in each beam is observed and removed separately. But, it
is better to have identical
diaphragms if the instrument has to be used
in the star-sky chopping photometric mode; if they are of
different sizes their relative sizes should be known
exactly for the removal of the sky background.

\begin{table}[htb]
\begin{center}
\caption{Details of the apertures available in the instrument.}
\medskip
\label{t:detaper}
\begin{tabular}{lll}
\hline
&&\\
Linear Diameter & Angular diameter & Nature\\
($mm$)  & ($arcsec$)&\\
&&\\
\hline
&&\\
1.3 & 20 & Twin\\
1.6 & 25 & Twin\\
1.3 & 20& single\\
4.5 & 70 & single\\
\hline
\end{tabular}
\end{center}
\end{table}

The annular opening on the chopper disc,
which is meant for the alternate blocking of one
of the images, has a width of 4~$mm$, and has been divided into four
segments, two alternate ones for each image. The diameter of the disc
is 158~$mm$ and that of the common
boundary of the slots that isolate the images
is 142~$mm$. The counting of the photomultiplier pulses should begin only
when the corresponding diaphragm is fully open, which means that the
counting should commence only when the chopper slot has advanced by the 
diaphragm size and the counting should end before the chopper starts
obstructing the diaphragm. Further,
the position sensors are about 2~$mm$ in size. The size of
the slots in the chopper during the rotation of which 
the pulse counting is done
subtends an angle of 82 degree at the centre;
the 8-degree blind stretch 
ensures that counting is done only
when the diaphragms are not blocked by the
chopper. The
observations can be made over close to 91 per cent of the rotation of the disc;
probably, this can be increased marginally by reducing the blind sector a 
little.  With a larger chopper disc the fractional size of the
blind stretch can be reduced; but,
because the alternate slots have a common boundary 
the blind stretch cannot be made very small either.
The size of the body of the instrument would increase, and hence,
also its weight, if we have to accommodate a larger
disc. Even with the present size for the chopper disc a portion
of it protrudes outside the body of the instrument; the protrusion
is covered with a cap to avoid any accidental contact with the chopper
while in rotation and to prevent the light leak through the slot 
on the side plate through which it protrudes.
The top surface of the chopper disc is separated
from that of the 
image-isolating apertures by slightly more than 3~$mm$. The centres of the
apertures are horizontally at 1.146~$mm$ from the common boundary of the
chopping slots. This corresponds to an angular distance of about
18 $arcsec$
from the axis of the telescope, and the fractional excluded energy
by a diaphragm of this radius at the focal plane is about 0.2 per cent.
When it is less than about 1 per cent, the excluded
energy is not a sensitive function of the vertical distance from the
focal plane for a fixed horizontal distance from the telescope axis
(Young 1970). Therefore, the chopper does not increase the
excluded energy even when the 1.6~$mm$ twins apertures are used.

The chopper disc is rotated directly
by using a servo motor with the Model No. 
Smartmotor\textsuperscript{\textregistered} SM2315 procured from Animatics.
The whole chopper unit can be
finely aligned so that the common boundary of
the slots that isolate the images
bisects the line joining the centres of the twin apertures.

The line joining the centres of twin diaphragms
is made parallel to the front side of the
polarimeter. The diaphragm and the chopper discs are mounted
on the two opposite side plates of the instrument.

The diaphragm viewing arrangement consists of a pair of achromats
of focal length 93~$mm$ and aperture 30~$mm$ for re-imaging,
and a Meade 3-element modified achromatic eyepiece of
12-$mm$ focal length. Both spherical and chromatic aberrations are
reduced considerably by using twin achromats with similar elements
facing each other while re-imaging. The eyepiece has a field of about 
40 degree and a field stop of about 8~$mm$. The starlight is brought into
the field of view of the achromats by flipping a flat mirror
of size 40~$mm$ $\times$ 40~$mm$ to
a 45 degree position. If the 4.5-$mm$ aperture has to viewed completely,
the image formed by the achromats at the focal plane of the
eyepiece should be less than its field stop.
This requires that the magnification produced by the achromats
should be less
than 1.8. With a distance of 75~$mm$ between the apertures and the first
achromat this can be achieved. The image formed by the
second achromat will then be at a convenient distance of about 125~$mm$
from it.

\subsection {Filters and detectors}

\begin{figure}[htb]
\centerline {\psfig{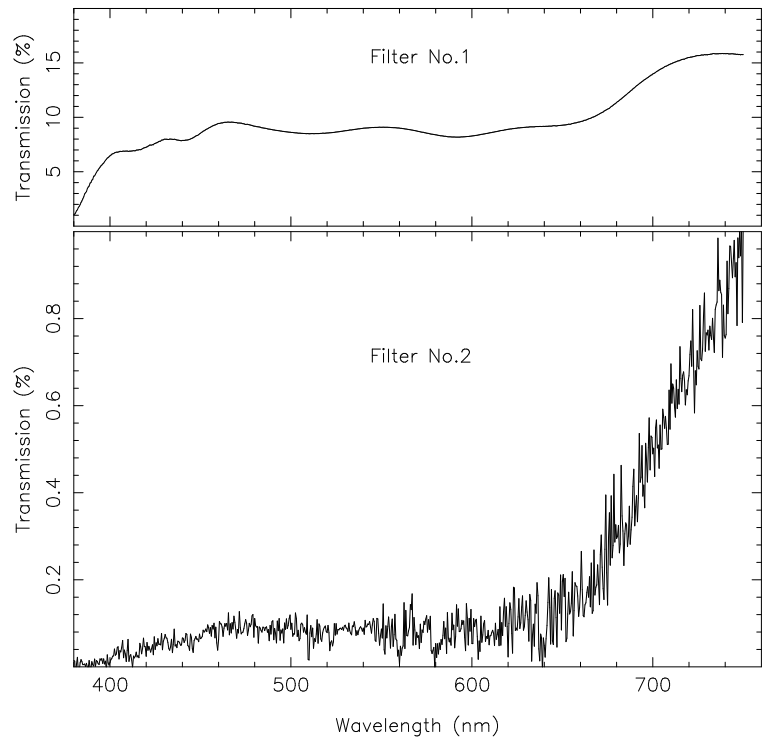}}
\caption {Wavelength dependence of the neutral density filters
available.}\label{f:ndfilter}
\end{figure}

If the objects to be observed could cause count rates in excess
of their maximum values allowed by the upper limits on the
anode currents, the incident
light intensities should be cut down to avoid any damage to the
photomultiplier tubes. For this purpose
two neutral density filters of optical densities 1.0 and 2.0 are made
available.
Since the optical density is the base ten logarithm of
opacity, which is simply the
reciprocal of the transmittance, the above densities translate
to 2.5 and 5.0 magnitudes, respectively. Both filters are of absorptive
type, and hence, do not produce any
scattered light when inserted in the light beam.
The wavelength dependencies of these filters are plotted in 
Figure~\ref{f:ndfilter}
which shows that these filters are not strictly neutral; the
changes in the mean wavelengths of the spectral bands
produced by these filters on inserting in the light path should
be taken into account when the effective wavelengths of observation
are computed, or when standard magnitudes of the programme
objects are derived. These
30-$mm$ diameter circular filters are mounted on a sliding holder
having an additional clear hole that allows the light to pass through
unattenuated while observing objects which are within safe brightness
limits.
These neutral density filters would also be useful to cut down the brightness
of objects by a known factor when observations are made to determine
the dead-time coefficients of the pulse counting setup.

\begin{figure}[htb]
\centerline {\psfig{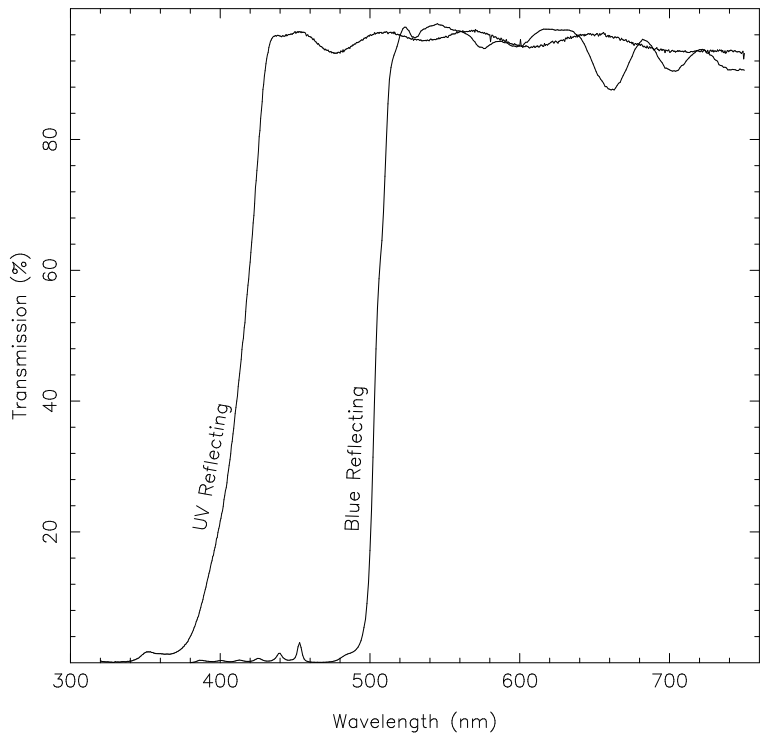}}
\caption {Wavelength characteristics of the dichroic filters
used.} \label{f:dichroic}
\end{figure}

\begin{figure}[htb]
\centerline {\psfig{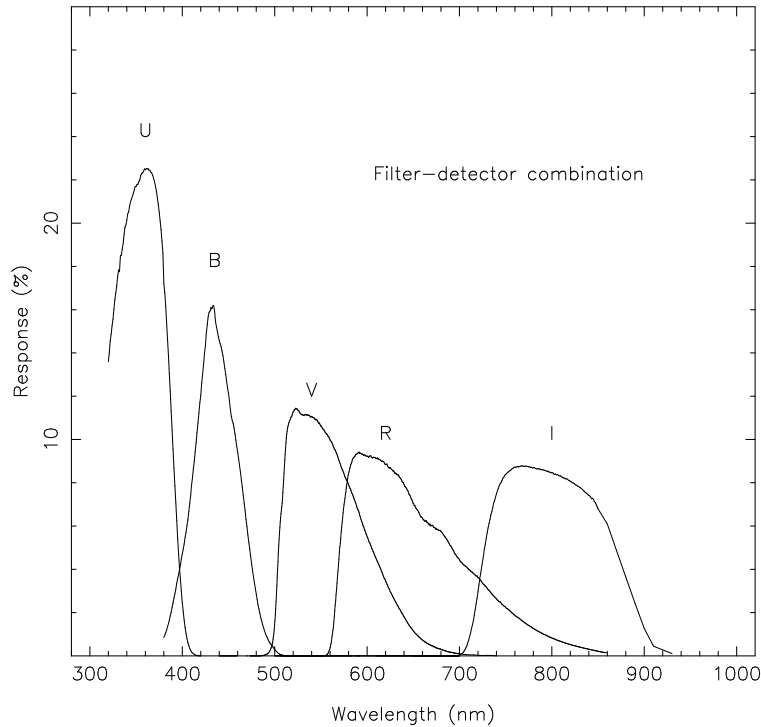}}
\caption {Combined spectral responses of the filter-detector combination.}
\label{f:filterdet}
\end{figure}

The isolation of the spectral regions into the three channels
of the polarimeter is achieved 
by using two dichroic filters, one to reflect the ultra-violet
part of the
incoming light and the other to reflect
the blue part of the spectrum from the light transmitted by the first. 
The filters are mounted such that the two beams are reflected
in mutually perpendicular directions,
with the $U$ band on the right side of the instrument and the $B$ 
band behind;
in this configuration the instrument becomes more compact.
The dichroic filters used are obtained from Custom Scientific, Inc.,
Arizona, USA.
The wavelength characteristics of these filters are given in 
Figure~\ref{f:dichroic}.
The reflective coatings required for the dichroism
are done on 2-$mm$ thick glass substrates.
The calcite block and dichroic filters are mounted such that
the vibrations of the two mutually
perpendicularly polarized beams make an angle of 45 degree with
the plane of incidence on the dichroic filters.
This ensures that the reflectivity of both
beams are the same, and hence, does not introduce any differential
change in the intensities of the beams.
\begin{figure}[htb]
\centerline {\psfig{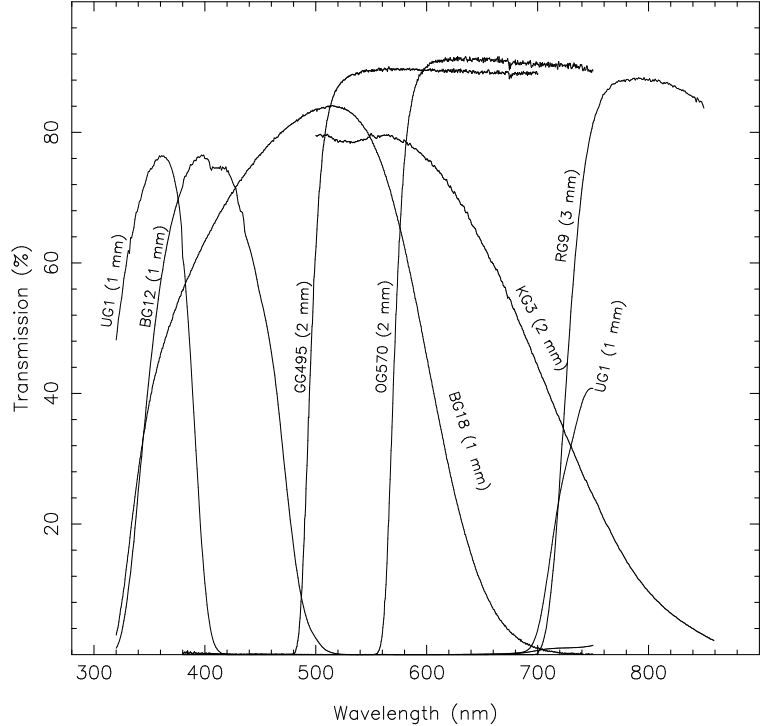}}
\caption {Transmission characteristics of the Schott glass
filters used.}\label{f:glassfil}
\end{figure}
\begin{figure}[htb]
\centerline {\psfig{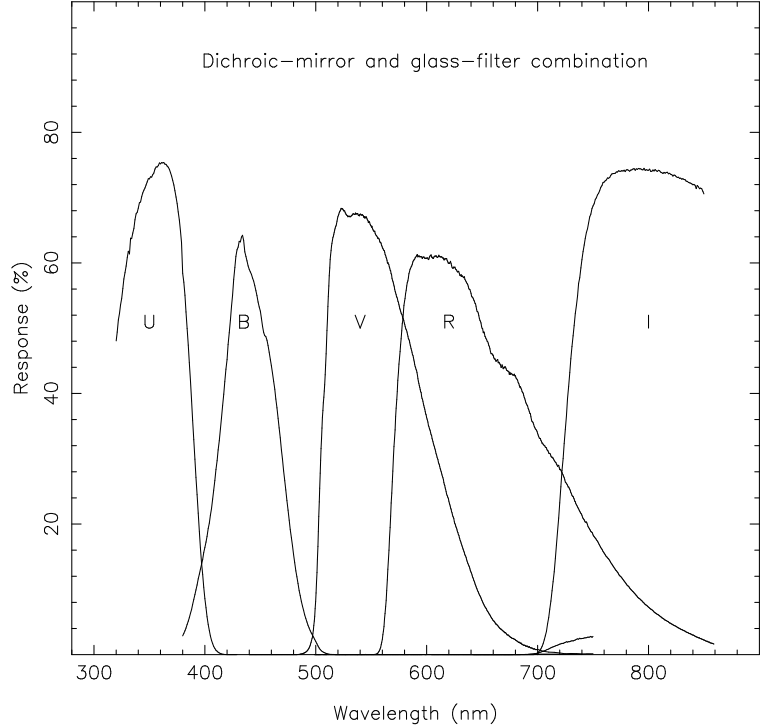}}
\caption {Combined spectral responses
 of the dichroic mirrors and glass filters.}
\label{f:dichroglass}
\end{figure}
The two reflected beams are detected by two separate uncooled 
photomultiplier tubes.
The bi-alkali photocathodes of these tubes along with
the Schott glass filters 
inserted in front of them produce
spectral bands that approximate the $U$ and $B$ bands of Johnson.
The light transmitted by both the dichroic filters, which
fall in the $VRI$ spectral bands, is detected by a
cooled photomultiplier tube with a GaAs photocathode. The observations are
made sequentially in $V$, $R$ and $I$ bands using suitable filter combinations
mounted on a sliding filter holder. These filters have clear apertures of 
25-$mm$ diameter. The transmission characteristics of the Schott glass
filters that are used are given in 
Figure~\ref{f:glassfil}. The combined responses of the
dichroic mirrors and the glass filters are plotted in
Figure~\ref{f:dichroglass} and the
final band passes that include the detector quantum efficiencies are
plotted in Figure~\ref{f:filterdet}. 
The $R$ and $I$ passbands approximate that of Cousins
(Bessel 1979, 1993),
while the $V$ passband approximates
that of Johnson. The selection of required glass filters are based
on the filter-detector combinations used by Piirola (1988) and Hough,
Peacock \& Bailey (1991); these authors also have employed dichroic
filters to isolate spectral regions in their multiband polarimeters.
The details of the filter-detector combinations used are given in
Table~\ref{t:filterdet}. The mean wavelength, which is defined as
$$ \lambda_0 = \frac{\int \lambda\,S(\lambda )\,\delta\lambda}
{\int S(\lambda )\,\delta\lambda},$$ calculated
from the data is also given in the table against the corresponding
spectral band; S($\lambda$) is the wavelength-dependent
response plotted in Figure~\ref{f:filterdet}.
\begin{table}[htb]
\begin{center}
\caption{Filter-detector combinations used in the instrument.}
\label{t:filterdet}
\medskip
\begin{tabular}{llll}
\hline
&&&\\
Spectral & Filter combinations & Mean wavelength & Photomultiplier tube\\
band & & $\lambda_0$\,($nm$) & \\
&&&\\
\hline
&&&\\
$U$  & UG1 (1~$mm$) & 357 & ETL 9893Q/350B\\
$B$  & BG12 (1~$mm$) & 437 & ETL 9893Q/350B\\
$V$  & BG18 (1~$mm$) + & 561 & Hamamatsu R943-02\\
& GG495 (2~$mm$) &   &\\
$R$  & OG570 (2~$mm$) + & 652 & $_{''}$\\
& KG3 (2~$mm$) &&\\
$I$  & RG9 (3~$mm$) & 801 & $_{''}$\\
&&&\\
\hline
\end{tabular}
\end{center}
\end{table}

\subsection {Fabry lenses}

The photocathode should be located
exactly at the exit pupil of the Fabry lens
so that 
the image of the primary mirror on the cathode will be in good focus. If not,
the two stellar images formed by the calcite block
will not fall at the same spot on the cathode, and
this will produce an
appreciable difference in the corresponding signals from
the photomultiplier because the sensitivity across the photocathode surface
is not uniform. The two images of the primary on the cathode, in general,
will not have the same brightnesses across them because the two beams
take different paths through the various optical elements, and the
surface inhomogeneities in the transmittance and reflectivity of these
elements modify them differently; this will give rise to a
difference in the signals even if the images fall at the same spot,
the ratio of the net responses, however, will be a constant.
The atmospheric scintillation also produces a distribution of brightness
across the images, but the distributions will be the same since the air
is birefringent.
If the cathode is kept at a distance
$\Delta F$ from the focus, the image-spots on the cathode
will be separated by a distance $d$
given by  (Serkowski \& Chojnacki 1969)
$$ d = \frac {x \, \Delta F}{F}, $$ where $x$
is the linear separation between
the images at the focal plane and $F$, the focal length of the Fabry lens.
If $F_{ratio}$ is the focal ratio of the telescope,
the size $s$ of the image is given by
$$ s = \frac {F}{F_{ratio}}. $$ 
If $r$ is the ratio of the separation of images to the image size, then
$$ r = \frac {x \, \Delta F}{s^2 F_{ratio}},$$
indicating that a larger image of the primary would give a smaller value
for $r$ and therefore a smaller effect due to an error in focusing.
The maximum sizes that the images can be of are decided by the effective
sizes of the cathodes. The cathodes of ETL~9893 and Hamamatsu~R943-02 are
of 9~$mm$$\times$ 9~$mm$ and 10~$mm$ $\times$ 10~$mm$, respectively.
 Normally, an image of diameter
about half the size of the cathode is used. 

Plano-convex lenses of 75-$mm$ focal length at 5876\AA \,
are used in all the three
channels. The lenses are made of synthetic fused silica, which is far purer
than fused quartz. The increased purity makes fused silica to have excellent
transmission from ultra-violet to infrared;
in the optical region its transmission is about 5 per cent better than
that of BK7. The focal length $F$ and radius of curvature
$R_c$ of a plano-convex
lens are connected according to the equation
$$ \frac{1}{F} = \frac{\mu - 1}{R_c}, $$
where $\mu$ is the refractive index. Because of the wavelength-dependence
of the refractive index the focal length, and hence, the
size of the image formed on the cathode
varies with wavelength. Table~\ref{t:flfabry} gives the focal
lengths and sizes of the images at a few selected wavelengths. The fractional
size of the image with respect to the size of the cathode is also given
in the table.
\begin{table}[htb]
\begin{center}
\caption{Focal lengths of the Fabry and the sizes of the image formed.}
\medskip
\label{t:flfabry}
\begin{tabular}{lllll}
\hline
&&&&\\
Wavelength & Refractive index & Focal length &
 Image size & Fractional size of\\
 ($nm$)  & $\mu$ &($mm$) & ($mm$) & the image (per cent) \\
&&&&\\
\hline
&&&&\\
351.1 & 1.47671 & 72.1 & 5.5 & 61\\
457.9 & 1.46498 & 73.9 & 5.7 & 63\\
546.1 & 1.46008 & 74.7 & 5.7 & 57\\
587.6 & 1.45846 & 75.0 & 5.8 & 58\\
643.8 & 1.45637 & 75.3 & 5.8 & 58\\
786.0 & 1.45356 & 75.8 & 5.8 & 58\\
&&&\\
\hline
\end{tabular}
\end{center}
\end{table}
The Fabry lenses used in the $U$ and $B$ channels can be focused properly.
Since the same Fabry is used in the $VRI$ channel it is impossible to get a
good focus in all the three bands simultaneously.
Assuming that a good focus is obtained for the $R$ band,
then the cathode will be away from the focal plane for the $V$ and $I$ bands
by $\pm$0.6~$mm$. With a separation of 2.292~$mm$ between the centres of the 
two images at the focal plane of the telescope, the above corresponds to
a separation of 18~$\mu$m between the centres of the
corresponding image-spots on the
cathode. The resulting ratio of the separation of image-spots
to their size, $r \approx$ 0.3 per cent.
To overcome the adverse effect of nonuniform sensitivity over the 
cathode surfaces the practical rule is that the relative shifts in the
image of the primary mirror formed by the Fabry lens on the detector
should not exceed 0.01 per cent of the diameter of the image (Serkowski 1974b).
The relative separation of the images will give rise to a small
difference
in the output signal, but with a constant gain-ratio.
 This will not be a serious problem because the
ratio of overall throughputs for the two beams can be treated as an
unknown in the data reduction procedure.
The images should remain at the same
positions with respect to the cathode over a full rotation of the
half-wave plate, which means that we should
be able to keep the image drift inside the diaphragm below
2 $arcsec$ which corresponds to a value of $\sim$ 0.01 per cent for
the ratio of drift in image-spots to their sizes, by accurately
guiding the object.
If the Fabry is in good focus, which can be
achieved in principle for the $U$ and $B$ channels, then a drift of the
images on the detector surface will not produce any significant effects
in the results since the fast chopping rate between the images
essentially ensures that
the effective sensitivity is the same for both the beams.
The problems associated with the nonuniform sensitivity of the cathode
can be minimized if the images are formed at the most sensitive
region of the cathode where the gradient of sensitivity is the least
(Young et al. 1991).
Since the position of maximum sensitivity
does not usually coincide with the
geometrical centre of the cathode such an alignment can be made
only with extra difficulties.

The actual sizes of the images of the primary mirror formed on
the cathodes will be slightly more than those given Table~\ref{t:flfabry}
because
of spherical aberration. All the aberrations depend on the way a
lens is used.
The plano-convex Fabry lenses are mounted with their curved surfaces
facing the primary mirror to reduce spherical aberration.

The longitudinal spherical aberration $a_l$, which is the distance along
the optical axis between the intercepts of the paraxial and marginal
rays, for a plano-convex lens of focal length $F$ with its convex
surface facing the infinite conjugate is given by (Subrahmanyam 1980)
$$ a_l= \frac{\rho^2\,(\mu^3 - 2\,\mu^2 + 2)}
{2\,F\,\mu\,(\mu - 1)^2}, $$
where $\rho$ is the distance between the marginal rays and the optical
axis, and $\mu$, the refractive index of the lens material. The distance
from the axis at which the marginal rays meet the paraxial focal plane
is the transverse spherical aberration, $a_t$, and it is given by
$$ a_t = \frac{a_l\,\rho}{F - a_l}. $$
It is this aberration which increases the spot-size on the photocathode.
The focal ratio at which the lens is operated is 2$\rho /F$, and it
is clear from the above relations that the transverse spherical aberration
increases with the focal ratio of the beam.

The lenses are about 220~$mm$ from the focal plane of the telescope,
and hence, are operated approximately at $F$/3.6. At this
$F$-ratio for the beam the increase in the
image diameter due to transverse spherical aberration is about 0.5~$mm$.

\subsection {PMT housings}

As indicated earlier ETL 9893Q/350 photomultiplier tubes are
used in the $U$ and $B$
channels; their quartz windows give extended ultra-violet
response. These tubes have bi-alkali photocathodes with small effective
diameters, and therefore produce very low dark counts; according to the
manufacturer's specification, at an ambient temperature of 20\degr~C
the tubes give around
40~counts~s$^{-1}$ when the cathode to anode voltage is 2250~V, and
on cooling the counts will be reduced only by a factor of about two.
Therefore, uncooled
ambient housings (ETL B2F-type), which are used for the above channels
mainly to reduce the size and total weight of the instrument,
are sufficient for the purpose.

The Hamamatsu R943-02 tube used in the $VRI$ channel has a
GaAs photocathode.
At ambient temperatures these tubes give a few thousand dark counts per
second under typical operating voltages.
To reduce the dark counts we have used a FACT~50 model housing,
which is of forced air-cooled type and cools 50\degr C below ambient
air at 20\degr C with a cool down time of about 1.5 hours.

We have fitted all the three housings with individual
shutter assemblies so that
the housings can be dismantled from the main structure
of the polarimeter without removing the photomultiplier tubes.
These mechanical shutters can be kept open during the night because
an electronically operated shutter, common to all the three
photomultiplier tubes, is provided. This shutter
is immediately after the neutral density filters in the light path
and is operated during the observations.
The plane of the window of the photomultiplier tube
is set 10.2~$mm$ behind the top of the mounting
flange of the B2F-type housing. In the case of ETL~9893 tubes the
photocathode is just behind the window of the 
tube, and hence, there is sufficient
space to accommodate a PR305 shutter assembly acquired from Electron
Tubes Limited (ETL),
which is about 22~$mm$ thick, between the photocathodes and the Fabry
lenses. In the case of FACT~50 housing the window of photomultiplier
tube is set about
32~$mm$ behind the face of the mounting flange. The cathode of R943-02
tube is 19~$mm$ behind its window. Since there is no space to
accommodate a similar shutter in the $VRI$ beam, the mechanical assembly 
interfacing the housing to the polarimeter is designed to incorporate
a shutter to control the light reaching the photomultiplier.
An offset of 1.6~$mm$ of the photocathode centre from the axis of the
R943-02 photomultiplier tube is taken care of by offsetting the
centre of the bottom mounting plate of the instrument.
The photomultiplier tube base is rotated such that the offset is in the
right direction.

A well-focused Fabry lens while keeping the image-spot at the same
position on the cathode transforms the spatial changes at the focal
plane of telescope to changes in the incident angles on the cathode.
The cathode of R943-02 is inclined at an angle of 15 degree with respect
to the input window surface. The Fresnel reflection losses will be
different for the two beams if their incident angles are different.
The tube is mounted such that the axis about which
the cathode is tilted is parallel to the line joining the two
images. This makes the incident angles of both the beams on the
cathode the same. If the star images drift inside the diaphragm,
the incident angles may become different depending on the direction
of drift on the cathode surface and may cause differential variation
in the signal. This is true even for the ETL~9893 tubes where the
photocathodes are parallel to the input window, though the associated
problems will be less than that for the GaAs tube. To reduce the 
polarimetric errors it is necessary to keep the images at the
same position inside diaphragm during the full rotation of the
half-wave plate by accurate guiding.  

\section {Polarized light} \label{s:pollig}
The underlying principle in the measurement of polarization of a beam
of light is that its state of polarization can be altered in a desired
way by introducing appropriate optical elements in its path.
In the present case it is the half-wave plates--calcite
block combination
that does the required modification. In order to know how
a particular optical element affects the state of polarization, the 
most important requirement is a way of quantitative parametrization
of the state of polarization of a beam of light.
In the following subsections we briefly describe the parameters
that are used to characterize a polarized beam of light and the effects
of introducing the half-wave plates and calcite block
in the path on the state of polarization of the incident light beam;
the different possible states of polarization of a beam of light
are also briefly discussed in the subsections that follow.

\subsection{Stokes parameters}
The state of polarization of a beam
of light, particularly in scattering problems, is most conveniently
described using the Stokes
parameters (Van de Hulst 1957) which are defined as
$$ I = E_{0x}^2 + E_{0y}^2, $$
$$ Q = E_{0x}^2 - E_{0y}^2, $$
$$ U = 2\, E_{0x}\, E_{0y}\,\cos \delta \quad \mbox{and} $$
$$ V = 2\, E_{0x}\, E_{0y}\,\sin \delta , $$
where $E_{0x}$ and $E_{0y}$ are the amplitudes of the electric vector in two 
orthogonal planes $x-z$ and $y-z$, and $\delta$ is the phase difference
between the $x-$ and $y-$vibrations; the instantaneous values of
the electric vector are
$$ E_x = E_{0x}\, \cos \omega t \quad \mbox{and} \quad
 E_y = E_{0y}\, \cos (\omega t+\delta ).$$ 
We can define a vector $E_0$ whose amplitude is
$(E_{ox}^2 + E_{0y})^{\frac{1}{2}}$
and which makes an angle $\theta$ with the positive $x-$axis such that
$E_{0x} = E_0\, \cos \theta$ and $E_{0y} = E_0\, \sin \theta$.
Usually, the angle $\theta$
is called the position angle of polarization. $E_{0x}$ and $E_{oy}$
can take both positive and negative signs, defining the quadrant in
which $\theta$ lies.
The Stokes parameters can be re-written in terms of the position angle
of polarization and the phase difference between the two orthogonal
components of the electric vector as
$$ I = E_0^2, $$
$$ Q = I\,\cos 2\, \theta ,$$
$$ U = I\,\sin 2\, \theta \, \cos \delta  \quad \mbox{and}$$
$$V = I\,\sin 2\, \theta \, \sin \delta . $$

The most general form of polarization is elliptical polarization where
the end points of the instantaneous electric vectors always lie
on an ellipse; all the other
states of polarization are its special cases.
It is clear that when $\delta$ = 0, 
$$ E_y = \frac{E_{0y}}{E_{0x}}\, E_x ,$$
i.e., the vibrations lie in a plane which makes an angle $\theta$ with
the $x-y$ plane. Similarly, when $\delta$ = $\pm \pi$ also the light
will be plane-polarized. When the phase difference is
$\pm \frac{\pi}{2}$ the major and minor axes of the ellipse traced
by the instantaneous electric vectors coincide with the $x-$
and $y-$axis; if the magnitudes of $E_{0x}$ and $E_{0y}$ are equal
the light is said to be circularly polarized.

When $\delta$ has any
value other than those given above the resultant electric vector will be
tracing an ellipse in the $x-y$ plane with the major axis arbitrarily
inclined to the x-axis. If we denote ($B/A$) by
$\tan \beta$, where $A$ and $B$ are the semi-major and minor axes of the
ellipse, and $\chi$ the angle between the
major axis of the ellipse and the $x-$axis, then
\begin{equation}
 \tan 2\chi = \tan 2\theta \, \cos \delta \quad \mbox{and} \quad 
 \cos 2\theta = \cos 2\chi \, \cos 2\beta .
\label{e:pa}
\end{equation}
Eliminating $\theta$ and $\delta$ from the expressions for the Stokes
parameters, we have
$$ I = E_0^2, $$
$$ Q = I \cos 2\beta \cos 2\chi ,$$
$$ U = I \cos 2\beta \sin 2\chi \quad \mbox{and} $$
$$ V = I \sin 2\beta .$$
\subsection{Degree of polarization}
The state of partially elliptical polarization
is the most general form of light, and
it can be described in terms of a combination of
two independent components: (i) natural unpolarized light
of intensity $I$\,($1 - p_e$) and (ii) fully elliptically polarized light
of intensity $I\,p_e$, where $p_e$ is the degree of polarization. The
Stokes parameters can then be written as
$$ Q = I\,p_e \cos 2\beta \cos 2\chi ,$$
$$ U = I\,p_e \cos 2\beta \sin 2\chi \quad \mbox{and}$$
$$ V = I\,p_e \sin 2\beta ,$$
where $$ I\,p_e = (Q^2 + U^2 + V^2)^{\frac{1}{2}}. $$
The factor $p_e\,\cos 2\beta$ is called the degree of linear polarization.
Denoting this by $p$, we have
$$ Q = I\,p\,\cos 2\chi ,$$
$$ U = I\,p\,\sin 2\chi ,\quad \mbox{and}$$
$$ V = I\,p_v ,$$
where $p_v = p_e\,\sin 2\beta$ is the degree of ellipticity, positive
for right-handed elliptical polarization and negative for left-handed.

It is obvious from the above that for any beam of light the parameters
$Q$ and $U$ take on particular values according to the chosen reference
axis since $\chi$ depends on the axis chosen while the other two, $I$ and
$V$, remain independent. In other words the three quantities $I$, ($Q^2 +
U^2$) and $V$ are invariant under a rotation of the reference coordinate
system. For a celestial object the equatorial coordinate system
provides a convenient system of reference with the positive $x-$axis in the
direction of the north defined by the meridian through the object
and the positive $y-$axis in the direction of east, the direction of
positive $z-$axis being along the line of sight towards the observer. 

The degree of linear polarization and the angle which the major
axis of the resultant ellipse makes with the $x-$axis are given by
$$ p = \frac{(Q^2 + U^2)^{\frac{1}{2}}}{I}  \quad \mbox{and}$$
$$ \chi = \frac{1}{2}\,\tan^{-1}\frac{U}{Q}.$$
Since $\delta$ = 0 for plane-polarized light equation (\ref{e:pa}) gives
$\chi$ = $\theta$, the position angle of polarization.

\subsection {Transformation of Stokes parameters}
As already mentioned the combination of half-wave plates and
the calcite block modifies the state of polarization of the incident
beam of light appropriately so that the Stokes parameters that
describe it can be determined by measuring the intensity of light
transmitted by the above combination.
The Stokes parameters $I'$, $Q'$, $U'$ and $V'$ of light transmitted through
a perfect analyzer with the principal plane, the plane containing the
optical axis and the incident beam, at position angle $\phi$ are
related to the Stokes parameters $I$, $Q$, $U$ and $V$ of the incident light
by the matrix of transformation equation (Serkowski 1974a)
\begin{eqnarray}
\left[
\begin{array}{c}
I'\\
Q'\\
U'\\
V'
\end{array}
\right]
=
\frac{1}{2}\,\left[
\begin{array}{cccc}
1 &\cos 2\phi & \sin 2\phi & 0\\
\cos 2\phi & \cos ^22\phi & \frac{1}{2}\sin 4\phi & 0\\
\sin 2\phi & \frac{1}{2}\sin 4\phi & \sin ^22\phi & 0\\
0 & 0 & 0 & 0\\
\end{array}
\right] \,
\left[
\begin{array}{c}
I\\
Q\\
U\\
V
\end{array}
\right]. 
\nonumber
\end{eqnarray}
The above equation shows that the Stokes parameters $I$, $Q$, and $U$ of
the incident light can be determined by measuring the intensities of
light transmitted by an analyzer with its optical axis
at several position angles. If the analyzer is rotated by an angle
$\phi$, the transmitted light will be modulated with an angle 2$\phi$.
Astronomical polarimeters that use a rotating analyzer for
measuring linear polarization exist at a few observatories
(Breger 1979; Jain \& Srinivasulu 1991; McDavid 1999).
Efficient polarizers for broadband applications, like, Glan-Taylor prism,
have thickness-to-aperture ratios close to unity, which implies
that the analyzers to be employed to accommodate the incident
beam-size will have to be appreciably thick. Because of the
inherent deviation of the emergent beam from the direction of
the incident beam such analyzers when
placed in the light path would cause significant lateral shift of
the star image,
severely affecting the accuracy of measurement in astronomical
polarimetry. Breger (1979) has reported a shift of about 0.1~$mm$
in the image position in the Glan-Air prism based
polarimeter used at McDonald Observatory.
The adverse effect of lateral shift of image will be considerably low
if the analyzer is inserted in the light path
after the focal plane diaphragm.
Plastic sheet polarizers are thin and do not cause
appreciable lateral shift of the image;
it is, therefore, possible to construct simple polarimeters
using these as analyzers (Jain \& Srinivasulu 1991) which could
give photon limited accuracy. The main
problems with the plastic sheet polarizers are that
they are efficient as polarizers in restricted wavelength regions
and have poor transmittances.
In principle we can properly place the image at each position
of the analyzer
and measure the transmitted intensity. However, the measurements over
a full rotation of the analyzer would take several minutes, and
hence would require excellent sky conditions; any variation in the
sky transparency during the intervening period would severely limit
the accuracy of the measurement. In differential photometry, which
is the basic technique of polarimetry, the accuracy in measurement
is decided by the frequency with which the comparison with the
reference light is made.

The spatial shift of the image, resulting from the
inherent deviation in the direction of the emergent beam from the
incident direction, would cause a rotation of the image
when the analyzer is rotated.
The image rotation could in turn introduce an instrumental polarization
that varies with wavelength (Breger 1979).
Any slight misalignment between the rotational
axis and the normal to the incident direction also would contribute
to the rotation of the image in the focal plane.
In any case the intensity of
light falling on the detector will be modulated
during the observations dealing with the measurement of the Stokes parameters.
The spatial shift in the image will be negligible
if the thickness of the optical element that modulates the light
is small. In most of the modern astronomical polarimeters, therefore,
retarders, which usually have thicknesses on the order of
a couple of millimetres, are used for the required modulation. An
added advantage of retarder-based polarimeters is that they can be easily
modified so as to measure circular polarization. 

A retarder is, in the simplest case, a plane parallel plate of uniaxial
crystal cut parallel to its optical axis.
The Stokes parameters $I'$, $Q'$, $U'$ and $V'$ of light transmitted
through a perfect retarder, with the position angle of
its optical axis at an angle $\psi$, is transformed according to the
following matrix equation:
\begin{eqnarray}
\left[
\begin{array}{c}
I'\\
Q'\\
U'\\
V'
\end{array}
\right] = 
\left[
\begin{array}{cccc}
1 & 0 & 0 & 0 \\
0 & G + H \cos 4\psi & H \sin 4\psi & -\sin \tau\,\sin 2\psi \\ 
0 & H \sin 4\psi & G - H \cos 4\psi & \sin \tau\,\cos 2\psi \\
0 & \sin \tau\, \sin 2\psi & -\sin \tau\,\cos 2\psi & \cos \tau\\
\end{array}
\right] \,
\left[
\begin{array}{c}
I\\
Q\\
U\\
V
\end{array}
\right]. \nonumber
\end{eqnarray}
Here $I$, $Q$, $U$ and $V$ are the parameters that characterize the incident light,
and $$ G = \frac{1}{2} (1 + \cos \tau), \quad
\mbox{and} \quad  H = \frac{1}{2} (1 - \cos \tau),$$
in which $\tau$ is the retardance, the phase difference introduced by
the retarder between the vibrations in the principal plane and those
in the plane perpendicular to it. 
If two retarders are kept in series the square  matrix in the above
equation should be replaced by the product of two such matrices.

\subsection{Rotational modulation by a half-wave plate}\label{ss:rotmod}
The determination
of the Stokes parameters, which describe the state of polarization
of a beam of light,
ultimately reduces
to  the measurement of the intensity of light
transmitted by the retarders and the analyzer in the
path for various position angles of the optical axis of the 
modulating retarder.
The above equations give the intensity of light transmitted by
two retarders with the optical axes at position angles $\psi_1$ and
$\psi_2$ followed by an analyzer as
\begin{eqnarray}
 I' = &&\frac{1}{2}\,\{I \pm Q\,[G_1\,G_2 + 
\,H_1\,H_2\,\cos 4(\psi_1 - \psi_2) + H_1\,G_2\,\cos 4\psi_1 +
G_1\,H_2\,\cos 4\psi_2 \nonumber\\
&& - \sin \tau_1\,\sin \tau_2\,\sin 2\psi_1\,\sin 2\psi_2]
\pm U\,[H_1\,H_2\,\sin 4(\psi_1 - \psi_2) +
H_1\,G_2\,\sin 4\psi_1 \nonumber\\
&& + G_1\,H_2\,\sin 4\psi_2 +
\sin \tau_1\,\sin \tau_2\,\cos 2\psi_1\,\sin 2\psi_2]
\mp V\,[H_2\,\sin \tau_1\,\sin (2\psi_1 - 4\psi_2) \nonumber\\
&& + G_2\,\sin \tau_1\,\sin 2\psi_1 +
\cos \tau_1\,\sin \tau_2\,\sin 2\psi_2]\}, 
\nonumber
\end{eqnarray}
where $\tau_1$ and $\tau_2$ are the retardance of the two retarders. The upper
signs correspond to the principal plane of the analyzer at position angle
$\phi$ = 0 degree and the lower signs to that at 90 degree.

The computed path difference of a superachromatic half-wave plate is
($\lambda$/2)$\pm$ 1.3 per cent in the wavelength interval 300$-$1100~$nm$,
and during manufacturing it is possible to achieve the theoretical
retardation to an accuracy of $\pm$3 per cent, giving a value of
$\tau$ = 180$\pm$6.5 degree for the half-wave plates used. The stationary
half-wave plate is mounted with its effective optical axis approximately 
parallel to the principal plane of the analyzer; hence all the terms
proportional to $\sin 2\psi_2$ and $\sin 4\psi_2$
become negligibly small in the above equation, and it reduces to
\begin{eqnarray}
I' = \frac{1}{2}\,[I \pm Q\,\cos 4(\psi_1 - \psi_2)
    \pm U\,\sin 4(\psi_1 - \psi_2) \pm c\,V\,\sin 2\psi_1] \nonumber.
\end{eqnarray}
For plane polarized light $V =$ 0, and therefore the last term in the above
equation vanishes.
When observations are made over a full cycle of rotation of
the half-wave plate
the contribution to the Stokes parameters $Q$ and $U$ arising from the
term containing $V$ is eliminated
since it produces a 2$\psi$ modulation in the data,
while the terms containing
$Q$ and $U$ terms produce a 4$\psi$ modulation. Therefore,
if the light is slightly
elliptically polarized, the 2$\psi$ modulation will not cause
any differences between the
true and derived $Q$ and $U$ values, but the derived
probable errors of their determination may
be larger because of the resulting larger residuals.
The value of $c$ is given by
$$ c= H_2\,\sin \tau_1\,\cos 2\psi_2 .$$
With $\tau_1 = \tau_2 = 180\pm$6.5 degree, $\mid c \mid$
$\approx$ 0.1, and the increase
in the probable errors in $Q$ and $U$ due to the 2$\psi$ modulation
will be negligible unless $V$ is appreciably large.
The intensities of the two beams emerging from the calcite
block are effectively given by
\begin{equation}
I' = \frac{1}{2}\,(I \pm Q\,\cos 4\psi \pm U\,\sin 4\psi) ,
\label{e:int}
\end{equation}
where $\psi$ is the angle between the effective optical axes of the
two half-wave plates. The upper signs correspond to the beam with
vibrations in the principal plane of the top plate of the calcite block
and the lower signs with vibrations perpendicular to the principal plane.
In the above relations it is assumed that the retarders and the calcite
block analyzer are perfect, in the sense that these elements do not cause
any attenuation in the two orthogonally polarized emerging beams.

The Stokes parameters $Q$ and $U$,
and hence, the degree of linear polarization
$p$ and the position angle $\chi$ of polarization can be determined by
measuring the intensity of the transmitted beam as a function of
different orientation $\psi$ of the rotating half-wave plate and calculating
the amplitude and phase of the 4$\psi$ modulation in the data. The variations
in the atmospheric transparency during the observations, and the 
transmittances and reflectivity of the various optical elements presented
to the two polarized beams will have to be taken into account in the actual
observational and data reduction procedures.

\section {Determination of linear polarization}\label{s:linpol}
In the following subsections we describe in detail the steps involved in
the determination of position angle and linear polarization
of light from an astronomical object from the raw data that consist
of counts proportional to the intensities of the two images at
different positions of the rotating half-wave plate.
In order to know how reliable these quantities are it is extremely important
that the probable errors in them are also estimated accurately.
The two main factors that contribute to the errors
in the measured values are the photon noise and the error in the determination
of the background brightness due to possible variations during the
observation of the object. The effects of these factors on the measured
quantities should be known exactly in order
to decide on the sequence of observations
to be followed and to distribute the available time between the
object and background brightness measurements for minimizing the errors.
We discuss these and several other relevant
points directly connected with the observations at the telescope
in the following subsections.

\subsection{Dead-time correction}
The single greatest drawback of the pulse counting system is its inability
to count photons with closely spaced arrival times
with accuracy; if two photons are closely
spaced in time, smaller than the response time of the
photomultiplier--pulse amplifier system, they may be detected as single
and as a result the number of pulses recorded will be always smaller
than the true number of photons detected.
The counting accuracy should be
better than 0.01 per cent if a similar accuracy has to be achieved
in polarimetry
when sufficient number of photons are available.
The error resulting from the finite time resolution of the
counting system is called pile-up or dead-time error.
The dead-time effects are present at all counting rates, but become
increasingly worse with increasing count rates, or increasing brightness
of the objects observed.
The photons from astronomical objects
do not arrive in evenly spaced time intervals. The manufacturers'
specifications on the photomultipliers, pre-amplifiers and counters
are based on evenly spaced, uniform
pulses from a signal generator and not on the ragged output
of a photomultiplier tube used in an astronomical photometer,
and hence, their role is solely restricted to provide guidelines
in the selection of the components of the pulse counting unit.
Observations show that the dead-time effects begin to be important
around counting rates of $10^5\,s^{-1}$ with a
pulse-amplifier-discriminator that gives output pulses of 20\,ns
wide in conjunction with a 100\,MHz counter (Henden \& Kaitchuk 1982).
Even the gain at which a photomultiplier tube is operated
could affect the response characteristics of a pulse
counting system (Moffett \& Barnes 1979).

When the half-wave plate is rotated the intensities of the
two orthogonally polarized beams
produced by the calcite block will vary if the incident
light is polarized. The degree of linear polarization
of the incident light is given in terms of the amplitude
and mean level of the intensity modulation of
each image by the relation (Hiltner 1962)
\begin{equation}
 p_{true} = \frac{N_{max} -  N_{min}}{N_{max} +  N_{min}}, \label{e:ptrue}
\end{equation}
where $N_{max}$ and $N_{min}$ are the true counts at the maximum and
minimum of the modulation. The dead-time effects will modify
both the amplitude and mean light level because the maximum counts
will be reduced by a larger fraction than the minimum counts.
The net effect of the non-linearity of the pulse counting
system is, therefore, a depolarization effect.

The most widely used formula in photometry for correcting
the observed counts, $N_{obs}$, to get the true counts, $N_{true}$,
is (Fernie 1976; Harris, Fitzgerald \& Reed 1981)
\begin{equation}
 N_{true} = N_{obs}\,(1 + \rho\,C_{true}) \approx 
 N_{obs}\,(1 + \rho\,C_{obs}), \label{e:ntrue}
\end{equation}
where $\rho$ is the dead-time coefficient that has the dimension of time,
and $C_{true}$ and $C_{obs}$ are the true and observed count rates.
At low polarization levels, the above formula gives
a simple relationship between the observed and true linear polarizations
(Hsu \& Breger 1982):
\begin{equation}
 p_{true} = p_{obs}\,(1 + \rho\,C_{obs}). \label{e:dead1}
\end{equation}

The dead-time coefficient as defined above is found to depend on the counting
rates themselves, brighter stars yielding higher values (Fernie 1976).
Henden \& Kaitchuk (1982) give the following equation,
which is probably valid even when the dead-time effects are significant,
connecting the observed and true counts:
\begin{equation}
 N_{true} = N_{obs}\,e^{\rho\,C_{true}}, \quad \mbox{with}\quad
C_{true} = \frac{N_{true}}{t}\,, \label{e:dead2}
\end{equation}
where $t$ is the time for which the counts are integrated.
The above transcendental
equation can be solved iteratively first by putting $N_{true}$ = $N_{obs}$
in the right hand side and getting a new value for $N_{true}$. The required
convergence can be obtained in two or three iterations.
The first step in the data reduction procedure is the correction for the
dead-time effects to the observed counts, using the known coefficients.

\subsection {Background sky subtraction}
The observed counts during object integration
contain the contribution from the background sky,
which should be obtained separately with sufficient accuracy
and subtracted from the observed
object counts.
Even though the background light
is unmodulated by the rotation of the half-wave plate,
it is desirable to observe it over a full rotational cycle at the
same positions of the half-wave plate as in the case of the object
brightness. The background counts thus obtained can be divided by the
number of angular positions of the half-wave plate to obtain the
required counts. The two beams should be considered separately
because of the possibility of the gain ratio $\alpha$ 
(see the next section, for its definition) being different from unity.

If $t_1$ and $t_2$ are the beginning and end of star observations,
the sky background to be actually subtracted is
\begin{eqnarray}
S(t_1,t_2) = \int_{t_1}^{t_2}{S(t) dt}\,, \nonumber
\end{eqnarray}
and if the background varies linearly
\begin{eqnarray}
S(t_1,t_2) = \frac{S(t_1) + S(t_2)}{2}\,(t_2 - t_1\,)\,.  \nonumber 
\end{eqnarray}
S($t_1$) and S($t_2$) are the background sky counts in unit time intervals
at time $t = t_1$ and $t = t_2$,
averaged over all the angular positions of the rotating
half-wave plate.
The background sky should be observed frequently if the brightness varies
drastically as is the case during moon-rise and moon-set, or if the
object observation lasts for a substantially
long period. It is always advisable to observe
the background brightness
before and after the object integration if moonlight
is present because of the possibility of its variation,
otherwise it could be observed either before or after
the object integration. 
The counts due to the object alone in the two channels
corresponding to the two light beams emerging from
the calcite block
at the $i^{th}$ position of the half-wave plate,
$N_1^i$ and $N_2^i$, are given by
\begin {eqnarray}
 N_1^i = N_1^{i(star+sky)}- N_1^{sky} \quad \mbox{and} \nonumber \\
 N_2^i = N_2^{i(star+sky)}- N_2^{sky},
\nonumber
\end{eqnarray}
 where $N_1^{sky}$ and $N_2^{sky}$
are the background sky counts S($t_1$,$t_2$) in the two beams calculated 
as above.  The observed counts are assumed to be corrected
for dead-time effects, with the already determined coefficients;
usually, only $N_1^i(star+sky)$ and $N_2^i(star+sky)$ may need corrections.

\subsection{The basic equation to be solved}
If $\psi^i$ is the difference in the position angles of the
effective optical axes of the two half-wave plates at the $i^{th}$ position
of the rotating half-wave plate, equation~(\ref{e:int}) 
gives the photon numbers
due to the two emerging beams from the calcite block in terms of the
Stokes parameters of the incident light as
\begin{eqnarray}
&N_1^i = \frac{1}{2}\,G_1\,(I + Q\,\cos 4\psi^i + U\,\sin 4\psi^i)\,T_{atm}
\quad \mbox{and} \nonumber \\
&N_2^i = \frac{1}{2}\,G_2\,(I - Q\,\cos 4\psi^i - U\,\sin 4\psi^i)\,T_{atm}, 
\label{e:N12}
\end{eqnarray}
where the subscripts 1 and 2 refer to the beams polarized in the
principal plane of the first plate of the calcite block and the plane
perpendicular to it, respectively; the former is taken as beam 1
and the latter as beam 2.
The parameters $G_1$ and $G_2$ in
the above equations
are the net efficiencies for the two light beams, which include the
transmittances and reflectivities of the various optical elements in the
light path and the quantum efficiencies of the cathode of the photomultiplier
tube. But for the sensitivities of the photocathode the efficiencies are
expected to be the same for the two mutually perpendicularly polarized
beams; the calcite block with two crossed plates is expected to have
the same net transmittance for the two beams and the dichroic filters
with the vibrations of the beams making an angle of 45 degree with the
plane of incidence are expected to affect both the beams similarly.
Since the plane of polarization are fixed with respect to the dichroic
filters and the photocathode, $G_1$ and $G_2$ will be constants and
will not depend on the position angle of the rotating half-wave plate.
The atmospheric transmittance $T_{atm}$ and the
scintillation pattern will be the same for both the
beams since the clouds, which are suspended water droplets
and tiny ice crystals, and air being non-birefringent; however, $T_{atm}$
could change during the integrations, which normally last several minutes,
at a particular position of the rotating half-wave plate. The chopping
frequency is high enough to ensure that the two beams are nearly
simultaneously measured. If the atmospheric transparency varies
during the observation, the
counts added to the two channels that register the counts due to the
two beams would vary, but both would be
affected by the same factor; the transmittance $T_{atm}$ appearing in
the above equation in that case should be replaced by an integral.
Taking the ratio of the counts accumulated in the two channels,
\begin{equation}
 Z^i = \frac{N_1^i}{N_2^i} = \frac{1 + q\,\cos 4\psi^i + u\,\sin 4\psi^i}
{\alpha\,(1 - q\,\cos 4\psi^i - u\,\sin 4\psi^i)},
\label{e:be}
\end{equation}
where $\alpha$ ($=G_2/G_1$) is the ratio of the efficiencies
of the two channels, and $q$ and $u$ are the normalized Stokes parameters
($Q/I$) and ($U/I$). 

The above gives the basic equation to be solved to get the normalized Stokes
parameters.

If there is no polarization produced by the telescope-instrument optics,
$\alpha$ can be determined by observing an unpolarized object at any
arbitrary position of the rotating halfwave plate, using
$$ \alpha = \frac{1}{Z^i} = \frac{N_2^i}{N_1^i} . $$
The gain-ratio can also be calculated from the observations of
polarized objects if observations are done over a full cycle of rotation
of the half-wave plate at equally spaced angular intervals using
$$ \alpha = \frac {\sum N_2^i}{\sum N_1^i} $$
because terms involving $Q$ and $U$ will cancel each other. The observations
should be carried out under steady sky conditions; variations in
the atmospheric transparency during observations
will give rise to net non-zero coefficients
for the $Q$ and $U$ terms.

\subsection{Data sampling}
In principle, if we have $N_1$ and $N_2$ measured at three different
values of $\psi$, we can get the required parameters. But, in practice, we
measure $N_1$ and $N_2$ at a number of values of $\psi$ in order to have
sufficient degrees of freedom for estimating properly
the errors involved in the determination of the parameters.
The error in the measured polarization due to statistical fluctuations
in the registered counts depends on the total number of counts
accumulated (see section \ref{ss:rough}).
This means that as far as the error in the
polarization due to photon noise is concerned there will be no difference
if we make observations at a large number of halfwave plate positions 
with a small
integration time at each step or at a small number of halfwave plate positions 
with a large integration time. Since three unknowns
are solved simultaneously, the thumb rule indicates that data should be
taken at a minimum of 15 positions of the halfwave plate
for a fairly good number of degrees of freedom for the error estimation.
In order to avoid any depolarizing effect, the data acquisition can begin
only after the halfwave completely stops at each position, and therefore
increasing the number of samples would increase the overhead time spent
in waiting for the rotating half-wave plate to stop. 
If the number of steps
chosen is not a submultiple of 400, then additional time will be needed to
bring the halfwave plate to the reference position after each rotation, 
increasing the overhead time; further, the summation method described in
section \ref{s:summ} will not work.
In order to give equal weightage to the observations in the data reduction,
the number of counts accumulated at all positions of the half-wave
plate should be of the same order. This requires that under poor sky
conditions the observations should be
repeated over several cycles of rotation of the half-wave plate with
a small integration time instead of a
few cycles with a large integration time.

Usually, observations in $U$ and $B$
bands will last longer than those in $V$, $R$ and $I$ bands.
The integrations in $U$ and $B$ can be continued till the integrations in
the other bands are completed successively.
For the number of counts accumulated at all positions of the
half-wave plate to be similar, again, the observations will have to be
divided into several cycles of half-wave plate rotation.

\subsection{Least square method}
The method of least squares is one of the most important tools for the
reduction of observational data in all fields, not just in astronomy
(Jefferys 1980).
According to the principle of least squares the best representative curve
given by equation~(\ref{e:be}) for the observational data,
which defines the most probable values of $q$ and $u$,
is that for which the sum of the squares of residuals is a minimum.
The residual at the $i^{th}$ position of the rotating half-wave plate
is given by
\begin{equation}
 R^i = Z^i - \frac{1 + q\,\cos 4\psi^i + u\,\sin 4\psi^i}
{\alpha\,(1 - q\,\cos 4\psi^i - u\,\sin 4\psi^i)}.
\label{e:r1}
\end{equation}

The gain-ratio $\alpha$ is normally a constant,
though it is expected to show long-term monotonic changes arising
from the slow changes in the transmittances
and reflectivities of the optical components in the path of the two beams.
The sensitivity variations in the associated electronics, especially,
that of the photomultiplier as a result of fluctuations in high voltage supply,
is not expected to produce any variations in $\alpha$ because both
the beams are detected by the same setup at a fast rate.
The factors that could possibly cause variations in $\alpha$ on
short time-scales or even from object
to object are: (i) presence of dust specks on the optical components
in the path of the two beams, (ii) errors in centring the images
in the diaphragms, thereby, including different levels
of light in the apertures, and (iii)
poor sky conditions, especially, passing clouds during moon lit
nights, while observing. 
A possible wavelength dependence of the surface
inhomogeneities in the transmittances
or reflectivities of the optical elements would also cause a variation
in $\alpha$ from object to object because of the related variations
in the isophotal wavelengths of observations.

The presence of a faint
component in one of the apertures would result in erroneous
normalized Stokes parameters. If $\alpha$ is also treated as an
unknown along with $q$ and $u$ its value also would be affected by the
presence of a faint component, and this fact can be advantageously
used to detect and remove the effects of extra light present in one
of the focal plane apertures. Therefore, we treat
$\alpha$ as an unknown each time $q$ and $u$ are derived from the data.

The residual $R^i$ is non-linear in $q$, $u$ and $\alpha$, and 
an iterative procedure has to be employed for the least square solution.

Expanding the residual about $q_0$, $u_0$ and $\alpha_0$, which
represent the approximate values of $q$, $u$ and $\alpha$,
using Taylor series for a function of several variables and taking
only the terms involving first derivative (Scarborough 1964; Murdin 1979)
we get
\begin{eqnarray}
   R^i\left(Z^i,q,u,\alpha\right) = R^i\left(Z^i,q_0,u_0,\alpha_0\right) +
  \Delta q \left(\frac{\partial R^i}{\partial q}\right)_0 +
  \Delta u \left(\frac{\partial R^i}{\partial u}\right)_0 +
  \Delta \alpha \left(\frac{\partial R^i}{\partial \alpha}\right)_0, \nonumber
\end{eqnarray}
where
$$ \left(\frac{\partial R^i}{\partial q}\right)_0 =
  \left(\frac{\partial R^i}{\partial q}\right), \quad \mbox{evaluated at
   $q = q_0, u = u_0, \alpha = \alpha_0$, etc.}, $$ and
\begin{eqnarray}
  &q = q_0 + \Delta q,\nonumber \\ 
  &u = u_0 + \Delta u \quad \mbox{and}\nonumber \\
  &\alpha = \alpha_0 + \Delta \alpha .
\label{e:start}
\end{eqnarray}
The above equation for the residual is linear in the differential
corrections
$\Delta $q, $\Delta $u and $\Delta \alpha$ to be applied to $q_0$, $u_0$
and $\alpha_0$. The sum of the squares is made a minimum with respect to
$\Delta $q, $\Delta $u and $\Delta \alpha$, and the conditions for
a minimum give the following normal equations.
\begin{eqnarray}
\Delta q\,\sum \left({\frac{\partial R^i}{\partial q}}\right)^2_0 + 
 \Delta u\, \sum \left(\frac{\partial R^i}{\partial u}\right)_0
  \left(\frac{\partial R^i}{\partial q}\right)_0 +
 \Delta \alpha\, \sum \left(\frac{\partial R^i}{\partial \alpha}\right)_0
  \left(\frac{\partial R^i}{\partial q}\right)_0 & \nonumber \\
\quad \quad \quad   = \sum -R^i\left(Z^i,q_0,u_0,\alpha_0\right)\,\left(\frac{\partial R^i}{\partial q}\right)_0, &\nonumber \\
\Delta q\,\sum \left(\frac{\partial R^i}{\partial q}\right)_0
  \left(\frac{\partial R^i}{\partial u}\right)_0 +
 \Delta u\,\sum \left(\frac{\partial R^i}{\partial u}\right)^2_0 +
 \Delta \alpha\, \sum \left(\frac{\partial R^i}{\partial \alpha}\right)_0
  \left(\frac{\partial R^i}{\partial u}\right)_0 & \nonumber \\
\quad \quad \quad   = \sum -R^i\left(Z^i,q_0,u_0,\alpha_0\right)\,\left(\frac{\partial R^i}{\partial u}\right)_0 &\nonumber 
\end{eqnarray}
and
\begin{eqnarray}
\Delta q\,\sum \left(\frac{\partial R^i}{\partial q}\right)_0
  \left(\frac{\partial R^i}{\partial \alpha}\right)_0 +
 \Delta u\, \sum \left(\frac{\partial R^i}{\partial u}\right)_0
  \left(\frac{\partial R^i}{\partial \alpha}\right)_0 +
 \Delta \alpha\,\sum \left(\frac{\partial R^i}{\partial \alpha}\right)^2_0 & \nonumber \\
\quad \quad \quad   = \sum -R^i\left(Z^i,q_0,u_0,\alpha_0\right)\,\left(\frac{\partial R^i}{\partial \alpha}\right)_0 \, , & 
\label{e:neq1}
\end{eqnarray}
where
 $$\left(\frac{\partial R^i}{\partial q}\right)_0 =  \frac{\cos 4\psi^i}{\alpha_0 g} -
\frac{f_1 \cos 4\psi^i}{\alpha_0 g^2}  \nonumber $$ 
$$ \left(\frac{\partial R^i}{\partial u}\right)_0 =  \frac{\sin 4\psi^i}{\alpha_0 g} -
\frac{f_1 \sin 4\psi^i}{\alpha_0 g^2}  \nonumber  \quad \mbox{and} $$ 
$$\left(\frac{\partial R^i}{\partial \alpha}\right)_0 = 
 \frac{f}{\alpha_0^2 g} \, ,\nonumber $$ 
with $$f = 1.0 + q_0 \cos 4\psi^i + u_0 \sin 4\psi^i \,\, \mbox{and} \,\, 
 g = 1.0 - q_0 \cos 4\psi^i - u_0 \sin 4\psi^i \, .$$
The value of $\alpha_0$ can be assumed to be equal to
either 1 or the average of the
values determined previously by observing unpolarized
objects or polarized objects through a full rotation cycle of
the half-wave plate under photometric sky conditions. We use a value of 1.0
for $\alpha_0$. The approximate values for the Stokes parameters, 
$q_0$ and $u_0$, as starting values
for the iterative procedure are derived as given below.

With $\alpha_0 = 1$, equation~(\ref{e:be}) can be written as 
\begin{eqnarray}
q_0\,(1 + Z^i)\,\cos 4\psi^i + u_0\,(1 + Z^i)\,\sin 4\psi^i
+ 1 - Z^i = 0 \, , \nonumber
\end{eqnarray}
which is linear in $q_0$ and $u_0$ and they can be calculated using the standard
least square procedure. 
The condition for minimum for the sum of squares of residuals gives
the following normal equations:
\begin{eqnarray}
q_0\,\sum \left({\frac{\partial R^i}{\partial q_0}}\right)^2 + u_0\,
 \sum \left(\frac{\partial R^i}{\partial u_0}\right)\left(\frac{\partial R^i}{\partial q_0}\right)
     = \sum (-1 + Z_i)\, \left(\frac{\partial R^i}{\partial q_0}\right)
\quad \mbox{and} \nonumber \\
q_0\,\sum \left(\frac{\partial R^i}{\partial q_0}\right)\left(\frac{\partial R^i}{\partial u_0}\right) +
    u_0\,\sum \left(\frac{\partial R^i}{\partial u_0}\right)^2
     = \sum (-1 +  Z^i)\, \left(\frac{\partial R^i}{\partial u_0}\right)\, ,
\label{e:neq2}
\end{eqnarray}
where
\begin{eqnarray}
 R^i = q_0\,(1 + Z^i)\,\cos 4\psi^i + u_0\,(1 + Z^i)\,\sin 4\psi^i
+ 1 - Z^i \, , \nonumber  
\end{eqnarray}

\begin{eqnarray}
\frac{\partial R^i}{\partial q_0} = (1 + Z^i)\,\cos 4\psi^i & \nonumber \quad \mbox{and} \\ 
\frac{\partial R^i}{\partial u_0} = (1 + Z^i)\,\sin 4\psi^i \, . & \nonumber  
\end{eqnarray}

The solution
of  normal equations~(\ref{e:neq2}) will yield values
of $q_0$ and $u_0$; these together with the $\alpha_0 = 1.0$, assumed
above will then provide the necessary starting values for the
iterative procedure. The corrections $\Delta q$, $\Delta u$
and $\Delta \alpha$ are calculated using normal equations~(\ref{e:neq1}),
and the new starting values for the iteration are obtained from
equation~(\ref{e:start}). The procedure is continued till convergence at
the required significant figure is obtained; usually this will
need four or five iterations since the starting values will
not be very much off from the final values.

The normal equations can be formed in the following way also.
Equation~(\ref{e:be})
can be re-written as
$$ q\,(1 + \alpha\,Z^i)\,\cos 4\psi^i + u\,(1 + \alpha\,Z^i)\,\sin 4\psi^i
+ 1 - \alpha\,Z^i = 0\, \nonumber ,$$
in which case the modified residual at the $i^{th}$ position of the 
rotating half-wave plate is given by
\begin{equation}
R^i = q\,(1 + \alpha\,Z^i)\,\cos 4\psi^i + u\,(1 + \alpha\,Z^i)\,\sin 4\psi^i
+ 1 - \alpha\,Z^i. 
\label{e:r2}
\end{equation}
The required partial derivatives are calculated using
$$ \left(\frac{\partial R^i}{\partial q}\right)_0 = 
(1 + \alpha_0\,Z^i)\,\cos 4\psi^i\, ,
\nonumber $$ 
$$ \left(\frac{\partial R^i}{\partial u}\right)_0 = 
(1 + \alpha_0\,Z^i)\,\sin 4\psi^i 
\nonumber  \quad \mbox{and} $$
$$\left(\frac{\partial R^i}{\partial \alpha}\right)_0 = Z^i (q_0 \cos 4\psi^i
+ u_0 \sin 4\psi^i -1)\,.$$
The two schemes (equations~\ref{e:r1} and
\ref{e:r2}), which are incorporated in the data reduction program,
are found to give results which mutually
agree within the probable errors.

The normal equations are solved using Cracovian matrix elimination
method (Kopal 1959). This method gives the same results as that
given by the matrix inversion method, but involves a fewer number
of operations. The coefficients of the normal equations
form a symmetric-square matrix and for the reduction of the values
of the unknowns a triangular matrix formed by the diagonal elements
and those below them are used. An additional column matrix with
the number of rows equal to the number of unknowns and initial
elements equal to $-1$ is also involved in the reduction procedure.
Finally the elements of column matrix give the values of
the unknowns directly. The standard deviation is calculated from
$$\sigma = \sqrt{\frac{\sum (R^i)^2}{n - L}}\, ,$$
where $n$ is the number of residuals $R^i$ calculated using the
values of $q$, $u$ and $\alpha$, and $L$ ($=$ 3), the number of unknowns.
The probable errors in the parameters are calculated using
$$ \mbox{p.e. }(\epsilon) = 0.6745\,w\,\sigma\, .$$
The elements of last row of the
triangular matrix after the final reduction give the squares of
the weight $w$ required to calculate the errors in the unknowns,
either $q$ and $u$, or $q$, $u$ and $\alpha$, which ever may be the case.

\subsection{Elimination method}
The normalized Stokes parameters $q$ and $u$, and the gain-ratio
$\alpha$ can be determined by the elimination technique 
in the following way from the
observations when $\psi^i$ is changed in steps of 22.5 degree.
The observations over a full rotation of the halfwave plate would yield
16 ratios of intensities of the type
given by equation~(\ref{e:be}), which form
4 groups, each having 4 ratios. Writing them down explicitly, we get
\begin{eqnarray}
Z^1 = Z^5 = Z^9 = Z^{13} = \frac{1 + q}{\alpha\,(1 - q)}, & \nonumber\\
Z^2 = Z^6 = Z^{10} = Z^{14} = \frac{1 + u}{\alpha\,(1 - u)}, & \nonumber\\
Z^3 = Z^7 = Z^{11} = Z^{15} = \frac{1 - q}{\alpha\,(1 + q)} \quad
\mbox{and}& \nonumber \\
Z^4 = Z^8 = Z^{12} = Z^{16} = \frac{1 - u}{\alpha\,(1 + u)}. \nonumber & 
\end{eqnarray}
Taking the averages of the above four sets of ratios, we have
\begin{eqnarray}
Z_1 = \frac{1}{4}\,(Z^1 + Z^5 + Z^9 + Z^{13}),  & \nonumber\\
Z_2 = \frac{1}{4}\,(Z^3 + Z^7 + Z^{11} + Z^{15}), & \nonumber\\
Z_3 = \frac{1}{4}\,(Z^2 + Z^6 + Z^{10} + Z^{14}) \nonumber
\quad \mbox{and}& \nonumber \\
Z_4 = \frac{1}{4}\,(Z^4 + Z^8 + Z^{12} + Z^{16}).& \nonumber 
\end{eqnarray}
If we define $A$ and $B$ such that 
$$  A^2 = \frac{Z_1}{Z_2}, \quad \mbox{and} \quad
 B^2 = \frac{Z_3}{Z_4}\, , $$
we get the normalized Stokes parameters and the mean gain-ratio as
\begin{eqnarray}
q = \frac{A - 1}{A + 1}, & \nonumber \\
u = \frac{B - 1}{B + 1}& \quad \mbox{and} \nonumber \\
\alpha = \frac{1}{2}\left(\frac{1}{\sqrt{Z_1\,Z_2}} +
\frac{1}{\sqrt{Z_3\,Z_4}}\right). \nonumber
\end{eqnarray}
The two sets of values which give $\alpha$'s are independent, and hence, the
arithmetic mean would be more appropriate to combine the individual vales
than the geometric mean as adopted by
Ramaprakash  et al. (1998), even though the two means may not be significantly
different.

The errors in $q$ and $u$ are given by
\begin{eqnarray}
\epsilon_q = \frac{2\,\epsilon_A}{(A + 1)^2} \quad \mbox{and} \quad
\epsilon_u = \frac{2\,\epsilon_B}{(B + 1)^2}\, , \nonumber
\end{eqnarray}
where
\begin{eqnarray}
\epsilon_A = \frac{1}{2\,A\,Z_2^2}\sqrt{(Z_2\,\epsilon_{Z_1})^2 +
(Z_1\,\epsilon_{Z_2}})^2 \quad \mbox{and} \quad
\epsilon_B = \frac{1}{2\,B\,Z_4^2}\sqrt{(Z_4\,\epsilon_{Z_3})^2 +
(Z_3\,\epsilon_{Z_4}})^2\, . \nonumber
\end{eqnarray}
The probable errors $\epsilon_{Z_1}$, 
$\epsilon_{Z_2}$, $\epsilon_{Z_3}$ and $\epsilon_{Z_4}$
in $Z_1$, $Z_2$, $Z_3$ and $Z_4$, respectively,
are calculated from the standard deviations in the corresponding sets
of ratios.

The least square method of solution is superior to the elimination
technique for the following reasons.
In the latter we solve $q$ and $\alpha$ using a set of 8 equations
and $u$ and $\alpha$ using another set of 8 equations. The normalized
parameters $q$ and $u$, and the errors in them are determined independently,
and the overall fit to the observational data with the parameters
are not taken into account while deriving their probable errors.
In the former method all the 16 equations are
used simultaneously to solve 3 unknowns, $q$, $u$ and $\alpha$
giving equal weightage to all the measurements. The removal
the effect of the presence of a component in one of the
beams on the normalized Stokes parameters is difficult using
the elimination technique, while it is rather easy to do that using
the least square method.

We find that both the least square method and the elimination method give
results agreeing mutually, within the
observational uncertainties. However, the
elimination method gives slightly
smaller uncertainties for the normalized Stokes parameters for
the observations made with the Glan-Taylor prism

\subsection{Summation method} \label{s:summ}
When observations are made at equal angular intervals over a full rotation
of the halfwave plate the counts observed in the two beams can be combined
to derive the polarization and position angle in a simpler
way, if no data samples with large deviations are present.
From equation~\ref{e:be} we get
\begin{equation}
\frac{\alpha N_1^i - N_2^i}{\alpha N_1^i + N_2^i} = q \cos 4\psi^i +
u \sin 4\psi^i = z^i
\label{e:comb}
\end{equation}
On multiplying the above equation, respectively,
by cos4$\psi^i$ and sin4$\psi^i$
and summing over all the positions of the halfwave plate we get
$$ q = \frac{2}{M} \sum z^i \cos 4 \psi^i\quad
\mbox{and}\quad u = \frac{2}{M} \sum z^i \sin 4 \psi^i \, .$$
Taking the differentials of the expressions for $q$ and $u$, and 
using the law of propagation of errors, we get
$$ (\sigma_ q)^2 = \left(\frac{2}M{}\right)^2 \sum (\sigma_{z^i})^2
 \cos^2 4 \psi^i  \quad \mbox{and} \quad 
(\sigma_ u)^2 = \left(\frac{2}M{}\right)^2 \sum (\sigma_{z^i})^2
\sin^2 4 \psi^i \, .$$
Assuming that the standard deviations in $z^i$\,'s are the same, the
above expressions reduce to
$$ (\sigma_ q)^2 = (\sigma_ u)^2 = 
\left(\frac{2}{M}\right)^2 \left(\frac{M}{2}\right) (\sigma_z)^2 \, ,$$
where $$ (\sigma_z)^2 = \frac{1}{M-2}\sum (z^i - z^i_{comp})^2 \, .$$
Using equation~\ref{e:comb} we have
$$ \sum (z^i - z^i_{comp})^2  = 
   \sum (z^i - q \cos 4\psi^i - u \sin 4\psi^i)^2 \,.$$
Since $$ \sum q z^i \cos 4\psi^i = q^2 \sum \cos^2 4\psi^i \quad
\mbox{and} \quad \sum u z^i \sin 4\psi^i = u^2 \sum \sin ^2 4\psi^i \,,$$
and
$$ \sum \cos 4\psi^i \sin 4\psi^i = 0 \,,$$
$$ \sum (z^i - z^i_{comp})^2
  =  -q^2 \sum cos^2 4\psi^i - u^2 \sum \sin^2 4\psi^i + \sum (z^i)^2
 =  \sum (z^i)^2 - \frac{M}{2} (q^2 + u^2) \, . $$
Substituting the above in the expression for the standard
deviations in $q$ and $u$ we get the probable errors in $q$ and $u$ as
$$ \epsilon_q = \epsilon_u = 0.6745
  \left\{\frac{1}{M-2} \left(\frac{2}{M} \sum (z^i)^2 - q^2 - u^2\right)\right\}
^{\frac{1}{2}} \, .$$
In this scheme of data reduction the 
value of $\alpha$ in equation~\ref{e:comb} is calculated first by taking
the ratio of the sums of counts in the two channels, $N_1^i$ and
$N_2^i$, as explained earlier. It may be noted that there is no easy way of
estimating the error in $\alpha$ in this scheme.

Under fairly good sky conditions the counts registered in the two channels
corresponding to the two beams can be used separately to derive the
polarization with fairly good accuracy. The procedure may be useful if
one of the diaphragms contains another image of a nearby star, which may
be difficult to exclude from the diaphragm even by rotating the polarimeter.
The counts registered in the first beam at the $i^{th}$ position of
the halfwave plate given by equation~(\ref{e:N12}) can be
written as
$$ N^i_1 = I_0 + A \cos 4 \psi^i + B \sin 4 \psi^i \nonumber \, ,$$
where $I_0 = G_1 I / 2$, $A = G_1 Q / 2$ and $B = G_1 U / 2$.
When the measurements are made at equal angular
intervals between 0\degr and 360\degr,
$$ I_0 = \frac{1}{M} \sum N^i_1\, ,\quad  
A = \frac{2}{M} \sum N^i_1 \cos 4 \psi^i\quad
\mbox{and}\quad B = \frac{2}{M} \sum N^i_1 \sin 4 \psi^i \, ,$$
where $M$ is the number of positions of the halfwave plate.
The normalized stokes parameters are given by
$q = A / I_0$ and $u = B / I_0$.
The differential in $A$ can be written as
$$ \delta A = \frac{2}M{} \sum \delta N^i_1 \cos 4 \psi^i \nonumber \, ,$$
from which we get the standard deviation in $A$ as
$$ (\sigma_A)^2 = \frac{2}{M} \left(\sigma_{N_1}\right)^2 \nonumber \, .$$
The standard deviation in $N_1$, $\sigma _{N_1}$, can be calculated from the
observed $N_1^i$ values and 
the corresponding computed values using $I_0$, $A$ and $B$.
The probable error in q will be 
$$ \epsilon_q = 0.6745 \frac{\sigma_{N_1}}{I_0}
 \sqrt{\frac{2}{M}} \nonumber .$$
Since $\sigma_A = \sigma_B$, we have $\epsilon_q = \epsilon_u$.

The counts in the other beam can be analyzed similarly to yield another
set of $q$ and $u$ values. The ratio of the
values of $I_0$ obtained for beams 1 and 2 will
give the gain-ratio.

\subsection{p, $\theta$ and their errors}
The percentage linear polarization $p$ is calculated from
\begin{eqnarray}
P\,(\%) = \sqrt{q^2 + u^2} \times 100\, . \nonumber
\end{eqnarray}
The position angle $\theta$ (assuming the Stokes parameter $V$ to be 0,
otherwise
we can obtain only the angle which the major axis of the
resultant ellipse makes with
the reference axis) that lies between 0 and 180 degree
is calculated as follows: \newline
\indent When $q$ = 0:
\begin{eqnarray}
\mbox{if }u \geq 0,\quad & 2\,\theta = \frac{\pi}{2}\quad & \mbox{and}
\nonumber \\
\mbox{if }u < 0, \quad &2\,\theta = -\frac{\pi}{2}.\quad &  \nonumber 
\end{eqnarray}
\indent When $q \neq $0:
\begin{eqnarray}
 & 2\,\theta = \tan^{-1}(\frac{u}{q}); \quad & \nonumber  \\
\mbox{if }q < \mbox{0 or }u < 0\,,\quad & 2\,\theta = 2\,\theta + \pi\quad & 
\mbox{and} \nonumber \\
\mbox{if }q \geq 0 \mbox{ and }u < 0\,, \quad &2\,\theta =
 2\,\theta + \pi .\quad &  \nonumber 
\end{eqnarray}
If the resulting $\theta$ is greater than 180 degree, 180 is
subtracted from it
and if it is less than 0 degree, 180 is added to it.
The probable errors in P (\%) and $\theta$ are calculated using
\begin{eqnarray}
\epsilon_P (\%) = \frac{\sqrt{(q\,\epsilon_q)^2 + (u\,\epsilon_u)^2}}{p}
\times 100 \quad \mbox{and}  \nonumber \\
\epsilon_\theta \,\mbox{(degree)} =
 \frac{28.65\,\sqrt{(q\,\epsilon_u)^2 + (u\,\epsilon_q)^2}}{p^2}\, . \nonumber 
\end{eqnarray}
If $\epsilon_q$ = $\epsilon_u$, then $\epsilon_p$ = $\epsilon_q$, and we get
\begin{eqnarray}
 \epsilon_\theta \,\mbox{(degree)} = \frac{28.65\,\epsilon_p}{p}. \nonumber
\end{eqnarray}

The equation~\ref{e:be} can be written as 
\begin{eqnarray}
Z^i = \frac {1 + p\, \cos \,(2 \theta - 4 \psi)}
      {\alpha \,\{1 - p\, \cos \,(2 \theta - 4 \psi)\}} \nonumber
\end{eqnarray}
and $p$ and $\theta$ and their errors can be determined directly by a non-linear
least square fit. We find that both the approaches give identical results. We
follow the  scheme of solving $q$ and $u$ from the data so that the 
instrumental polarization can be directly removed from the observed
normalized Stokes parameters. 

 At low polarization levels the normalized Stokes parameters of the
 instrumental and intrinsic polarizations add approximately linearly as
\begin{eqnarray}
q^{obs} = q^{intr} + q^{inst} \quad \mbox{and} \quad
 u^{obs} = u^{intr} + u^{inst}\, ,\nonumber
\end{eqnarray}
from which the intrinsic Stokes parameters, and
hence, the linear polarization and position angle can be determined.
The errors in intrinsic Stokes parameters are determined from
\begin{eqnarray}
\epsilon_{intr}^q = \sqrt{(\epsilon_{obs}^q)^2 + (\epsilon_{inst}^q)^2}
\quad \mbox{and} \nonumber\\
\epsilon_{intr}^u = \sqrt{(\epsilon_{obs}^u)^2 + (\epsilon_{inst}^u)^2}\, .
\nonumber
\end{eqnarray}
Here the subscripts $obs$, $intr$ and
$inst$ denote the observed, intrinsic and instrumental, respectively.
Once the probable errors in $q$ and $u$ are known the probable errors
in $p$ and $\theta$ can be calculated as given above.

\subsection{Effect of the presence of extra light in one of the beams}
If the programme object has an unpolarized
optical companion that cannot be isolated by the
focal plane aperture the derived degree of polarization will be less
than the actual value because of the increased unpolarized total
light. It may so happen that the position of the
companion in the sky is such that it contributes light
only to one of the beams. It is then possible to avoid the component from
the apertures if the entire polarimeter is rotated so that the line
joining the apertures is perpendicular the line joining the images
of the two objects at the focal plane.

The presence of extra-light in one of the beams can be detected and
removed, probably, fairly accurately from the data in the following way.
Assuming that the component is present in the first beam, 
equation~(\ref{e:be}) can be written for 
the case when $\psi^i$ equals 0 degree explicitly as,
\begin{eqnarray}
Z^0 = \frac{N^0_1}{N^0_2} = \frac{n_p + Q + n_f}{\alpha(n_p - Q)}
\nonumber
\end{eqnarray}
where $n_p$ is the intensity due to the program star and $n_f$ that
due to the faint component. The above ratio can be re-written
successively as follows:
\begin{eqnarray}
&Z^0  = & \frac{(n_p + n_f)\left(1 + \frac{Q}{n_p + n_f}\right)}
{\alpha \,n_p\left(1 - \frac{Q}{n_p}\right)} \nonumber \\
&~~  =& \frac{1 + q'}{\frac{\alpha \,n_p}{n_p + n_f}(1 - q)}
\quad \mbox{with}\quad q = \frac{Q}{n_p}\quad \mbox{and} \quad
q' = \frac{Q}{n_p + n_f} \nonumber \\
&~~  =&  \frac{1 + q'}{\alpha '\,(1 - q)}  \quad \mbox{with} \quad
\alpha ' = \frac{\alpha \,n_p}{n_p + n_f} \nonumber \\
&~~  =&  \frac{1 - \epsilon_1 + q' + \epsilon_1}
  {\alpha '\,(1 - \epsilon_2 - q + \epsilon_2)} \nonumber \\
&~~  =&  \frac{1 + q''}{\alpha ''\,(1 - q'')}  \quad \mbox{with}
\quad \alpha '' = \frac{\alpha '\,(1 - \epsilon_2)}{(1 - \epsilon_1)}
\quad \mbox{and} \quad q'' = \frac{q' + \epsilon_1}{1 - \epsilon_1}
 = \frac{q - \epsilon_2}{1 - \epsilon_2}, \nonumber
\end{eqnarray}
indicating that the values of the gain ratio $\alpha$ and the Stokes parameters
will be scaled down, though by different factors.
The values of $\epsilon_1$ and $\epsilon_2$  will be nearly equal
to each other, and appreciably
smaller than the value of $q$; therefore,
$$ \frac{1 - \epsilon_2}{1 - \epsilon_1} \approx 1,
 \quad \mbox{and hence,} \quad
\alpha '' \approx \frac{\alpha\,n_p}{n_p + n_f}, $$
if a component is present in the first beam.
A similar reasoning will show that if the
faint component is present in the second beam
$$\alpha '' \approx \frac{\alpha\,(n_p + n_f)}{n_p}. $$
When the gain ratio is also determined along with the Stokes parameters
from the data, its value will indicate whether a faint component is
present in one of the beams. If $\alpha$ is less than the average value
already determined,
which normally will be close to unity,
then the faint component is in the first beam and if it is greater than
the average value then the component is in the second beam, since we have taken
$$ Z^i = \frac{Intensity~of~the~first~beam}{Intensity~of~the~second~beam}. $$

\subsection{Solution for $q$ and $u$ with extra-light in one of the
beams}
Sometimes it may be possible to keep the component outside the diaphragms
by rotating the polarimeter, by rotating the Positional Angle Device, and
carry out the observations in the usual way. The
position angle of polarization determined should be offset by the same angle
through which the PAD is rotated from its normal position.

If the sky is good then the data registered in the component-free beam
can be used to determine the polarization with limited accuracy. A provision
to compute the polarization in each beam 
separately has been provided as an option in the data reduction program.

As already mentioned, the gain-ratio $\alpha$ would be significantly
different from its average value if any extra-component is present in one
of the beams. The fractional
contribution to the observed light by the component can then be determined,
by the least square method, if we set the value of $\alpha$ 
to its already known average value.
Equation~(\ref{e:r1}) for the residual can be written as
\begin{equation}
 R^i = Z^i - \frac{1 + q\,\cos 4\psi^i + u\,\sin 4\psi^i + \beta}
{\alpha\,(1 - q\,\cos 4\psi^i - u\,\sin 4\psi^i)} \,
\label{e:r1f1}
\end{equation}
if the component is in the first beam and 
\begin{equation}
 R^i = Z^i - \frac{1 + q\,\cos 4\psi^i + u\,\sin 4\psi^i}
{\alpha\,(1 - q\,\cos 4\psi^i - u\,\sin 4\psi^i +\beta)} \, ,
\label{e:r1f2}
\end{equation}
if the component is in the second beam; here
$\beta$ ($= n_f/n_p$) is the ratio of the brightness of the component
to that of the programme object.
When $\alpha$ is treated as known  the normal equations to be solved
are similar to equation~(\ref{e:neq1}) with $\Delta \beta$'s replacing 
the $\Delta \alpha$'s
in it. The required partial derivatives with respect to $q$ and $u$
are calculated using the equations given earlier with
f or g modified depending on in which beam the component is present. The
partial derivatives with respect to $\beta$ is calculated using
\begin{equation}
\left(\frac{\partial R^i}{\partial \beta}\right)_0 = 
 \frac{-1}{\alpha_0 g} \, ,
\label{e:pdb1}
\end{equation}
if the component is in the first beam and using
\begin{equation}
\left(\frac{\partial R^i}{\partial \beta}\right)_0 = 
 \frac{-f}{\alpha_0 g^2} \, ,
\label{e:pdb2}
\end{equation}
if it is in the second beam.

Equations (\ref{e:r1f1}) and (\ref{e:r1f2}) 
for the residuals can be rewritten as
\begin{equation}
 q\,(1 + \alpha\,Z^i)\,\cos 4\psi^i + u\,(1 + \alpha\,Z^i)\,\sin 4\psi^i
+ 1 + \beta - \alpha\,Z^i = 0, \quad \mbox{and}
\label{e:r2f1}
\end{equation}
\begin{equation}
 q\,(1 + \alpha\,Z^i)\,\cos 4\psi^i + u\,(1 + \alpha\,Z^i)\,\sin 4\psi^i
+ 1 + (-\beta) \,\alpha\,Z^i - \alpha\,Z^i = 0 \, .
\label{e:r2f2}
\end{equation}
When $\alpha$ is treated as known the residual becomes linear in
$q$, $u$ and $\beta$, and the normal equations to be solved
are the following:
\begin{eqnarray}
q_0\,\sum \left({\frac{\partial R^i}{\partial q}}\right)_0^2 +
  u_0\,\sum \left(\frac{\partial R^i}{\partial u}\right)_0\left(\frac{\partial R^i}{\partial q}\right)_0 + 
  \beta_0\,\sum \left(\frac{\partial R^i}{\partial \beta}\right)_0\left(\frac{\partial R^i}{\partial q}\right)_0 \nonumber & \\
= \sum (-1 + \alpha Z^i)\,\left(\frac{\partial R^i}{\partial q}\right)_0 \, ,
\nonumber  &\\
q_0\,\sum \left(\frac{\partial R^i}{\partial q}\right)_0\left(\frac{\partial R^i}{\partial u}\right)_0 +
    u_0\,\sum \left(\frac{\partial R^i}{\partial u}\right)_0^2 + 
  \beta_0\,\sum \left(\frac{\partial R^i}{\partial \beta}\right)_0\left(\frac{\partial R^i}{\partial u}\right)_0  \nonumber &\\
 = \sum (-1 + \alpha Z^i)\,
\left(\frac{\partial R^i}{\partial u}\right)_0  \, , \nonumber&
\end{eqnarray}
and
\begin{eqnarray}
q_0\,\sum \left(\frac{\partial R^i}{\partial q}\right)_0\left(\frac{\partial R^i}{\partial \beta}\right)_0 +
  u_0\,\sum \left(\frac{\partial R^i}{\partial u}\right)_0\left(\frac{\partial R^i}{\partial \beta}\right)_0 +
  \beta_0\,\sum \left(\frac{\partial R^i}{\partial \beta}\right)_0^2 \nonumber & \\
 = \sum (-1 + \alpha Z^i)\,\left(\frac{\partial R^i}{\partial
\beta}\right)_0 \, ,\nonumber &
\end{eqnarray}
where
\begin{eqnarray}
 \left(\frac{\partial R^i}{\partial q}\right)_0 = & (1 + \alpha\,Z^i)\,\cos 4\psi^i 
 \quad \mbox{and} \nonumber \\ 
 \left(\frac{\partial R^i}{\partial u}\right)_0 = & (1 + \alpha\,Z^i)\,\sin 4\psi^i, 
 \quad \mbox{as before},\nonumber  
\end{eqnarray}
and
$$ \left(\frac{\partial R^i}{\partial \beta}\right)_0 =  1 \, ,$$
if the component is in the first beam and
$$ \left(\frac{\partial R^i}{\partial \beta}\right)_0 = \alpha\,Z^i \, , $$
if the component is in the second beam. In equations~(\ref{e:pdb1}) 
and (\ref{e:pdb2}) for the
partial derivatives we have changed the sign so that the sign of
$\beta$ obtained as the solution
will indicate the beam in which the extra-light due to the component
is present; if $\beta$ is positive it is in the first beam, and if
negative it is in the second beam.

The $q_0$, $u_0$ and $\beta_0$ evaluated using the above equations will
be the final values, and no further significant
improvement can be made by using
these values as the starting values for solving equations~(\ref{e:r1f1}) 
and (\ref{e:r1f2})
in an iterative way. Since we are minimizing different quantities when
equation (\ref{e:r1f1}) or (\ref{e:r1f2}) and the corresponding 
modified residuals (equations~\ref{e:r2f1} or \ref{e:r2f2})
are used for
solving these parameters, there is a possibility of marginal differences
in the solutions obtained by the two approaches. For consistency these
values are used as starting values for the iterative procedure in the
data reduction program, if
the residuals are defined using equations~(\ref{e:r1f1}) and 
(\ref{e:r1f2}).

\subsection{Rejection of erroneous data samples}
Large departures, as indicated by the corresponding residuals,
in some of the intensity ratios $Z^i$ obtained
at different positions of the rotating half-wave plate may occur
as a result of a drift of the star images inside the apertures or
random noise pickup in the associated electronic circuits.
In such cases the Stokes
parameters should be solved for after eliminating such ratios.
From the derived values of $q$, $u$
and $\alpha$ the resulting residuals can be calculated using
equations~(\ref{e:r1}), (\ref{e:r1f1}) or (\ref{e:r1f2}), 
at each position of the half-wave plate, depending on whether a faint
component is present and if present, in which beam it is present, if the sum of
squares of the residuals in the ratios of the intensities are made a minimum
(scheme~1),
or equations~(\ref{e:r2}), or (\ref{e:r2f1}) or (\ref{e:r2f2}), 
if the sum of squares of the modified residuals are made a minimum
(scheme~2). 
If the modulus of the observed residual is
a specified times larger than the standard
deviation of the over all fit, that particular pair of values
is removed from the
reduction procedure and fresh normal equations are formed for the solution.
After obtaining a fresh solution the process of deleting largely
deviating values is repeated, if there is a need. Every time a fresh solution
is obtained the residuals are computed at all positions of the halfwave
plate afresh.

The statistical fluctuations in observed $N^i_1$ and $N^i_2$ will result
in a distribution in the observed ratio, having a standard
deviation given by
\begin{eqnarray}
(\sigma_{Z^i})^2 = \frac{(\sigma_{N_1^i})^2}{(N_2^i)^2} +
        \frac{(N_1^i)^2\,(\sigma_{N_2^i})^2}{(N_2^i)^4}\, .
\label{e:sdZ}
\end{eqnarray}
The standard deviations in  $N^i_1$ and $N^i_2$ are related to 
those in star-$plus$-sky and sky counts by the following:
\begin{eqnarray}
(\sigma_{N_1^i})^2 = (\sigma_{N_1^{i(star+sky)}})^2 + (\sigma_{N_1^{sky}})^2 \nonumber \\
(\sigma_{N_2^i})^2 = (\sigma_{N_2^{i(star+sky)}})^2 + (\sigma_{N_2^{sky}})^2. \nonumber
\end{eqnarray}
Since the counts registered obey Poisson statistics,
$\sigma_{N_1^{i(star+sky)}} = \sqrt{N_1^{i(star+sky)}}$ and 
$\sigma_{N_2^{i(star+sky)}} = \sqrt{N_2^{i(star+sky)}}$.
The standard deviations in the sky counts used will depend on the
average sky counts ($\bar{s}$), the number of positions of half-wave plate
(M), and the times
spent on the sky ($t_b$) and star-$plus$-sky integrations ($t_o$) and are given
by
\begin{eqnarray}
 (\sigma_{N_1^{sky}})^2 =\frac{t_o^2 \bar{s_1}}{M t_b} \quad \mbox{and}
 \quad (\sigma_{N_2^{sky}})^2 =\frac{t_o^2 \bar{s_2}}{M t_b}. \nonumber 
\end{eqnarray}
When the degree of polarization is small, $N_1^i \approx N_2^i$ and
$\sigma_{Z^i} \approx \sqrt{\frac{2}{N}}$, in the absence of any
sky background; here, N is the average
counts recorded per channel at a particular position of the half-wave plate.

The expected residuals (equations~\ref{e:r1} or
\ref{e:r2}, or the corresponding equations,
if any extra light is present) 
at each position of the halfwave plate is calculated
using the expected statistical fluctuations in $Z^i$ using 
equation~(\ref{e:sdZ})
and the probable errors in the derived parameters. These values are
plotted along with the observed residuals at each position of the
halfwave plate by the data reduction program. 
The  plots show the standard deviation of the over all
fit to the data also. A provision to reject largely deviating data based
on their departures from the expected residual is also made. 
There is a possibility that this scheme may result in rejecting a large
number of data when observations are made with the Glan-Taylor prism because
the ratio of the intensities will vary by a large amount over a full
rotation of the halfwave plate.

\subsection{Magnitude of the object}
In most cases it is desirable to know the brightness of the
object at the time of polarimetric observations, particularly,
if it is a variable star. Further, a clear idea of
the prevailing sky conditions
can be obtained by monitoring the brightness of the object being
observed. No separate observations are needed;
the data available are sufficient to derive
the magnitude of the object observed. 
If we sum the counts in the two channels at all M positions of the
half-wave plate, the $Q$ and $U$ terms cancel each other since
observations are made at equal angular intervals over full rotations of
the halfwave plate, and we get
$$ N = 2\, \frac{\sum N_1^i}{M} \quad \mbox{and} \quad
\alpha\,N = 2\, \frac{\sum N_2^i}{M}, $$
which give 
$$ N = 2\, \frac{\sum N_1^i + \sum N_2^i}{M\,(1 + \alpha )}\, .$$ 
The magnitude of the object can be calculated using the relation
$$  m = -2.5\,\log \frac{N}{t}\, ,$$ 
where $t$ is the total time of integration.
Taking the differential of the above expression, we get
$$\partial m = -1.0857\,\frac{\partial N}{N},$$
from which we obtain $$\epsilon_m = \frac{1.0857}{N}\,\epsilon_N =
 \frac{0.7323}{N}\,\sigma_N. $$ 
It is to be remembered that if the sky transparency changes during a
cycle of rotation of the half-wave plate
the various $Q$ and $U$ terms will not exactly cancel and the magnitude
derived will be inaccurate.

When one of the data points is rejected because of its large deviation
the remaining 
$Q$ and $U$ terms do not cancel exactly. Referring to equation~(\ref{e:N12}),
if the data point corresponds
to the position of the half-wave plate which gives the value of $Q$
then the error in $N$ in not including that point is given by
$$ \Delta N = \pm\frac{Q\,(1 - \alpha)}{M-1}. $$
Assuming that (1 $- \alpha$) is $\pm$0.1 and $Q$ is ($N$/10), which
are the maximum
possible values, with M = 16, ($\Delta N/N$) 
turns out to be less than $10^{-3}$, and
the corresponding error in the magnitude derived will be less than
1 millimag. When an even number of data points are rejected
resulting error could be even zero;
the maximum possible error in such a case will be only twice that given above.
Similarly, the maximum possible error when an odd number of
data points are neglected will be on the order of the value given
above. The value of $\alpha$ is expected to be very close to unity and the
stellar polarization, in general, much smaller than 10 per cent, making the
error caused in the magnitude by the rejection of any data point
almost non-existent.

If the extra light due to the presence of a component
is not negligible the magnitude of the object is calculated using
$N$ obtained from
$$  N = 2\, \frac{\sum N_1^i + \sum N_2^i}{M\,(1 + \beta + \alpha )}\, ,$$ 
if it is present in the first beam, and from
$$ N = 2\,\frac{\sum N_1^i + \sum N_2^i}{M\,(1 + \alpha - \alpha\,\beta)}\, ,$$ 
if it is present in the second beam.
As mentioned earlier $\beta$ is made
positive if the component is present in the
first beam and negative if present in the second beam.

If $r_c$ is the number of revolutions that the chopper makes per second,
the integration time for each channel during one rotation of the chopper
is $\gamma$/(2\, $r_c$). The parameter
$\gamma$ depends on the fractional angular size of
each slot during which counting is done, 
and is close to 0.90 in the present case. 
The total time spent on each channel
will be the product of the number of rotations of the chopper ($n_{rc}$),
the number of rotations of half-wave plate ($n_{rhwp}$) and the above
quantity. The total integration time that is needed for the computation of
magnitude is given by
$$ t = \frac{n_{rhwp}\,  n_{rc} \gamma}{r_c}\, .$$
The value of $\gamma$ is not needed; since it is a constant it will be
absorbed in the zero point of magnitude. Since the basic unit for counting is
the time of crossing of a slot over the diaphragm without blocking the
latter, the program computes the actual time of integration at each position
of the halfwave plate from the chopper frequency and the size of the
blind sectors in the chopper disc. 

\subsection{Optimization of observational time in polarimetry}
\label{s:opt}
The value of $\alpha$ is assumed to be 1 in the following error analysis
and rough estimations of the various quantities; such an assumption
will not affect the conclusions at all as $\alpha$ is expected to be
very close to unity.

If there are M positions over a full
circle of rotation of the halfwave plate at which measurements are made,
from equation~(\ref{e:be}) we have
\begin{eqnarray}
z_i = \frac{1 + q\, \cos \theta_i + u\, \sin \theta_i}
{1 - q\, \cos \theta_i - u\, \sin \theta_i}, \quad \theta_i = 4\,\psi^i, 
\quad \mbox{and} \quad i = \mbox{1, 2, ..., M}. \nonumber
\end{eqnarray}
The conditions for the sum of squares of the modified
residuals to be a minimum give
\begin{eqnarray}
q  =&& \frac{\sum(z_i^2 -1)\, \cos \theta_i\, \sum(1 + z_i)^2\, \sin^2\theta_i -
 \sum(z_i^2 -1)\, \sin \theta_i\, \sum(1 + z_i)^2\, \sin \theta_i\, \cos \theta_i}
  {\sum(1 + z_i)^2\, \cos^2\theta_i\, \sum(1 + z_i)^2\, \sin^2\theta_i - 
   (\sum(1 + z_i)^2\, \sin \theta_i\, \cos \theta_i)^2} \nonumber \\
  =&& \frac{A}{B}\, .  \nonumber
\end{eqnarray}
and a similar relation for $u$. Taking the differentials of the above we get
\begin{eqnarray}
B\, \delta q = \delta A - q\, \delta B. \nonumber
\end{eqnarray}
If we keep only the linear terms in $q$ and $u$ the following approximations
will result:
\begin{eqnarray}
z_i = 1 + 2\, q\, \cos \theta_i + 2\, u\, \sin \theta_i, \nonumber \\
z_i^2 = 1 + 4\, q\, \cos \theta_i + 4\, u\, \sin \theta_i, \nonumber \\
z_i^3 = 1 + 6\, q\, \cos \theta_i + 6\, u\, \sin \theta_i \nonumber \quad \mbox{and} \\
(1 + z_i)^2 = 4\, (1 + 2\, q\, \cos \theta_i + 2\, u\, \sin \theta_i). \nonumber 
\end{eqnarray}
Since the measurements are done over a full circle at equal angular intervals,
\begin{eqnarray}
\sum z_i^2\, \sin \theta_i = 2\, M\, u, \quad
\sum z_i^2\, \cos \theta_i = 2\, M\, q, \quad 
\sum(1 + z_i)^2\, \sin^2 \theta_i = 2\, M, \nonumber \\
\sum(1 + z_i)^2 \cos^2 \theta_i = 2\, M, \quad \mbox{and} \quad
\sum(1 + z_i)^2 \,\sin \theta_i\, \cos \theta_i = 0. \nonumber
\end{eqnarray}
With the above substitutions,
\begin{eqnarray}
\delta A = \sum \{4\, M\, z_i\, \cos \theta_i + 4\, M\, q\, (1 + z_i)\, \sin ^2 \theta_i -
4\, M\, u\, (1 + z_i)\, \sin \theta_i\, \cos \theta_i\}\, \delta z_i, \nonumber\\
\delta B = \sum 4\, M\, (1 + z_i)\, \delta z_i \nonumber  \quad \mbox{and}\quad
B = 4\, M^2. \quad \quad \quad \quad \quad \quad \quad \quad \quad \nonumber 
\end{eqnarray}
If $n_1^i$ and $n_2^i$ are the observed
star $plus$ background sky counts at the $i^{th}$ position of the halfwave
plate,
and $s_1$ and $s_2$ are the scaled background sky counts due to the two beams,
\begin{eqnarray}
N_1^i =  n_1^i - s_1 \, \, \mbox{and} \, \, N_2^i = n_2^i - s_2. \nonumber
\end{eqnarray}
With 
\begin{eqnarray}
z_i = \frac{n_1^i -s_1}{n_2^i -s_2} \quad \mbox{and} 
\quad \frac{1}{n_2^i - s_2} = \frac{2}{n_*} (1 + q\, \cos \theta_i + 
u\, \sin \theta_i), \nonumber
\end{eqnarray}
the differential in $z_i$ is given by
\begin{eqnarray}
\delta z_i = \frac{1 + z_i}{n_*}\, (\delta n_1^i - \delta s_1 -
z_i\, \delta n_2^i + z_i\, \delta s_2)\, 
\label{e:dz}
\end{eqnarray}
where $n_*$ is the counts corresponding to the intensity of the object alone.
The expression for the differential in $q$ can be written as
\begin{eqnarray}
 M\, n_*\, \delta q = \sum x_i\, (1 + z_i)\, \delta n_1^i  - 
 \sum x_i\, (1 + z_i)\, z_i\, \delta n_2^i - \delta s_1\, \sum x_i\, (1 + z_i) +
  \nonumber \\
  \delta s_2\, \sum x_i\, (1 + z_i)\, z_i \, , 
\label{e:dif}
\end{eqnarray}
where
\begin{eqnarray}
x_i = z_i\, \cos \theta_i - q \,(1 + z_i)\, \cos^2 \theta_i -
u\, (1 + z_i)\, \sin \theta_i\, \cos \theta_i. \nonumber
\end{eqnarray}
From equation~(\ref{e:dif}) we derive a few useful results in the following
subsections.

\subsubsection{Error in $p$ due to photon noise}
The measurements giving the counts $n_1^i$'s, $n_2^i$'s,
$s_1$ and $s_2$ are all independent. Hence, according to the
law of propagation of errors (Scarborough 1964),
if $\sigma_{n_1^i}$, $\sigma_{n_2^i}$,
$\sigma_{s_1}$ and  $\sigma_{s_2}$ are the
corresponding standard deviations in these quantities, equation~(\ref{e:dif}),
which gives the differential for $q$, can be written as 
\begin{eqnarray}
(M\, n_*\, \sigma_q)^2 = & \sum \left\{x_i\,(1 + z_i)\,
\sigma_{n_1^i}\right\}^2 +
\sum \left\{x_i\,(1 + z_i)\, z_i\, \sigma_{n_2^i}\right\}^2 + \nonumber\\
&  (\sigma_{s_1})^2\, \left\{\sum x_i\,(1 + z_i)\right\}^2 +
(\sigma_{s_2})^2 \left\{\sum (x_i\,(1 + z_i) z_i\right\}^2\, , \nonumber
\end{eqnarray}
where $\sigma_{q}$ is the standard deviation in the normalized Stokes
parameter $q$.
Again keeping only linear terms in $q$ and $u$, we get
\begin{eqnarray}
&x_i\, (1 + z_i) = 2\, \cos \theta_i\, (1+ q\, \cos \theta_i + u\, \sin \theta_i)
\nonumber \quad \mbox{and}\\
&x_i\, (1 + z_i)\, z_i = 2\, \cos \theta_i\, (1+ 3\,q\, \cos \theta_i + 3\, u\, \sin \theta_i).
\nonumber
\end{eqnarray}
The gain ratio of the beams 
$\alpha$ is close to unity  and generally, $q$ $\ll$ 1, 
making the counts $n_1^i$'s and
$n_2^i$'s, and the standard deviations in their determination
similar. Since photons obey Poisson statistics, we have $\sigma_n
= \sqrt{n}$. Assuming $\sigma_{n_1^i} = \sigma_{n_2^i} =
\sigma_{n}$  and $\sigma_{s_1}
= \sigma_{s_2} = \sigma_{s}$, the above equation simplifies to
\begin{eqnarray}
(\sigma_q)^2 = \frac{4}{M\, n_*^2} \left\{(\sigma_n)^2 + 2.5\, M\,
 q^2\, (\sigma_s)^2\right\}. 
\label{e:sq}
\end{eqnarray}
If $\bar{n}$ and $\bar{s}$, respectively, are the averages of the observed
object $plus$ background sky and
background sky counts at various positions of the rotating half-wave plate
in one second in each beam,
and if the object is observed for
$t_o$ and background for $t_b$ seconds at each position of the half-wave
plate, then
\begin{eqnarray}
 & n_* = 2\,(n - s) = 2\,t_o\,(\bar{n} - \bar{s}),& \nonumber \\
& (\sigma_n)^2 = t_o\,\bar{n} \quad \mbox{and} \quad 
(\sigma_s)^2 = (\frac{t_o}{M\,t_b})^2\,M\,t_b\,\bar{s}  = 
\frac{t_o^2}{M\,t_b}\,\bar{s}. & \nonumber
\end{eqnarray}
The average counts are assumed to be the similar in the two channels
corresponding to the two beams, which is the case when the
polarization of the incident light is low.
The total number of counts accumulated in each of the
M$\times$2 of the channels during the object integration
is approximately $t_o\,\bar{n}$ and 
that during the background brightness integration it
is approximately M\,$t_b$\,$\bar{s}$.
While removing
the background contribution from the observed object counts, the
background counts should be scaled by the factor ($t_o/M\,t_b$).
Making use of these values we get
\begin{eqnarray}
\epsilon_q = \frac{0.6745}{\sqrt{M}\,(\bar{n} - \bar{s})}
\left(\frac{\bar{n}}{t_o} + 2.5\,q^2\,\frac{\bar{s}}{t_b}\right)^{\frac{1}{2}}
\, ,
\nonumber
\end{eqnarray}
where $\epsilon_q$ is the probable error in $q$.
There will be a similar relation for the other Stokes parameter $u$.
It is clear from the above that the errors in $q$ and $u$ due to statistical
fluctuations in the accumulated counts depend on the values
of $q$ and $u$ themselves.
We have seen that when $\epsilon_q = \epsilon_u$,
then $\epsilon_p = \epsilon_q$. 
According to the above relation for the error
due to photon noise $\epsilon_q = \epsilon_u$, when
$q$ = $u$, i.e.,  when
$q$ = $p/\sqrt{2}$. This implies that for a given
linear polarization $p$, the observations would yield a minimum error in 
$p$  if $q$ = $u$, all other parameters being the same.
The highest possible error due to photon fluctuations
in $p$, which occurs when either $q$ = $p$,
and $u$ = 0, or $u$ = $p$ and $q$ = 0,
is obtained by replacing $q$ by $p$ in the above equation as
\begin{eqnarray}
\epsilon_p = \frac{0.6745}{\sqrt{M}\,(\bar{n} - \bar{s})}
\left(\frac{\bar{n}}{t_o} + 2.5\,p^2\,\frac{\bar{s}}{t_b}\right)^{\frac{1}{2}}.
\label{e:ep}
\end{eqnarray}

\subsubsection{Rough estimations of total required counts}
\label{ss:rough}
When the background brightness is negligible, i.e., when $\bar{s} \approx 0$,
then equation (\ref{e:sq}) gives 
$$ \sigma_q = \sigma_u = 
 \frac{1}{\sqrt{M}\,\bar{n}}
\left(\frac{\bar{n}}{t_o}\right)^{\frac{1}{2}},$$ and hence,
$$ \sigma_p = \frac{1}{\sqrt{M\,t_o\,\bar{n}}}. $$
If $N$ is the total counts accumulated in all the M$\times$2 channels
then $$ M\,t_o\,\bar{n} = \frac{N}{2} \quad \mbox{and}
\quad \sigma_p = \sqrt{\frac{2}{N}}. $$
For a standard deviation of 0.01 per cent in the measured linear polarization,
$N$ should be of the order of 2$\times 10^8$ counts.
The corresponding integration time will
depend on the brightness of the object.

The above condition can be easily understood the following way. In order
to determine $q$ two measurements of intensities are needed with the
analyzer at position
angles oriented at 0 and 90 degree, respectively. If $I_0$ and
$I_{90}$ represent counts corresponding to
these intensity values, then $q$ is given by
$$ q = \frac{I_{90} - I_0}{I_{90} + I_0}, $$
and the standard deviation in $q$ by
$$ \sigma_q = \sqrt{\left(\frac{2\,I_{90}\,\sigma_{I_0}}
{(I_{90} + I_0)^2}\right)^2+
\left(\frac{2\,I_0\,\sigma_{I_{90}}}{(I_{90} + I_0)^2}\right)^2}. $$
At low polarization levels the two counts will be nearly equal,
and if $I$ is the average intensity, then
$$ \sigma_q = \frac{\sigma_I}{\sqrt{2}\,I} = \frac{1}{\sqrt{2\,I}},
\quad \mbox{with} \quad \sigma_I = \sqrt{I}.$$
A similar number of counts will be required to determine the other Stokes 
parameter $u$. The total counts to be registered $N$ = 2$\times$2\,$I$.
Substituting the value of $I$ from this in the above expression, we get
$$ \sigma_p = \sqrt{\frac{2}{N}}, \quad \mbox{as found before, since} \quad
\sigma_p = \sigma_q \quad \mbox{when} \quad \sigma_q = \epsilon_u.$$

\subsubsection{Optimum background observation}
Equation~(\ref{e:ep}) shows that the probable
error in the linear polarization contains two
terms, one that depends on the time spent to observe the object and the
other on the time
spent to observe the background sky, $t_o$ and $t_b$, respectively.
The second term also depends on the
square of the polarization, indicating that the contribution from an
error in the background sky brightness will be smaller
at lower polarization
levels. The contribution to the total error by
both the terms are inversely proportional to the respective times
spent on observations. We should distribute the total time available between
the object and background observations such that the signal-to-noise
ratio is a maximum, or equivalently the error in $p$ is a minimum, for
an efficient use of the telescope time.
Differentiating the expression for ($\epsilon_p$)$^2$ with
respect to $t_o$, the time spent on object integration, we have
$$ \frac{\partial (\epsilon_p^2)}{\partial t_o} =
 \left(\frac{67.45}{\sqrt{M}\,(\bar{n} - \bar{s})}\right)^2\,
\left(-\frac{\bar{n}}{(t_o)^2} - \frac{2.5\,p^2\,\bar{s}}{(t_b)^2} \,
\frac{\partial t_b}{\partial t_o}\right). $$
Substituting (${\partial t_b}/{\partial t_o}$)$ =  -1$
in the above expression,
since $t_{total}$ (= $t_o + t_b$) is a constant,
and equating $\left({\partial (\epsilon_p)^2}/{\partial t_o}\right)$
to zero to get the condition for maximum or minimum, we have
$$ \frac{t_b}{t_o} = p\,\sqrt{\frac{2.5\,\bar{s}}{\bar{n}}}\,.$$
The second derivative of ($\epsilon_p$$)^2$ with respect to $t_o$ is
always positive, indicating that the above actually represents
the condition 
for $\epsilon_p$ to be a minimum.

In polarimeters where the beam separation is large,
the background sky will be modulated by the rotating half-wave plate
because it will be polarized. The intensities of the
ordinary- and extraordinary-components
at each position of the half-wave plate
should be measured separately
and removed from the object $plus$ background data. When we deal with
such a situation,
$s_1$ and $s_2$ appearing in equation~(\ref{e:dz}) should be replaced by
the corresponding $s^i_1$'s and $s_2^i$'s, which
have uncorrelated statistical errors.
The differential in $z_i$  then will be modified to
\begin{eqnarray}
\delta z_i = \frac{1 + z_i}{n_*}\, (\delta n_1^i - \delta s_1^i -
z_i\, \delta n_2^i + z_i\, \delta s_2^i) \nonumber 
\end{eqnarray}
and equation~(\ref{e:dif}) will become
\begin{eqnarray}
 M\, n_*\, \delta q = \sum x_i\, (1 + z_i)\, \delta n_1^i - 
 \sum x_i\, (1 + z_i)\, z_i\, \delta n_2^i -
\sum x_i\, (1 + z_i)\delta s_1^i\, \nonumber  &\\
 +  \sum x_i\, (1 + z_i)\, z_i\, \delta s_2^i\, , &
\end{eqnarray}
and the expression for the standard deviation in $q$ will be reduced to
\begin{eqnarray} 
(\sigma_q)^2 = \frac{4}{M\,(n_*)^2}\left\{(\sigma_n)^2 + (\sigma_s)^2\right\}.
\nonumber
\end{eqnarray} 
Again we have used $\sigma_{n_1^i} = \sigma_{n_2^i} =
\sigma_{n}$  and $\sigma_{s_1}
= \sigma_{s_2} = \sigma_{s}$ and the approximations given in \S\, \ref{s:opt}.
With 
\begin{eqnarray} 
(\sigma_s)^2 = \left(\frac{t_o}{t_b}\right)^2\,t_b\,\bar{s}, \nonumber
\end{eqnarray} 
the expression for $\sigma_q$ becomes
\begin{eqnarray} 
\sigma_q = \frac{1}{\sqrt{M}\,(\bar{n} - \bar{s})}
\left(\frac{\bar{n}}{t_o} + \frac{\bar{s}}{t_b}\right)^{\frac{1}{2}},
\nonumber
\end{eqnarray} 
independent of the value of $q$.
There will be a similar expression for the error in $u$.
Since $\epsilon_q$
= $\epsilon_u$, the probable error in $p$ is given by
\begin{eqnarray} 
\epsilon_p = \frac{0.6745}{\sqrt{M}\,(\bar{n} - \bar{s})}
\left(\frac{\bar{n}}{t_o} + \frac{\bar{s}}{t_b}\right)^{\frac{1}{2}},
\nonumber
\end{eqnarray} 
and we get the condition for minimum error in $p$ for a given
available time for observation as
\begin{eqnarray} 
\frac{t_b}{t_o} = \sqrt{\frac{\bar{s}}{\bar{n}}}.
\nonumber
\end{eqnarray} 

\begin{figure}[htb]
\centerline {\psfig{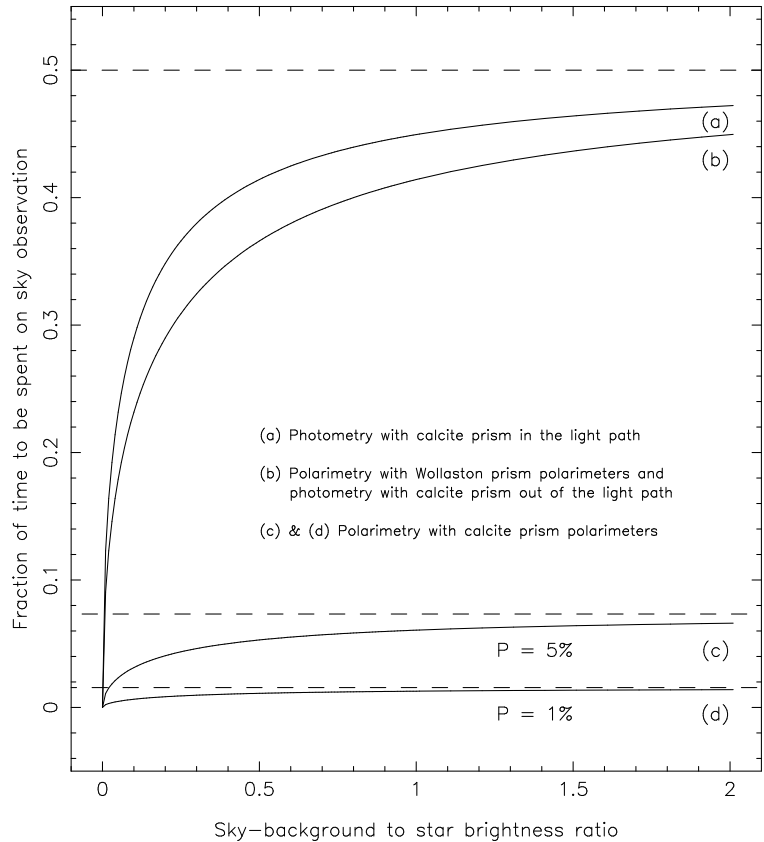}}
\caption {Plot of the optimum fraction of time to be spent on background
sky observation
with  beam displacement- and Wollaston-prism-based polarimeters against the
sky-background to star brightness ratio.
Optimum fractions of time to be spent for photometry with
the beam displacement prism
in and out of light path are also shown. The dashed lines show the
corresponding asymptotic values.}
\label{f:optsky}
\end{figure}

Figure~\ref{f:optsky} 
shows the fraction of time to be spent on background sky
observations
as a function of the ratio of background brightness to the object
brightness with two different types of polarimeters, one with
well separated ordinary- and extraordinary-beams as in the case
of Wollaston or Foster prism-based polarimeters, and the other
with overlapping ordinary- and extraordinary-beams as in the
case of beam displacement prism-based polarimeters. In the case
of overlapping beam polarimeters the fraction of time to be spent
on background observation increases with the polarization of the
object and at low polarization levels, which is usually encountered
in stellar polarization measurements, it is negligibly small,
indicating that most of the available time can be spent on
observing the object, and thereby, partially compensating for the loss
in efficiency in not utilizing 50 per cent of the
light collected by the telescope.

As a specific example, when p = 1 per cent and $\bar{s} = \bar{n_*}$,
in the case of the beam displacement prism-based polarimeters,
the time to be spent on measurement of background brightness, $t_b$,
is less than 2 per cent
of the total time available for observation for 
the error in the polarization derived to be a minimum.
In the case of Wollaston or Foster prism-based polarimeters if the same
object is observed the background brightness will be half of that
observed with the above mentioned type of polarimeters,
since only one component is included in the measurement along with
the object, and the time to be spent on background observations is
about 37 per cent. 

\begin{figure}[htb]
\centerline {\psfig{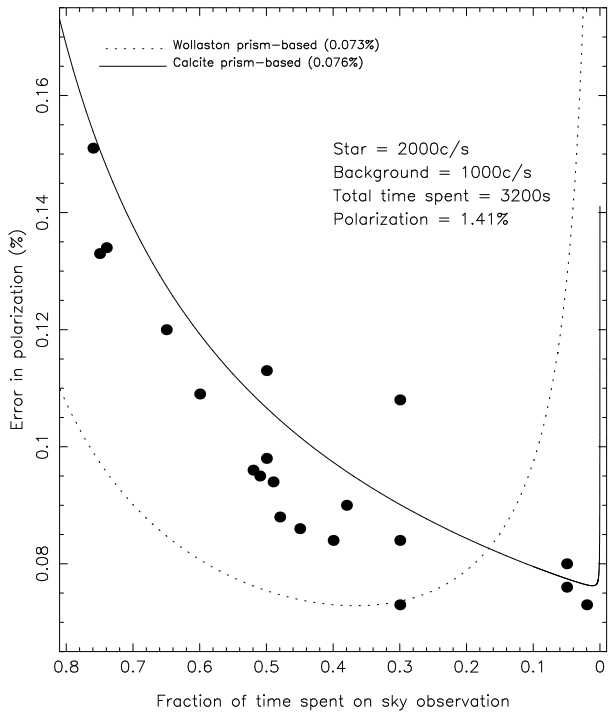}}
\caption {Plot of the error in polarization due to photon noise
against the fraction of time spent on background sky observation.
The continuous curve shows the expected error in polarization for
overlapping beam polarimeters and the dashed curve that
for well-separated beam polarimeters. The figures in the brackets indicate
the minimum error that can be achieved if the object and background sky
observations are optimized. The filled circles show the
results of numerical experiments for a beam displacement prism-based
polarimeter.}
\label{f:numexp}
\end{figure}

The variation of expected error in polarization as a function of
the fraction of time spent on background observation from a total
available time of 3200\,$s$ is plotted in Figure~\ref{f:numexp}. 
The continuous curve represents the 
case of a beam displacement prism-based polarimeter, and the dashed curve
the case of a Wollaston or Foster prism-based polarimeter. The total
counts due to the object is taken as 2000\,$s^{-1}$ and that due to
the background as 1000\,$s^{-1}$.
The filled circles in the figure, which show the results of
a numerical experiment for the case of beam displacement prism-based
polarimeter with counts having a Poisson distribution,
follow the expected curve, but lie systematically below. This is due
to the departure of the randomly generated counts,
which do not include counts larger than three times the standard deviation, 
from a strict Poisson distribution.  Both $q$ and $u$ were assumed
to be equal to 1.0 per cent for the numerical exercise, and the
number of halfwave positions was taken as 16.
It is clear that in the case of beam displacement prism-based polarimeters
almost the entire time available can be
spent on observing the object even when the background brightness
is the same as the brightness of the object in one of the beams.

\subsubsection{Error in $p$ due to variations in the background brightness}
During moonlit nights if passing clouds are present the background
brightness can vary randomly.
If the sky brightness fluctuates during the observations an accurate
determination of the background counts which are to be subtracted from
the object $plus$ background sky data is extremely difficult, in which case
the background
counts used in the data reduction may significantly differ from their
true values. However, the differences between these,
$\epsilon_{s_1}$ and $\epsilon_{s_2}$,
which may be considered as errors in the background brightness determination,
for the two channels due to the two beams will be correlated, the reason
being the in-phase variation in sky brightness in the two beams.
In the case of such a correlation equation~(\ref{e:dif}) becomes
\begin{eqnarray}
 M\, n_*\, \delta q =  \sum x_i\, (1 + z_i)\, \delta n_1^i 
 -  \sum x_i\, (1 + z_i)\, z_i\, \delta n_2^i - \nonumber\\ 
   \delta s\, \sum \left\{x_i\, (1 + z_i) -
  \,  x_i\, (1 + z_i)\, z_i\right\} \, , \nonumber
\end{eqnarray}
with $\delta s_1$ =  $\delta s_2$ = $\delta s$. Setting the standard deviation 
$\sigma_{n_1^i}$ and $\sigma_{n_2^i}$ equal to
$\sigma_n$, and $\sigma_{s_1}$ and $\sigma_{s_2}$ to $\sigma_{s}$,
we get
\begin{eqnarray}
(M\,\sigma_q)^2 = \frac{4 M}{(n_*)^2}\left\{(\sigma_n)^2 +
M\,q^2\,(\sigma_s)^2\right\},
\nonumber
\end{eqnarray}
giving the maximum possible probable error in $p$  as
\begin{eqnarray}
\epsilon_p = \frac{2\times 0.6745}{\sqrt{M}\,n_*}\left\{(\sigma_n)^2 +
M\,p^2\,(\sigma_s)^2\right\}^{\frac{1}{2}}.
\nonumber
\end{eqnarray}

In order to estimate the error in the derived polarization
due to an error in the determination of
background brightness owing to fluctuations in it, possibly caused
by passing clouds during the observation, we 
assume the standard deviation $\sigma_n$ to be negligibly small.
The above equation
then reduces to
$$ \epsilon_p = \frac{2\, p\,\epsilon_s}{n_*}\, ,$$
where $\epsilon_s$ is the probable error in the background brightness used.
If the sky brightness fluctuates by 10 per cent during the observation,
the average change over the intervening period will be a maximum of 5 per cent.
It is clear from the above expression that the error in the polarization
depends on the relative brightnesses of the object and the background.
If s = ($n_*$/2) then $\epsilon_p$ = ($p$/20), i.e., the maximum
uncertainty in the observed polarization will be 5 per cent 
of its actual value because of
using a wrong value for the sky background. If the sky brightness varies
linearly during the observation of the object
it can be accurately removed from the data by
taking the average of the measurements before and after the
object integration.
During moon-rise and moon-set, when the background changes rapidly,
the sky brightness should be observed frequently so that a linear
variation for the background can be assumed
for the intervening periods when the object
integrations are made. It may be noted that the position angle of polarization
will be unaffected by variations in sky brightness, since there is no
background polarization involved in the measurements.

\subsubsection{Instrument in the photometric mode}
The  instrument  can be used in the photometric mode also. Ideally, the
calcite block should be  removed  from  the  beam  for  photometry.  At
present  there is no provision in the instrument for doing so
easily. For photometry,
observations are made over a full  rotation  of  the  half-wave
plate  and  the  counts  obtained  at each position are added together.

The condition for optimum of background brightness observation is
different when the instrument is used in the photometric mode. If we
sum all the counts registered at M positions of the half-wave plate,
from equation~(\ref{e:N12}) we get
\begin{eqnarray}
M\,n_* = \sum n_1^i + \sum n_2^i - M\,s_1 - M\,s_2.
\nonumber
\end{eqnarray}
Taking the differential of the above equation
and writing down the probable error in $n_*$, we get
\begin{eqnarray}
(M\,\epsilon_{n_*})^2 = \sum (\epsilon_{n_1^i})^2 +
\sum (\epsilon_{n_2^i})^2  + M^2\,(\epsilon_{s_1})^2 +
M^2\,(\epsilon_{s_2})^2. 
\nonumber
\end{eqnarray}
With $\epsilon_{n_1^i}$ = $\epsilon_{n_2^i}$ =
  $\epsilon_n$, and
 $\epsilon_{s_1}$ = $\epsilon_{s_2}$ = $\epsilon_{s}$, the above
equation reduces to
$$ (\epsilon_{n_*})^2 = \frac{2}{M}\,\left\{(\epsilon_n)^2 +
M\,(\epsilon_s)^2\right\}, $$
and for the probable error in magnitude we get
\begin{eqnarray}
 (\epsilon_{mag})^2 = \left(\frac{0.7323}{n_*}\right)^2\,(\sigma_{n_*})^2
 =  \frac{2}{M}\,\left(\frac{0.7323}{n_*}\right)^2\,
\left\{(\sigma_n)^2 + M\,(\sigma_s)^2\right\},\nonumber 
\end{eqnarray}
which on substituting
$$n_* = 2\,(n - s), \quad n - s = t_o\,(\bar{n} -\bar{s}), \quad
(\sigma_n)^2 = t_o\,\bar{n} \quad  \mbox{and} \quad
(\sigma_s)^2 = \left(\frac{t_o}{M\,t_b}\right)^2\,M\,t_b\,\bar{s}, $$
gives
$$ (\epsilon_{mag})^2 = \frac{(0.7323)^2}{\sqrt{2 M}\,(\bar{n} - \bar{s})}
\, \sqrt{\frac{\bar{n}}{t_o} + \frac{\bar{s}}{t_b}}\, . $$
When $\bar{s} \approx$ 0, the above gives  
$$ \epsilon_{mag} = \frac{0.7323}{\sqrt{2 M \,t_o\,\bar{n}}}
= \frac{0.7323}{\sqrt{Total~number~of~counts~registered}}\, . $$ 
The condition for optimum background observation can be obtained 
by differentiating the expression for the error in magnitude
with respect to $t_o$ and setting
it to zero as done earlier. The optimum time to be
spent on background brightness observations is given by
\begin{eqnarray}
\frac{t_b}{t_o} = \sqrt{\frac{\bar{s}}{\bar{n}}}\, . \nonumber
\end{eqnarray}
The above condition is the same as that for Wollaston or Foster prism
based polarimeters. If photometry is done with the beam displacement
prism in the light path then about 42 per cent of the total time should be
spent on background observation when $\bar{s} = \bar{n}$ to
minimize the error. If the same
object is observed with the prism out of the beam then, as in the
case of Wollaston prism polarimeters, the corresponding
time turns out to be about 37 per cent, since the object brightness
would become double of that with the prism in the light path.

\section{Control electronics}\label{s:conele}
\begin{figure}
\centerline {\includegraphics[width=13cm]{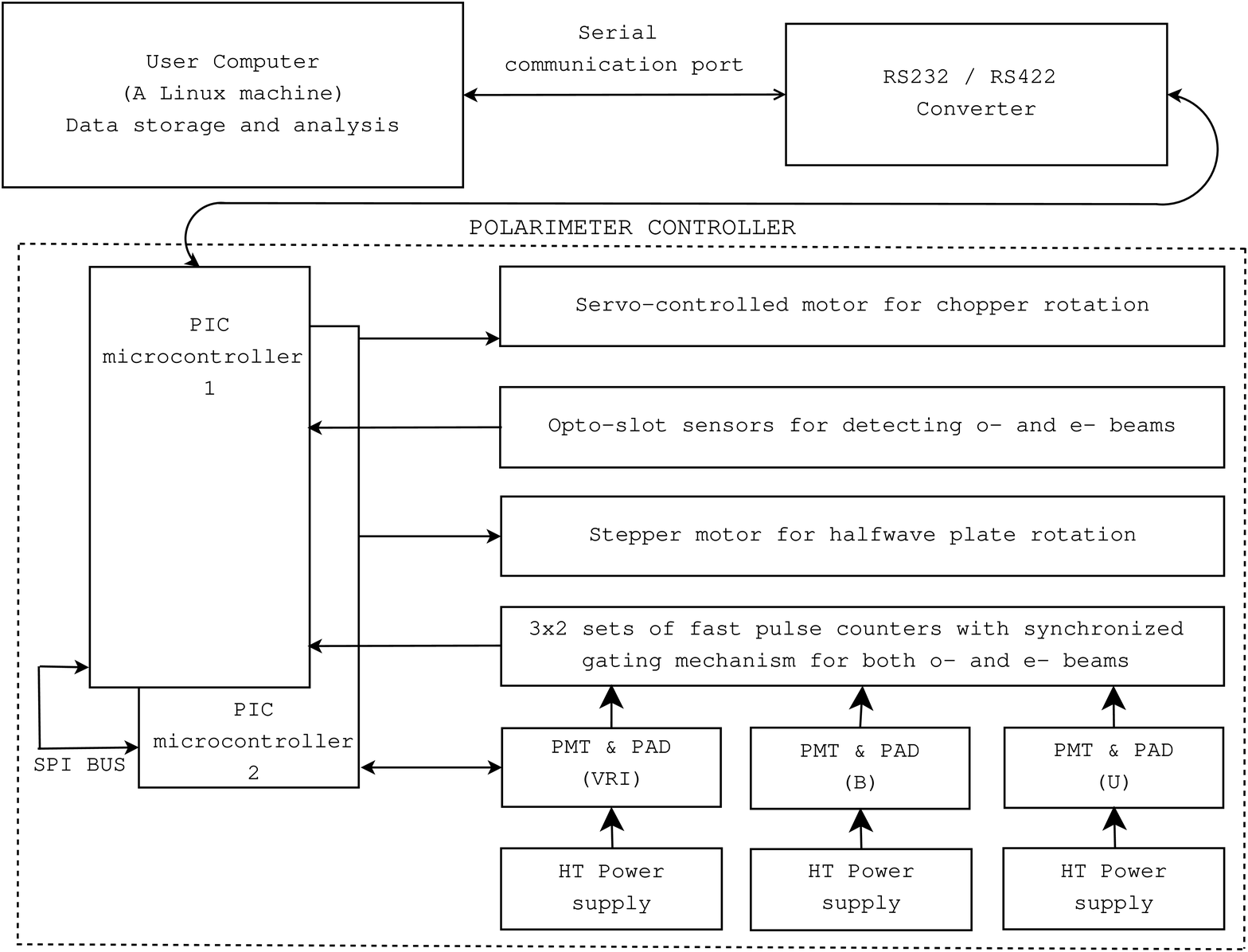}}
\caption {Schematic block diagram of the control and counting electronics.
\label{f:inte}}
\end{figure}

The operation of the polarimeter is done using a Linux machine and
all the polarimeter functions are performed by two PIC microcontrollers. 
One microcontroller controls  both the step motor
coupled to the half wave plate and the servo motor coupled to the
chopper wheel. The second microcontroller interfaces with the
computer with Linux
loaded  through serial communication and interprets the commands from the
computer and executes them in an orderly way. We have used the 
inexpensive and widely  available PIC family PIC16F877A1 microcontrollers in 
the control electronics.  The schematic  block diagram given in
Figure~\ref{f:inte}  shows the overall scheme of  the implementation of the
polarimeter controller unit, which is mounted onto the polarimeter.    
The communication between the user computer and the microcontrollers is through
a standard serial communication port available in the computer in  an
asynchronous mode with the RS232C protocol. In order to have a larger
operating distance between the computer and the controller,
the RS232 signals are converted to RS485 signals through an adapter 
placed at the computer end.

We have used a command-based communication protocol, 
wherein the computer acts as the master and the microcontroller as a slave
unit.  Always the commands are generated from the computer and sent to the
microcontrollers, which
interpret and execute them, and then respond indicating the
status of the execution to the computer. 
We followed this scheme to avoid any communication clash between the computer
and the microcontrollers, and to identify the errors, if any, quickly.
The entire operation of the polarimeter, including the data
acquisition, is based on a set of predefined commands.
The two microcontrollers in the instrument 
communicate between them through the built-in SPI Bus.

The firmware of the microcontrollers was written in the PIC microcontroller
assembly language using the Microchip\textsuperscript{\textregistered}
developer tools, and then using the
ICD2 programmer, it was uploaded to the microcontrollers.

\subsection{Chopper and halfwave plate rotations}
The chopper disc used for isolating the ordinary and extra-ordinary beams
is made of a light weight aluminium composite material.
The data acquisition timing accuracy mainly depends on the
stability of the chopper rotational frequency.
We have used a servo motor with the model number Smartmotor\textsuperscript{TM}
SM2315, procured from  Animatics,  Santa Clara, California, USA.
The  motor has a built-in encoder and a DSP controller, and can be configured
and   controlled  by communicating with the DSP controller through the RS232
serial communication port provided on the motor body.
The motor is capable of rotating up to 10000 revolutions
per minute,  maintaining the same accuracy in speed
throughout.  
The chopper is directly attached to the servo motor shaft to avoid any
possible differential motion. 
The opto-coupled sensors associated with the rotating chopper detect the
slots corresponding to the  ordinary and extra-ordinary beams and generate
out of phase square wave pulses of 5~V  height. 
The rotational speed of the chopper can be set to any value between 60 and
6000~rpm through the command from the computer accurately.
When the chopper rotation command 
along with the required rpm as the data is received
from the computer by the first PIC controller, it transfers 
the command data to the second PIC controller  through the 
internal SPI bus. The
 second PIC controller then communicates with servo motor's DSP controller
through a sequence of commands and sets the required rpm. In order to make sure
that the motor is really spinning at the set rpm before the starting of the
observation, the controller measures the
rpm of the motor through the opto-slot sensor waveform and gives the feedback
to the first controller, and thereby, to the polarimeter operating computer.

The halfwave plate is rotated using a step motor of model number MO61-FC02,
 procured from  Superior Electric. It has a step resolution of 0.9 degree
when operated in the half-step mode and completes one full rotation in
400 half-steps. The driver for the motor is designed
using the standard step motor driver ICs LM297 \& LM298 pair. An arbitrary
starting mark is created with an opto-coupler arrangement on the gear fixed
to the shaft of the step motor. Every time when the halfwave plate starts 
rotating for observations for the given number of position angles, it 
always starts from the marker position and on completion of a full
rotation, it is brought back to it. This way the proper
positional accuracy of the halfwave plate is ensured.          

\subsection{Pulse counting system}
\begin{figure}
\centerline {\includegraphics[width=13cm]{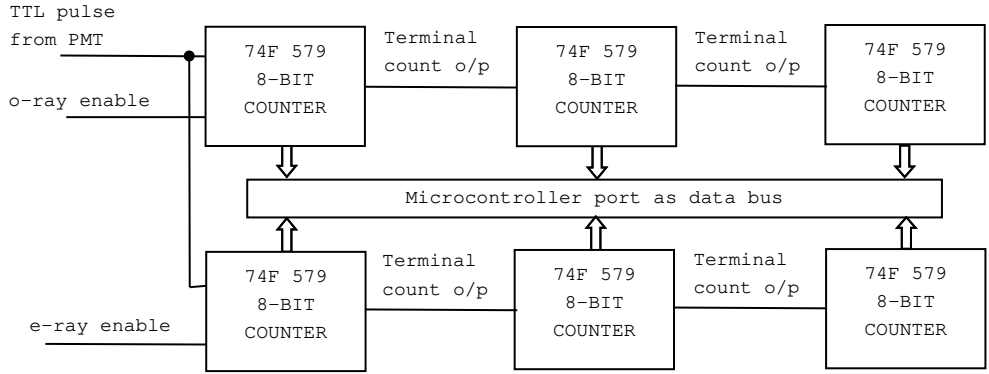}}
\caption {Schematic block diagram of a set of PMT pulse counters.
\label{f:puls}}
\end{figure}

The pulse amplifier-discriminators of Model ETL AD6 and Hamamatsu
Model C9744 are used to interface the ETL~9893Q/350 and Hamamatsu R943-02
photomultiplier tubes, respectively, to the pulse counters.
The PADs amplify the photon generated current pluses
above the set threshold level to TTL voltage pulses for counting.
The pulse pair resolution of Model AD6
is 20\,ns while that of C9744 is 25\,ns.
The pulse counter is a 24-bit TTL logic counter designed with three numbers of 
74F7592, an 8-Bit bidirectional counter with 3-state output.
The maximum counting frequency of the ICs is typically 100~MHz, and
the pulse pair resolution is about 10~ns.
They also have parallel outputs with tri-state control, and
therefore, it is easy to integrate the ICs with the bus without any
additional logic. One PMT requires two of these 24-bit counters,
one for the ordinary beam and the other for the extra-ordinary beam. 
Therefore, three sets of dual counters are required for the three PMTs used.
The gating pulses for the selection of the counter for the ordinary and
extra-ordinary beam pulses
come from the opto-sensors associated with the chopper disc.
These gate pulses are monitored by the 
microcontroller for the count-integration purpose. The counters can store up to
16,777,216 pulses without getting saturated.
A schematic block diagram of a set of pulse counter is shown in
Figure~\ref{f:puls}.

\subsection{Rough estimates of maximum allowed count rates}
The operating stability of a photomultiplier tube depends on the
magnitude of the average anode current drawn from it.
If a tube is used continuously over a long period of time
at a high average anode current,
variations in the output current occur even without
any change in the operating conditions.
A high anode current may cause even fatigue effects.
It also reduces the life-time of a photomultiplier tube;
the half-life, where the gain for a particular fixed
high voltage drops by a factor of two, of a photomultiplier tube is
proportional to the amount of total charge drawn from the anode.
The charge drawn before the half-life is reached,
which is on the order of 100 Coulomb, varies for individual tubes.
The recommended anode current for stable performance
for ETL tubes is 10\,$\mu$A, while for Hamamatsu tubes it is 
less than 1\,$\mu$A.

In the case of Hamamatsu R943-02, the anode current averaged
over any interval of 30 $s$ maximum should not exceed 1\,$\mu$A,
otherwise permanent damage may be caused to the last dynode by
the heavy electron bombardment. For ETL 9893B/350 tube the
maximum specified current is 200\,$\mu$A. This current can be
sustained only for a brief time because ionization will eventually
lead to catastrophic electrical break down in the multiplier
causing permanent damage. The recommended maximum anode current
for ETL tube is about 100\,$\mu$A. The photomultiplier tubes
should be protected from any accidental bright illumination
during the observations. Peak anode currents exceeding 100\,mA, but
lasting only on the order of 100\,ns, produced by short pulsed
light sources, usually,
do not cause any permanent damage to the photomultiplier
tubes. Since the tubes are operated in the pulse
counting mode the maximum permissible counting rates should be
known so that either the electronic shutter can be closed, or
the high voltage supply to the tubes can be turned down immediately 
in the case of any accidental illumination of the photocathode
by bright sources.

If $G$ is the photomultiplier current gain, the charge contained
in a pulse is given by
$$ Q = 1.6\times 10^{-19}\,G, $$
which produces a current
$$ I_a = \frac{Q}{\tau} = \frac{1.6\times 10^{-19}\,G}{\tau}. $$
In the above expression $\tau$ is the full width at half maximum (FWHM)
of the current pulse reaching the anode.
The charge and current are expressed in Coulomb and Ampere, respectively.

\begin{table}[htb]
\begin{center}
\caption{Rough estimates of the maximum 
allowed pulse count rates for the safe operation
of the photomultiplier tubes at different operating voltages.}
\medskip
\label{t:maxall}
\begin{tabular}{ccc|ccc}
\hline
&&&&&\\
\multicolumn{3}{c|}{ETL 9893/350} & \multicolumn{3}{c}{H R943-02R}\\
\multicolumn{3}{c|}{$I_a$ (max) = 100 $\mu$A}     &  
       \multicolumn{3}{c}{$I_a$ (max) = 0.5 $\mu$A}\\
&& Maximum &&& Maximum\\
 Voltage (V) &  Gain  & counts s$^{-1}$ & Voltage (V)  & 
Gain  & counts s$^{-1}$ \\
&&&&&\\
\hline
&&&&&\\
1600 &  3$\times 10^6$ & 2.1$\times 10^8$ &  1500 & 5$\times 10^5$ & 6.2$\times 10^6$\\
1800 & 10$\times 10^6$ & 6.3$\times 10^7$ &  1700 & 1.6$\times 10^6$ & 2.0$\times 10^6$\\
2000 & 36$\times 10^6$ & 1.7$\times 10^7$ &  1900 & 4$\times 10^6$ & 7.8$\times 10^5$\\
2200 & 60$\times 10^6$ & 1.0$\times 10^7$ &  2100 & 9$\times 10^6$ & 3.5$\times 10^5$\\
2300 & 110$\times 10^6$ & 5.6$\times 10^6$ &  & &\\
&&&&&\\
\hline
\end{tabular}
\end{center}
\end{table}

The number of counts registered per second by the
photon counting system when the average anode current is $I_a$
is given by
$$ n_{counts} \approx \frac{I_a}{1.6\times 10^{-19}\,G}\, . $$
The anode current depends on the current gain of photomultiplier, and
hence, on the operating voltage.
The maximum allowed count rates for different operating
voltages given in Table~\ref{t:maxall}
 are calculated after
setting $I_a$ to the values of the maximum allowed anode current 
for the safe operation of the photomultiplier tubes. 
The tubes should not be illuminated
for an appreciable time by light
sources that would produce count rates above that given in the table
at the corresponding operating voltages.
There will be a small contribution to the anode current from the
leakage currents flowing into the anode because of the applied
biasing voltages of the dynodes. The contribution to the anode
current from the current pulses below the threshold level of the
pulse amplifier-discriminator is neglected in the rough estimation
of the allowed maximum count rates.

\section{Data acquisition and analysis program}\label{s:datacq}

\begin{figure}[htb]
\centerline{\psfig{figure=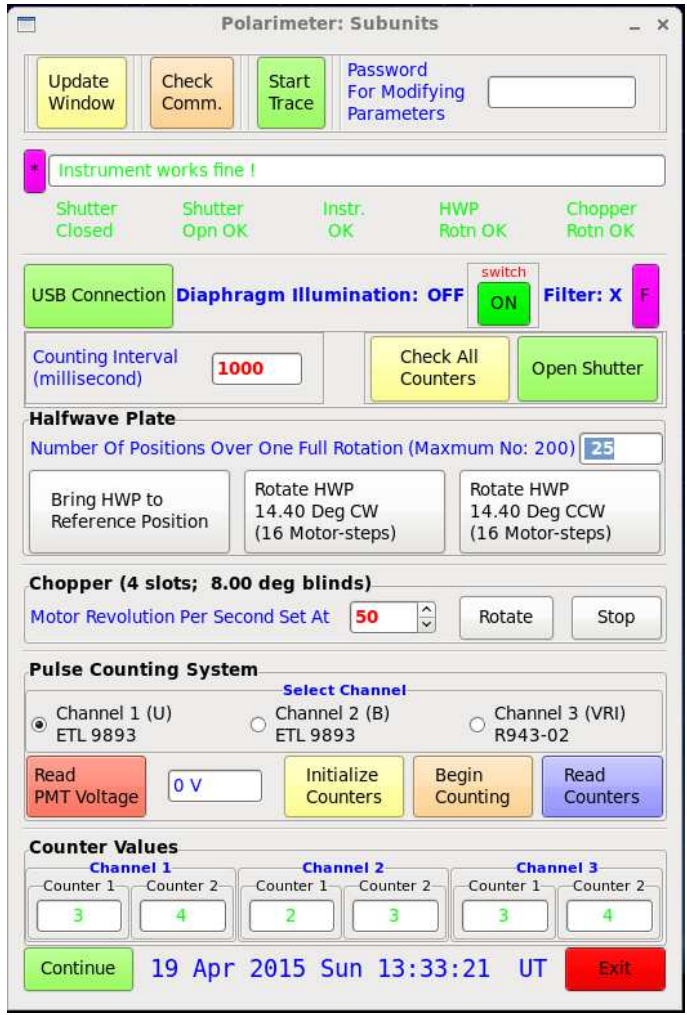,width=9.0cm,angle=0}}
\caption {GUI for the instrument set-up and check-up of subunits.}
\label{f:instrset}
\end{figure}

A fully automated
data acquisition procedure is not feasible at all; several
objects will have to be observed during a night; 
switching between the object and background sky integration will have
to be done a few unspecified number of times
if the background brightness varies during the
observation of an object; observations will have to be discontinued if
the prevailing sky conditions become bad;
spectral bands of observation will have
to be changed. There are several more tasks that may have to be
performed, requiring keyboard operations or mouse button clicks under very low
ambient illumination, if not total darkness, that is possible
when observations are in progress.
Definitely, the whole procedure of data acquisition will
have to be done interactively; it is extremely difficult to
define in advance the sequence of observations that has to be
followed.
\begin{figure}[htb]
\centerline{\psfig{figure=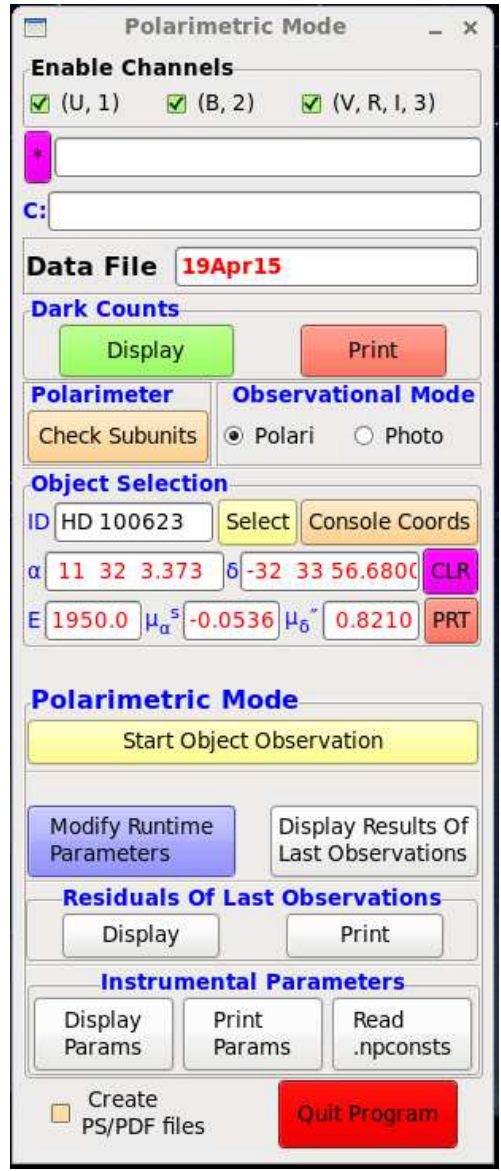,width=6.50cm,angle=0}}
\caption {GUI for the beginning of observations with the polarimeter,
in the polarimetric mode.}
\label{f:polmode}
\end{figure}

For an efficient use of the telescope time, it is highly
imperative that the number of keyboard operations or mouse button clicks
that is necessary to operate the instrument
is kept to a minimum during observations, and thereby, avoiding
any waste of time because of searching for keys or making corrections
for the wrong keys entered. The acquisition program
at the same time
should have the provisions to interactively change the sequence
of execution of its various sub-functions, executing additional
functions, like, displaying intermediate values which are
usually suppressed, should there be a need. 

Most of the tasks in the data acquisition and reduction program
that we have developed are performed with a single key-stroke or
mouse-click to avoid any waste of time;
only those tasks which cannot be easily undone or may cause 
loss of data already acquired are performed after getting confirmation.
The program has the provision to store and process,
including the correction for instrumental polarization, 
the data simultaneously in eight spectral bands. 
The system time required for the data reduction
is negligibly small when compared to the observation
time. The final results, linear polarization, position angle of 
polarization and the
magnitude of the object, can be displayed almost immediately as soon
as a full rotation of the half-wave plate is complete. If the instrumental
polarization and the zero offset of the position angle are known
appropriate corrections can also be made, and the final results
which do not require any further processing can be obtained
as soon as the object integrations are over.

\begin{figure}[htb]
\centerline{\psfig{figure=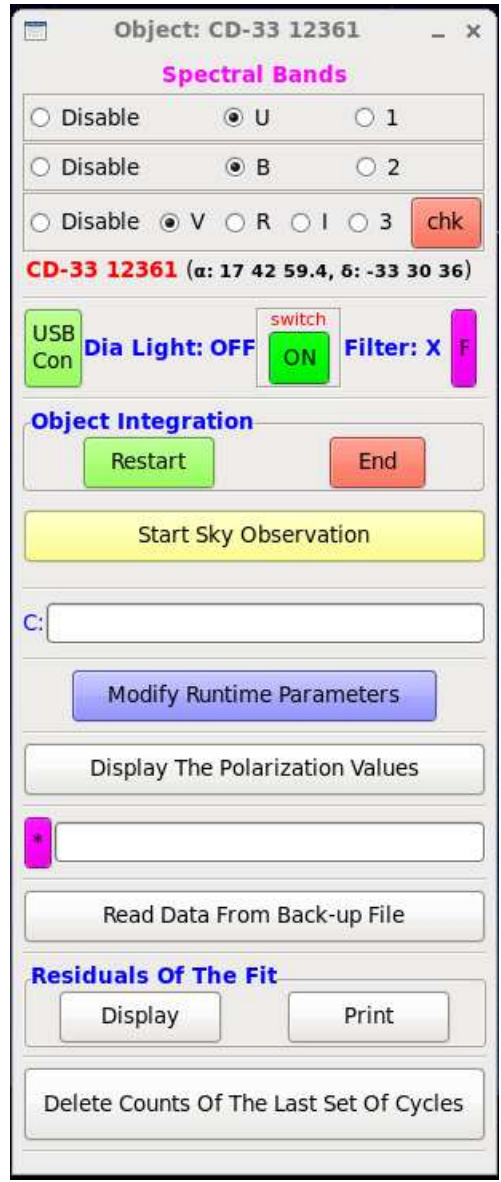,width=6.50cm,angle=0}}
\caption {GUI for the object observations in the polarimetric mode.}
\label{f:objobs}
\end{figure}

\begin{figure}[htb]
\centerline{\psfig{figure=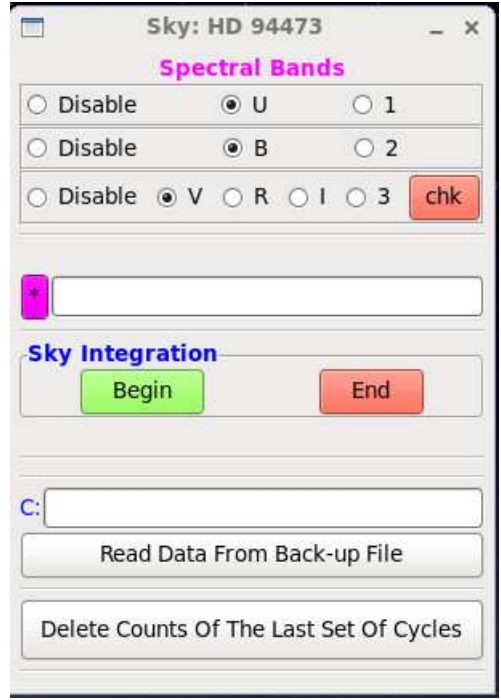,width=6.5cm,angle=0}}
\caption {GUI for the sky observations in the polarimetric mode.}
\label{f:skyobs}
\end{figure}

\begin{figure}[htb]
\centerline{\psfig{figure=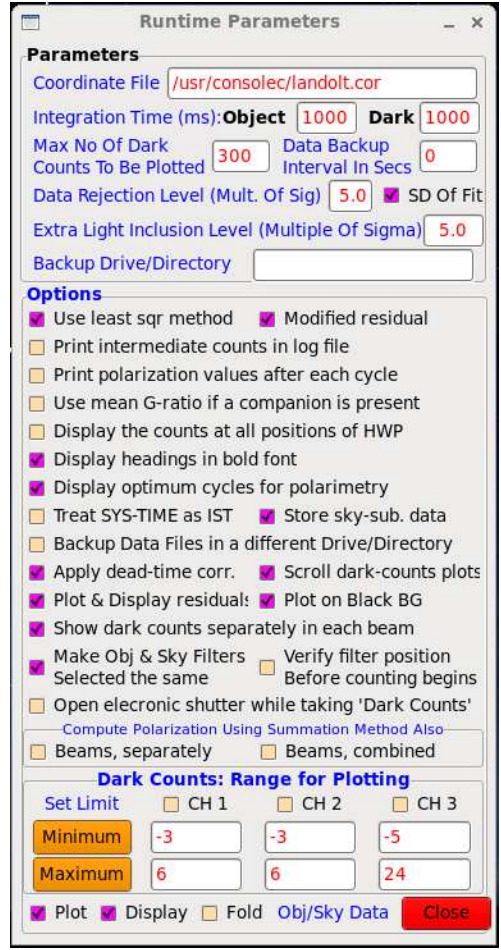,width=6.50cm,angle=0}}
\caption {GUI for the modifications of runtime parameters.}
\label{f:modpar}
\end{figure}

\begin{figure}[htb]
\centerline{\psfig{figure=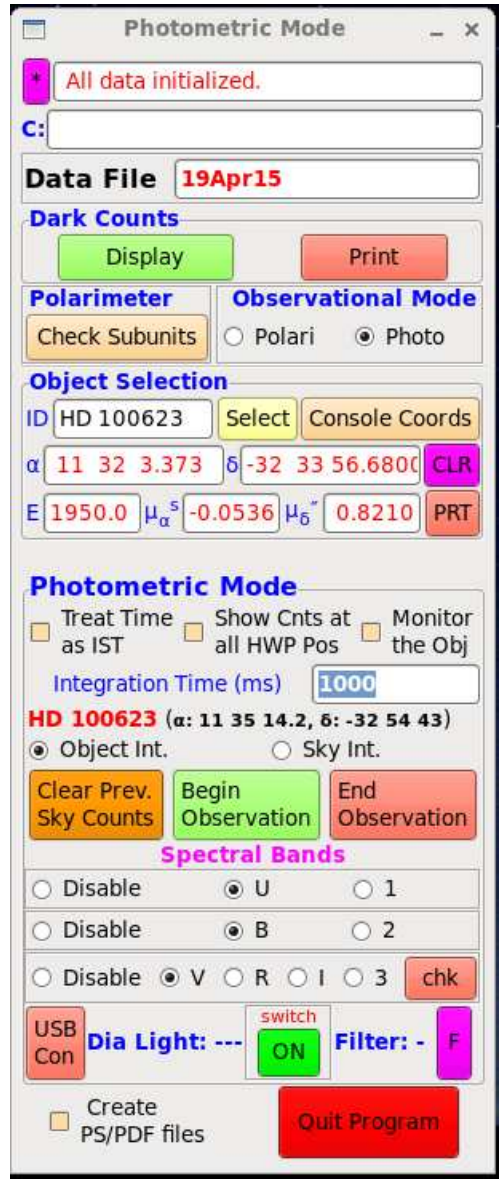,width=6.50cm,angle=0}}
\caption {GUI for the polarimeter in the photometric mode.}
\label{f:phomode}
\end{figure}

The Graphical User Interfaces, which are created using GIMP Tool Kit routines,
are self explanatory.
There are seven main GUI's: (1)~for checking the subunits of the
polarimeter and displaying its operational status, 
(2)~for operating the instrument in the polarimetric mode,
(3)~for operating the instrument in the photometric mode, 
(4)~for object observation in the polarimetric mode, (5)~for sky observation
in the polarimetric mode, (6)~for
modification of the run-time parameters and (7)~for offline polarimetric
data reduction; they are shown Figures~\ref{f:instrset}, \ref{f:polmode},
\ref{f:objobs}, \ref{f:skyobs}, \ref{f:modpar},
\ref{f:phomode} and \ref{f:offline}. 

\begin{table}[htb]
\begin{center}
\caption{Instrumental constants and runtime parameters.}
\medskip
\label{t:instcon}
\begin{tabular}{lr}
\hline
&\\
Parameter & Default value \\
&\\
\hline
&\\
Dead-time coefficients of the three channels & 0.00 \\
Instrumental $q$ (\%) and $u$ (\%) and their errors in all the bands & 0.00 \\
Zero offsets in position angle (degree) in all bands & 0.00 \\
Average gain-ratios ($\alpha$) of the two beams in all bands & 1.00 \\
Standard deviations of average $\alpha$'s in all bands & 0.05\\
Zero offsets of magnitude in all bands & 0.00 \\
Polarization efficiency in all bands & 1.00\\
&\\
\hline
\end{tabular}
\end{center}
\end{table}

We describe briefly the interactively run data acquisition and
reduction programme in the following subsections. For further details
see the manual page of the program.

\subsection{Tasks performed at the startup}
As soon the program is initiated the polarimeter initial set-up and check-up
procedure is run, reporting the status of the various sub-units.
The following tasks are executed at the start-up after making sure that
the communication link with the interface is functioning properly.

\begin{enumerate}
\item {Proper rotation of the motors that control the chopper
and the half-wave plate are checked.}
\item {The functioning of the electronic shutter is checked.}
\item {Outputs from  the photomultipliers are checked by counting
for a specified time interval.}
\item {The chopper is rotated at the pre-set frequency.}
\item {System date and time at the start-up are displayed.}
\end{enumerate}
If any fault is reported in the functioning of the subunits
the program can be either terminated or it can be
continued after rectifying the fault reported.

The time of observation is important when polarimetry of variable stars
are made. If the prevailing sky conditions are good, the magnitudes
computed from the data can be utilized for the photometry
of the object, and for the purpose of computing the airmass of the observed
object, the correct time of observation is needed.
If the displayed time is wrong the system time should be properly set.

The instrumental parameters listed in Table~\ref{t:instcon}
 are read from a file,
if such a file exists.
While reading the file, if an error is encountered the program execution
is terminated after giving a suitable message. If the parameter
file does not
exist in the specified directory, the default values given
in the table are assumed for the program execution.
The instrumental Stokes parameters $q$ and $u$ and their errors
are in percentage and they should be listed for all the spectral bands used
in the observations.

\begin{table}[htb]
\begin{center}
\caption{Program runtime parameters initialized at the startup.}
\medskip
\label{t:prorun}
\begin{tabular}{lr}
\hline
&\\
Parameter & Value \\
&\\
\hline
&\\
Chopper rotational frequency & 50 Hz\\
Number of positions of HWP over one full rotation & 25\\
Counting time to check the pulse counters & 1000\,$ms$\\
Object integration time  & 1000\,$ms$ \\
Dark counts integration time  & 1000\,$ms$ \\
Maximum number of dark count values plotted at a time & 300 \\
Data back-up frequency during observation & 0\,$s$ \\
Output file without extension &  DDMmmYY  \\
Backup drive/directory & None \\
Data rejection level & 5\,$\sigma$ \\
Extra-light inclusion level & 5\,$\sigma$ \\
&\\
\hline
\end{tabular}
\end{center}
\end{table}
In addition to the above, the parameters listed in Table~\ref{t:prorun}
are initialized with the values given against them. All these values
can be altered interactively, if there is any need.
The parameters used in the data acquisition,
the instrumental constants used in the data reduction, and the system date
and time are displayed on the terminal,
indicating that the instrument is ready for carrying out observations.
\begin{figure}[htb]
\centerline{\psfig{figure=offline.ps,width=6.50cm,angle=0}}
\caption {GUI for the offline reduction of the polarimetric data.}
\label{f:offline}
\end{figure}

Several check-buttons that can be set active (On) or inactive (Off)
interactively
have also been provided; these are listed in
Table~\ref{t:toggle1}. 
The initial settings of these toggling options, which can be
either $\lq$On' or $\lq$Off', 
at the startup are indicated in the table against each of them.
On setting some of the toggling option
$\lq$On' those particular options are executed 
while setting them $\lq$Off' the execution of those
options are suppressed.
The other toggling options when set $\lq$On' or $\lq$Off', alternately
activates one of the two different available options;
these are listed in Table~\ref{t:toggle2}. The first option
is executed if the corresponding check button is set $\lq$On' and
the second option, if it is set $\lq$Off'.

\begin{table}[htb]
\begin{center}
\caption{Toggling options initialized at the startup.}
\medskip
\label{t:toggle1}
\begin{tabular}{lr}
\hline
&\\
Toggle option & Status \\
&\\
\hline
&\\
Use least square method to solve the unknowns  & On \\
Minimize the sum of the squares of the modified residuals & On \\
Use standard deviation of fit to remove erroneous data & On\\
Print intermediate counts in the log file  & Off \\
Print polarization values after each cycle  & Off \\
Use average gain-ratio if a companion is present  & Off \\
Display the counts at all half-wave plate positions  & Off \\
Display headings in bold font  & On \\
Display optimum cycles for polarimetry  & On \\
Treat system time as Indian Standard Time  & Off \\
Backup data on a USB drive after observations of each object  & Off \\
Store sky-subtracted data for reduction later  & On \\
Apply dead-time correction to the observed counts  & On \\
Scroll the dark counts window (Max 300 values) & On \\
Plot and display the residuals of the computation & On\\
Plot on black background & On\\
Show/plot dark counts separately in each beam & On \\
Make object filter bands the same as those selected for sky & On\\
Verify filter position before counting begins & On\\
Compute polarization using summation method (beams, separately) & Off\\
Compute polarization using summation method (beams, combined) & Off\\
Fold object/sky data while plotting & Off\\
Use the given maximum/minimum limits for plotting dark counts & Off \\
Plot object/sky data at each position of the HWP & On\\
Display object/sky data at each position of the HWP & On\\
&\\
\hline
\end{tabular}
\end{center}
\end{table}

\begin{table}[htb]
\begin{center}
\caption{Options selected when the check buttons are On and Off.}
\medskip
\label{t:toggle2}
\begin{tabular}{p{7.0cm} p{7.0cm}}
\hline
&\\
Check button active & Check button inactive\\
&\\
\hline
&\\
Use least square to solve the unknowns & Use elimination method to solve\\
Minimize the sum of the squares of the modified residuals &
Minimize the sum of the squares of residuals of intensity ratios\\
Use standard deviation of fit to reject erroneous data & Use the expected
values at each channel to reject erroneous data\\
Use $\alpha_{ave}$ if extra-light is present & Neglect the extra-light \\
Display counts at all HWP positions & 
   Display counts only at the last position \\
Display headings in bold Font & Display headings in normal font to reduce the
terminal brightness\\
Display optimum times for polarimetry & Display those for photometry\\
Treat system time as IST & Treat system time as UT \\
Scroll dark counts window & Fold and overplot with a different
   colour once the maximum number of specified counts are reached\\
Plot and display the residuals of the computation & Plot and display the
 raw data\\
Plot on black background & Plot on white background\\
Show/plot dark counts separately in each beam &
Add the dark counts in the two beams and then display/plot \\
Make object filter bands the same as those selected for the sky &
 Make sky filter bands the same as those selected for the object\\
Fold object/sky data over 360 degrees while plotting & 
  Plot the data sequentially, channel-wise\\
Use the given maximum/minimum limits for plotting dark counts &
Choose automatically from the values available already\\
\\
\hline
\end{tabular}
\end{center}
\end{table}

The dark currents of the photomultiplier tubes
can be monitored when the instrument is in the Polarimetric
or Photometric mode.
The counts are simultaneously displayed and plotted.
The terminal size (25$\times$80) set 
allows the display of a maximum of around 50 values
in each channel at a time; however, a maximum of 300 values can be plotted
at a time.

\subsection{Tasks performed during object integration}
As soon as the observational procedure is initiated four output files
are created in the append mode:
(i) a log-file, for writing a summary of the various tasks
performed and some of the important information that may be needed later
while analyzing the data, (ii) a file to write the polarization and position
angle of the objects observed,
(iii) a file to write the observed
normalized Stokes parameters Q(\%) and U(\%), and (iv) a file to
write the sky-subtracted counts, which can be processed later, if needed.
During the object integration the minimum tasks that should be carried
out are listed in Table~\ref{t:objtask}.

\begin{table}[htb]
\begin{center}
\caption{Minimum tasks to be performed during the object integration.}
\medskip
\label{t:objtask}
\begin{tabular}{l}
\hline
\\
Tasks to be performed \\
\\
\hline
\\
Input the name of the object \\
Select filter-bands for observation\\
Begin object integration \\
Temporarily halt for centring of the object in the diaphragm, etc.\\
Terminate integration in a band, if necessary\\
Perform background sky integration\\
End the object integration to observe another object \\
\\
\hline
\end{tabular}
\end{center}
\end{table}

\begin{table}[htb]
\begin{center}
\caption{Minimum tasks to be performed during the background sky integration.}
\medskip
\label{t:skytask}
\begin{tabular}{l}
\hline
\\
Tasks to be performed \\
\\
\hline
\\
Select filter-bands for observation\\
Begin background sky integration \\
Temporarily halt integration for changing the filters, if needed\\
Terminate integration in a band, if necessary\\
End background sky integration to begin or continue the object integration \\
\\
\hline
\end{tabular}
\end{center}
\end{table}

The name of the object is essential to identify the
polarization measurements later
because the results for all the objects are stored in the same files.
The name of the object, as soon as it is entered,
is compared against the names of unpolarized
and polarized objects contained in two separate lists
to ascertain whether it is one of them. If the
object is found to be in one of the lists, the appropriate designation,
Z for unpolarized and P for polarized,
is appended to the object name. This information can be used later while
sorting the objects from the files created.

The background sky observation can be performed before or after the object
integration or in between if the brightness of the background sky is
suspected to be variable. It is better if the sky observations are done
first so that the polarization values displayed will be the final
values. If the sky is bright it is advisable to observe the sky before and
after the observation of the star.
The minimum tasks to be performed during background
sky observation, which are similar to those performed during the object
integration, are listed in Table~\ref{t:skytask}.

A complete set of observations may take several tens of minutes
if the object being observed is faint. Therefore, it is necessary
to know the status of the observation at any instant of time.
All the relevant details are displayed on the monitor so that
the observer will have a clear cut understanding of what is going
on, how long the process would take and what has to be done next.

The displays during both the sky and object integrations are updated after
each cycle of the half-wave plate rotation so that
the build-up of respective counts can be easily monitored and the integration
terminated when sufficient counts are accumulated.
Once the observation is stopped,
either to centre the object or to discontinue the integration, along with
the results of polarization measurements, the plots of
the residuals of the fit are also displayed.
Once the observation is ended the final results are written in the
respective files.

\subsection{Output files}

From the mean time of observation in each spectral band,
determined from the
beginning and end of object integrations, the Julian day of observation
is computed and recorded along with the results
of polarimetry. In variable star polarimetry a continuously running time
is needed; since the conversion of the civil time
to Julian day is made immediately it saves the trouble of doing so later. 
Brief descriptions of the four output files which  are created
are given below.

\subsubsection{File containing the log of observations}
The log file contains
a record of the various tasks performed during the observations. Since
all the required information connected with the observations
are automatically recorded or can be entered in the log-file
at any time during the observation, there is
no necessity of noting down anything in a register manually. 
This  procedure, in addition to avoiding a rather cumbersome job
to be performed in the ambient dim light, which is generally the case,
also reduces any waste of observational time.
All the instrumental constants used for the
data acquisition and data reduction during the particular
observing run, the beginning and end of the integrations,
the number of rotations of the half-wave plate during the
integration, etc. are also recorded in the log-file.
Any changes made in the reduction procedure or any parameter
connected with the data acquisition or reduction procedure
by toggling the options
are also recorded along with the times at which these changes are made.
The background sky brightness used in the data reduction, and the results
of the polarimetric observations are also recorded.

Intrinsic polarization in many late-type stars has been found to be
closely tied-up with their light variation (Serkowski 1970a; Shawl 1975;
Raveendran 1999b).
A knowledge of the photometric behaviour at the time of polarimetric
observations is essential for a meaningful interpretation of the polarimetric
data of variable stars.
If the sky transparency changes during the integration as a result of thin
passing clouds, the magnitudes derived
in various spectral bands will not be accurate. Since the extinction by
clouds is nearly independent of wavelength in the $UBVRI$ passbands
(Serkowski 1970b), the effect of transparency variations would be negligible
in the colours if they are
derived from simultaneous multiband observations. These
colours could be effectively used in determining the photometric
phases of the variable stars in the absence of accurate magnitudes. For this
purpose the star-$plus$-sky counts in simultaneously observed bands are written
in the log file, and if the sky observations are already available the counts
due to the star are calculated and converted into magnitudes.
These magnitudes should
be corrected by adding $-2.5 \log \{2/(1 + \alpha)\}$
if the ratio of efficiencies
of the two channels ($\alpha$)
is significantly different from unity.
In those cases where a
faint component is present in one of the beams
the magnitudes should be re-calculated from the listed counts.
The colours of the object could be determined from
these data if there is a need. Such a procedure ensures that no
information available in the data collected is lost.

\subsubsection{File containing polarization and position angle}
This file contains the polarization (\%) and position angle (degree)
 of the observed objects.
If the instrumental polarization, polarization efficiency and position angle
offset are available appropriate corrections are made before writing
to these values. The errors
in polarization and position angle contain the errors in the determination
of instrumental polarization also. The mean Julian day of observation and
the magnitude, including the gain-ratio, are also written in the file.

\subsubsection{File containing normalized Q(\%) and U(\%)}
The observed Q(\%) and U(\%), without any correction, including even the
polarization efficiency, are written in a file; these values are stored
with the idea of using
a different set of corrections to the instrumental polarization
and polarization efficiency
at a later date, if there is a necessity of doing so when
improved values of corrections available. The
mean Julian day of observation, the airmass of observation and
the magnitude, including the gain-ratio, are also written in the file.

\subsubsection{File containing the acquired data}
The sky-subtracted counts are written in this file. All the relevant
information which are needed for processing the acquired data
later are also written in the file along with the counts.
With the data stored, the polarization values may be derived using an
algorithm different from that used at the time of observation,
if the situation demands.

\subsection{Options available additionally}
It is convenient to store the results, sometimes, in different
files, for example while objects under different observational
programmes or for different observers are observed in a night.
For this purpose the provision to change the name of
the output files interactively is also made available.

Several additional tasks can be performed to either save the data already
acquired, or prevent it from being corrupted by improper data.
After every predefined time interval, which can be interactively altered,
all the data present in the computer memory are stored in a backup file.
If there is any interruption in the data
acquisition as a result of power failure
or any other reason, the raw data
backup file can be read and the observational
procedure can be re-started from that stage onwards. It may so happen
that after a re-centring of the images
the object integration is re-started with
the diaphragm illumination kept on.
If the counts acquired then is co-added to
that already existing in the memory the final data
would be corrupted. If such a
situation arises the ongoing
integration can be discontinued and the counts co-added
during the intervening period can be deleted. 

Provisions are also made in the program to write the dark counts in the
log-file, if required. If we want to visually inspect the residuals in
the data after solving the unknown or print in the log-file that
also can be done. The data points that are excluded from the solution
because of large deviations are indicated against the corresponding
position of the half-wave plate.
Instead of the residuals the data can be displayed and plotted
with an approximate fit by toggling the corresponding option.
Since a full rotation of the halfwave plate produces four cycles of
a sine wave,
a provision to fold the data and plot it as a single sine wave cycle
is also made available.  
It is also possible to disengage any desired channel
and carry out observations in one or two spectral bands. Any
relevant remarks that
would be helpful in assessing the quality of the data later
can also be written in the log-file at anytime; along with the remarks so
entered the time of its entry is also recorded in the file.
If there is a need, the dark counts can be monitored visually
any time.

After the end of the integration of each object, backup of all the
result files are taken on a USB drive automatically if the
corresponding check button in the GUI is set active and the
USB drive specified. If any problem is encountered in taking the backup
a message is displayed to that effect. 

\section{Observational validation}\label{s:obsres}

The two main parameters of a polarimeter that determine its suitability
for observations are the polarization it produces for an unpolarized beam
and its ability to measure correctly the degree of
polarization of a polarized beam without causing any depolarization.
The first parameter is referred to as the instrumental polarization,
and is usually determined by observing unpolarized stars with the instrument.
The second parameter is the polarization efficiency, which is
the numerical value obtained by the instrument for an input beam that is
100\% polarized. The required beam is produced by inserting the Glan-Taylor
prism in the telescope beam when an unpolarized star is observed.
Ideally, the telescope-instrument combination should produce zero polarization
and the instrument should have a 100\% polarization efficiency. The observed
data can be corrected properly for these two parameters, only if, they deviate
slightly from their ideal values. In order to check whether these two parameters
are determined accurately, a few polarization standards are also normally
observed.

\subsection {Observational procedure}
The photomultiplier pulses corresponding to the
intensities of the two emergent beams from the beam-displacement prism
are counted separately 
over a full rotation of the halfwave plate at the specified
equal angular intervals, starting from a reference position.
The actual counting time interval for each beam over a rotational
cycle of the chopper depends on its
frequency of rotation since the latching pulses for the electronic pulse
counters are derived from the positional sensors of the chopper.
When the integrated counting
time over several rotational cycles of the chopper
equals to what is specified, the counting is stopped, and the resulting
counts are stored. The halfwave plate is then moved to its next position, and
the process is repeated at all the required positions. The entire
procedure is counted as one cycle. The cycle can be repeated till the required
accuracy in the measurement is achieved. After each cycle, the counts are
added to the previously stored counts at the respective position of the
halfwave plate.  Before a new cycle begins
 the halfwave plate is always brought to its reference position.
In order to give equal weightage to the observations in the data reduction,
the number of counts accumulated at all positions of the half-wave
plate should be of the same order. This requires that under poor sky
conditions the observations should be
repeated over several cycles of rotation of the half-wave plate, with a
smaller integration time at each position of the halfwave plate.
Usually, the observations in $U$ and $B$
bands will last longer than those in $V$, $R$ and $I$ bands.
The integrations in $U$ and $B$ can be continued till integrations in
the other bands are completed successively.

The procedure is the same for the object and the sky background
that has to removed from the object $plus$ sky background
counts before the data reduction.
As shown in the section on optimum background sky observation,
 the time to be spent on sky integration is usually a small fraction of the
time spent on the object integration. The optimum number of sky cycles for
an observed number of object cycles and the optimum number of object cycles
for an observed number of sky cycles, which are computed from the relative
brightnesses of the object and sky, are displayed on the monitor.

After each cycle, the linear polarization, position angle, gain-ratio and
their probable errors are displayed on the monitor. If the sky background is
observed first, the displayed values will represent the actual values,
otherwise,
they will only be approximations. Once the integration of the object is
terminated, the final values of the Stokes parameters Q(\%) and U(\%),
the polarization (P\%), the position angle ($\theta$\degr)
and the mean Julian day of observation
are stored in appropriate files.

\subsection {Observations}
The polarimeter was mounted onto the 1-m Carl Zeiss telescope at
Vainu Bappu Observatory, Kavalur, and observations of several polarized and 
unpolarized stars were made during 14~April--30~May~2014, to determine its
suitability for efficient astronomical
observations. The latter group of stars were observed with and
without the Glan-Taylor prism in the light path.
Due to the prevalent poor sky conditions, the instrument could be
used effectively only on a few nights during this period.
An analysis of the observational data indicated a very
high degree of mechanical stability for the instrument.
We found the instrumental polarization to be very small ($<$ 0.05\%). 
However, the
polarization efficiency of the instrument was found to be 94.72\%, against
a normally expected value of 98-99\%.
It may be noted that this value is not abnormally low
when we consider the fact that a measured value of 97\% has been reported
by Prescott et al. (2011) for the polarimeter they used.

In the data reduction program
four schemes of determination of the polarization are included:
(i) least square method where the sum
of the squares of residuals of intensity ratio is made a minimum
(equation~\ref{e:r1}),
(ii) least square method where the sum of the squares of the modified
residuals is made a minimum
(equation~\ref{e:r2}), (iii) elimination method and (iv) summation
method. The last two schemes are restricted in their use: the elimination
method requires that the data are sampled at 22.5 degrees of the rotation
of the halfwave plate, where a full rotation is completed in 16 steps, and
the summation method will be useful if all the data samples
are free from any large deviations. 
The four reduction schemes were applied to the observations of the
unpolarized star HD~100623
through the Glan-Taylor prism obtained on 27~Apr~2014. We found that
the polarizations and the position
angles agreed well with each other
when the two beams were used simultaneously irrespective of 
the reduction scheme used. However, there were differences in the probable 
errors derived; the first scheme gave the largest
probable errors and the elimination
method, the smallest errors. The smaller errors in the case of the latter
method is probably because the gain-ratios are solved separately. The larger
errors in the case of the first scheme is because the ratios involved when
a fully polarized beam is analyzed show a very large range.
All the four methods
gave nearly identical results for the observations without the Glan-Taylor prism
since the polarization involved was comparatively very low.
As expected, the polarization derived using the beams separately showed a large
difference. However, the average of the polarizations obtained separately agreed
well with those derived using the other schemes.

During the April--May~2014 observing run we had used a chopper
disc without blackening since
 the original one had gone bad. We also found that the positional
sensors of the chopper were producing a few thousand counts in the $R$ and $I$
spectral bands. We blackened the chopper and modified the light-shields of
the sensors apart from making slight modifications in the mounting unit of
the chopper for still better stability. Observations were carried out
again during
February--April~2015 with the polarimeter attached to the 1-m telescope.
In the following subsections we present an analysis of these observations
and the results obtained.

Observations during April--May~2014 were made
with different settings of the
chopper speed, number of positions of the
halfwave plate over a full rotation, integration times
and diaphragm sizes, in order to look for any dependency on their values.
Since we could not find any obvious differences in the results of
the measurements,
the observations during February--April~2015 were done with a
chopper frequency of 50~Hz, 25 positions of halfwave plate
over a full rotation,
integration time of 1-s, and diaphragms of 20~$arcsec$ diameter.
In addition to the $UBVRI$ bands, another broad spectral band which included
both the $R$ and $I$ bands was also used for the observations. We refer to
this band, which has a mean wavelength of 712~nm, in this report as $R'$.
\begin{figure}[htb]
\centerline {\psfig{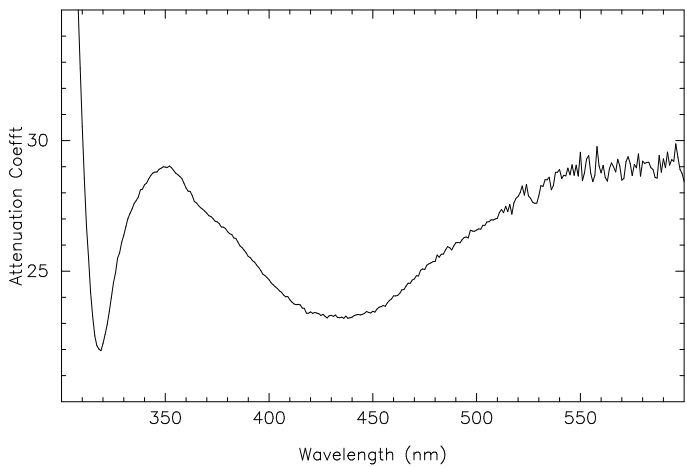}}
\caption {Spectral response of the neutral density filter used for the
determination of the dead-time
coefficient of the counting electronics associated with the $VRI$ channel.}
\label{f:ndfilm}
\end{figure}

\subsection{Determination of dead-time coefficient}

Equation~\ref{e:dead2} shows that the true count rates should be known to
derive the dead-time coefficient ($\rho$) of the counting electronics
from the observed count rates.
The true count rate is actually computed by observing the same object 
through a neutral density filter of known attenuation
coefficient ($\alpha$), which is the ratio of the 
incident light flux to the transmitted light flux.
The counts observed through the filter
should be free from any dead-time effects, which requires that $\alpha$
should be around 25--30.
For this purpose, we made a suitable neutral density filter
by trial and error using a photographic film.
The spectral characteristic of the filter which we used is given in
Figure~\ref{f:ndfilm}.

\begin{figure}[htb]
\centerline {\psfig{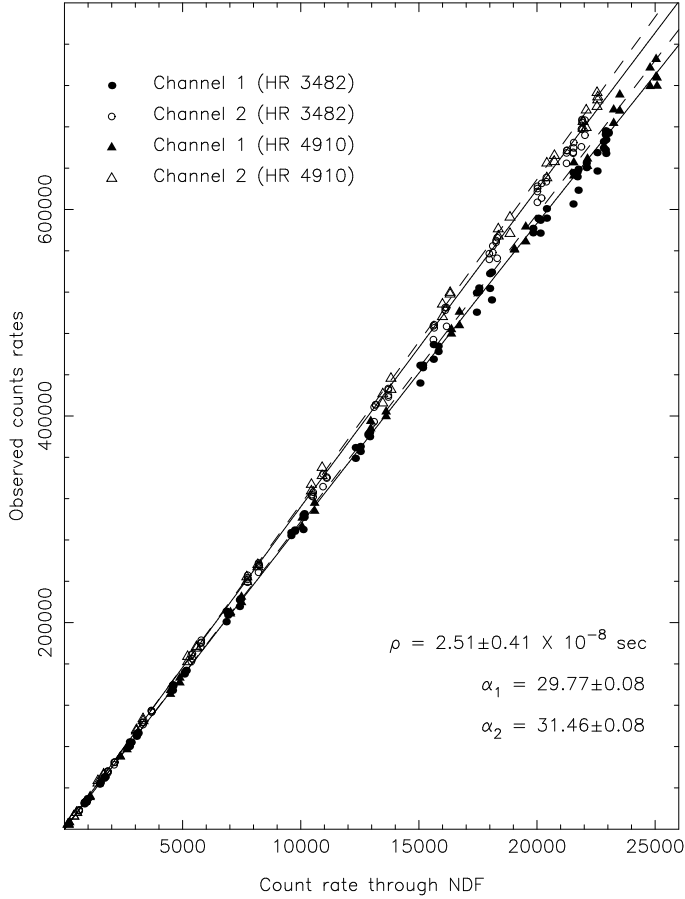}}
\caption {Plots of the count rates at different positions of the
rotating halfwave plate observed without the neutral density
filter against the corresponding count rates observed with the filter. The
dashed lines represent the case of zero dead-time effects for the two
orthogonally polarized beams. The solid lines show the least square fit with
the dead-time effect.}
\label{f:dtvri2}
\end{figure}
\begin{figure}[htb]
\centerline {\psfig{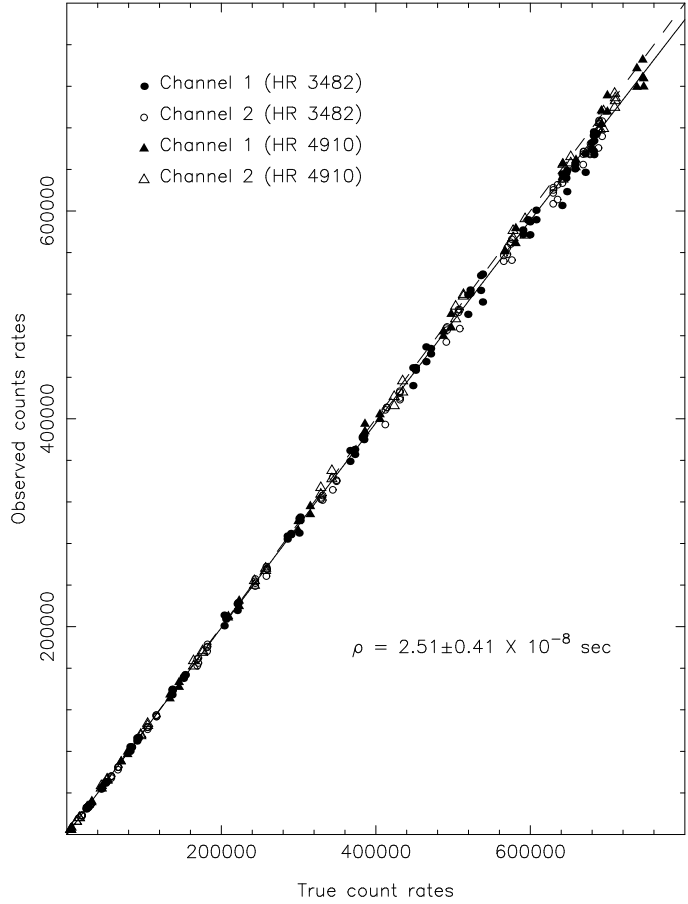}}
\caption {Plots of the observed count rates 
without the neutral density at different positions
of the rotating halfwave plate 
against the corresponding true count rates computed from the count rates
observed with the filter.
The true count rates were computed using the already computed attenuation
coefficients for the two orthogonally polarized beams.
The dashed line represents the case of zero dead-time effects, and
the solid line shows the least square fit with the dead-time effect.}
\label{f:dtvri1}
\end{figure}

A large range of count rates,  which is needed for
determining both the attenuation coefficient and
the dead-time coefficient, can be easily obtained by observing
a suitably bright star through the Glan-Taylor prism at
different positions of the rotating halfwave plate.
We observed HR~3482 (G5 III) and HR~4910 (M3 III) on two different nights
through the Glan-Taylor prism
to determine the dead-time coefficient of the counting electronics associated
with the $VRI$ channel. The observational sequence was: without the filter
(1~cycle), with the filter (30~cycles), without the filter (1~cycle),
without the filter (1~cycle), with the filter (30~cycles) and without the
filter (1~cycle) for the first object, and without the filter (2~cycles),
with the filter (31~cycles) and  without the filter (1~cycle) for the second
object. In Figure~\ref{f:dtvri2} we have plotted the count rates at
different positions of the rotating halfwave plate observed
without the neutral density filter against the corresponding
count rates with the filter observed immediately before or after. It is clear
from the figure that the attenuations of the filter for the two orthogonally
polarized beams are significantly different. Even though the two stars
have very different spectral types, the attenuation coefficients seem
to be be nearly the same for both the stars, indicating that the variation
in the transmittance with wavelength seen in Figure~\ref{f:ndfilm} has little
effect in the $V$ band for the observed objects. 

\begin{figure}[htb]
\centerline {\psfig{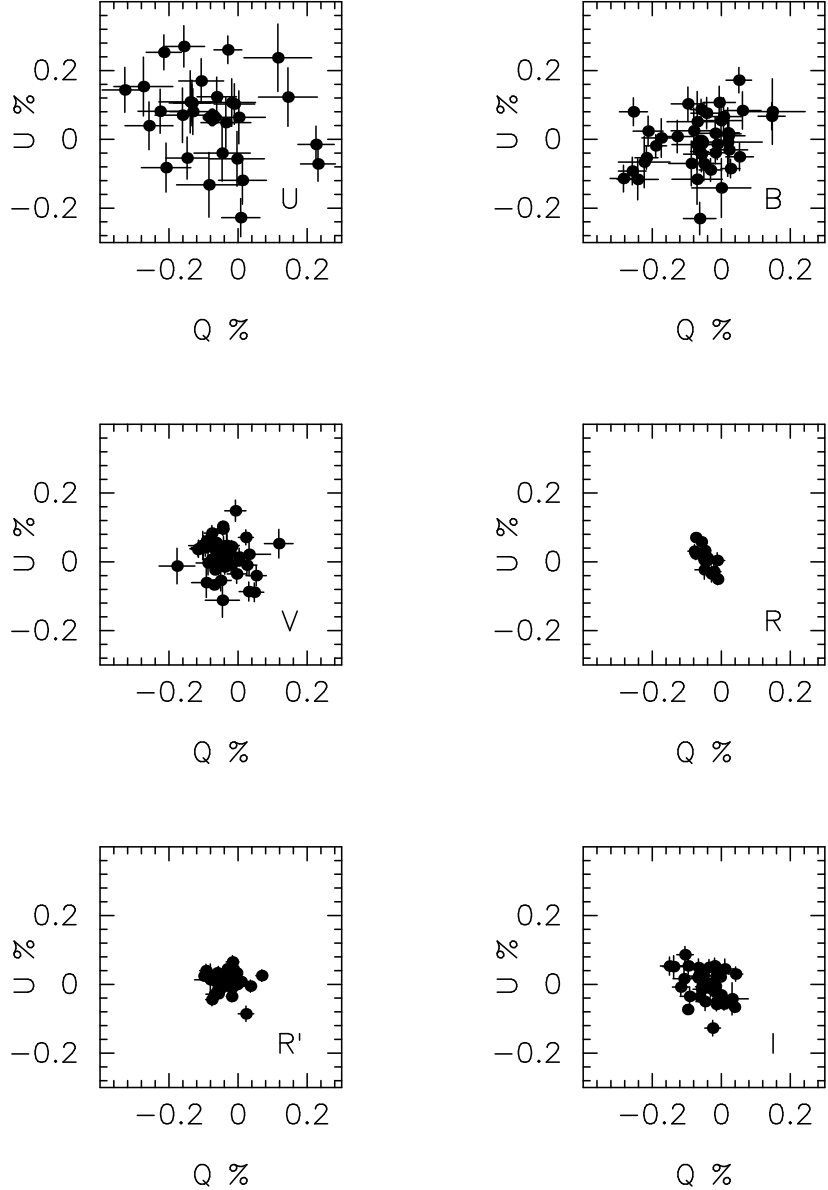}}
\caption {Observed instrumental polarization in $UBVRR'I$ bands plotted
in the (Q,U) plane. The thick symbols and lines in the plots
indicate the averages for the respective filter bands.}\label{f:instpol}
\end{figure}

Equation~\ref{e:dead2} can be written as 
\begin{eqnarray}
 C_{obs} = C_{true}\,e^{-\rho\,C_{true}}, \nonumber
\end{eqnarray}
where $C_{obs}$ and $C_{true}$ represent the observed and true count rates.
If we denote the count rates with and without the neutral density as
$C^{nd}$ and $C$, with $\alpha_1$ and $\alpha_2$ as the attenuation coefficients
of the neutral density filter for the two orthogonal beams, the above equation
can be written as
\begin{eqnarray}
C_1 = \alpha_1 C_1^{nd} e^{-\rho \alpha_1 C_1^{nd}} \quad \mbox{and}\quad 
C_2 = \alpha_2 C_2^{nd} e^{-\rho \alpha_2 C_2^{nd}}, \nonumber
\end{eqnarray}
for the two beams. The above equations were solved simultaneously
using the observed data by the method of non-linear least-squares, yielding
$\alpha_1 = $29.771$\pm$0.077, $\alpha_2 = $ 31.457$\pm$0.080 and
 $\rho = $ 2.51$\pm$0.41 $\times 10^{-8}$ s. In Figure~\ref{f:dtvri2} the solid and
dashed lines show the solutions with and without the dead-time effects for
the two polarized beams.
We have plotted in Figure~\ref{f:dtvri1}
the observed counts rates against the true counts computed 
using the above values for the attenuation coefficients.
 The solid and dashed lines again show 
the solutions with and without the dead-time effects.
The dead-time coefficient $\rho =$ 2.51$\times 10^{-8}$ s derived for the
counting electronics of the $VRI$ channel is essentially the pulse pair
resolution of 25~ns quoted for the pulse amplifier-discriminator C9744
used in that channel; the
counters are able to handle much higher counting rates without causing any
dead-time effects.

We could not determine the dead-time coefficients for the counting electronics
associated with the other two channels because of the non-availability of a
suitable star under clear sky conditions during the observing runs.
The pulse pair resolution of pulse amplifier-discriminators AD6 used in the
$U$ and $B$ channels is 20~ns, and hence, the dead-time coefficient is
expected to be smaller than that of the $VRI$ channel, given above.
\begin{table}
\caption{Instrumental polarization in the
instrument's coordinate system. The reading on the Position Angle Device of
the telescope was kept at 300 degrees.} \label{t:ipol}
\medskip
\begin{center}
\begin{tabular}{cccccc}
\hline
&&&&&\\
            &        &        &        &  & Number of\\
Band & Q (\%) & U (\%) & P (\%) & $\theta$ (\degr) &Observations\\
&&&&&\\
\hline
&&&&&\\
$U$ & $-$0.075$\pm$0.017 &  0.064$\pm$0.016 &  0.098$\pm$0.017& 69.78$\pm$4.76 &29\\
$B$ & $-$0.061$\pm$0.011 & $-$0.009$\pm$0.009 &  0.061$\pm$0.011& 93.99$\pm$4.02 &41\\
$V$ & $-$0.042$\pm$0.006 &  0.015$\pm$0.006 &  0.045$\pm$0.006& 80.12$\pm$3.52 &43\\
$R$ & $-$0.045$\pm$0.006 &  0.008$\pm$0.009 &  0.046$\pm$0.006& 84.86$\pm$5.42 &10\\
$R'$ & $-$0.032$\pm$0.004 &  0.010$\pm$0.003 & 0.033$\pm$0.004& 81.04$\pm$3.00 &35\\
$I$ & $-$0.044$\pm$0.005 & $-$0.001$\pm$0.005 &  0.044$\pm$0.005& 90.43$\pm$3.12 &40\\
&&&&&\\
\hline
\end{tabular}\\[5pt]
\end{center}
\end{table}
\begin{figure}
\centerline {\includegraphics[width=10.90cm]{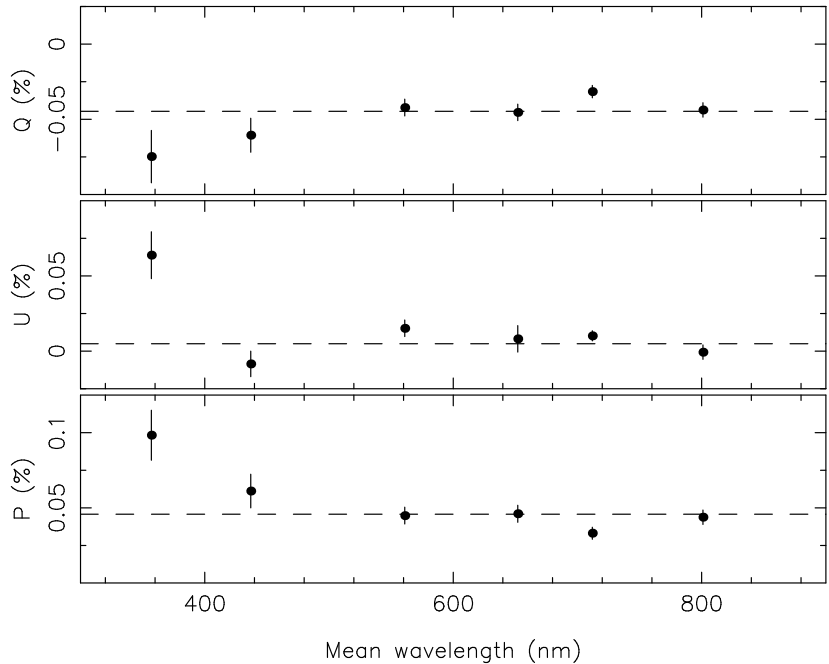}}
\caption {Plots of instrumental Q (\%), U (\%) and
p (\%) against the mean wavelength
of the corresponding spectral band. The dashed lines show the averages of the
corresponding quantities in the $BVRR'I$ bands.
\label{f:instqup}}
\end{figure}

\subsection{Instrumental polarization}
We observed the unpolarized stars, HD~42807, HD~65583, HD~90508, HD~98281,
HD~103095, HD~100623, HD~125184 and HD~144287, on several occasions.
Figure~\ref{f:instpol} shows the 
results of the observations in the ($Q$, $U$) plane, where the
individual values are plotted.
The average values of the observed Q (\%), U (\%) and P (\%)
in the $UBVRR'I$ bands are given in Table~\ref{t:ipol}
and plotted against the mean wavelengths of the
spectral bands in Figure~\ref{f:instqup}.
The observed values were corrected for the wavelength dependent polarization
efficiency (see section~\ref{ss:poleff}).
The table also gives the position angles (\degr) of the instrumental
polarization in the instrument's coordinate system, and the offset 20.\degr 072
(see section~\ref{ss:offset})
should be added to those angles to convert them to the
equatorial coordinate system.
It is clear from the table and the figure that
the polarization produced by the
telescope-polarimeter combination is small. The polarization
is nearly constant in the $V-I$ spectral region, and
apparently, it increases slightly towards the ultraviolet.

\begin{figure}[htb]
\centerline {\psfig{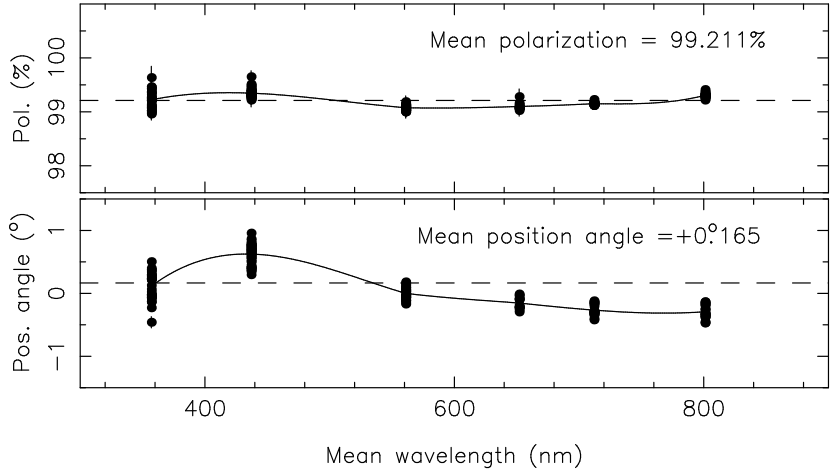}}
\caption {Plots of polarization efficiency and position angle against
the mean wavelength of the spectral band. The dashed lines represent the
middle of the respective maximum and minimum values, and the continuous
line connects the mean value in each spectral band.}\label{f:poleff}
\end{figure}

The polarization given in Table~\ref{t:ipol}
is, most likely,
produced by the telescope because a set of observations in $V$ band with a
light source fed using an optical fibre the polarimeter
gave almost zero polarization. This possibility
can be checked by rotating the polarimeter
using the Positional Angle Device and observing unpolarized stars.

\begin{table}
\caption{Wavelength dependence of polarization efficiency and
position angle.}
\medskip
\label{t:poleff}
\begin{center}
\begin{tabular}{ccc}
\hline
&&\\
Spectral Band & Polarization (P\%) & Position angle ($\theta$\degr)\\
&&\\
\hline
&&\\
 $U$ &  99.222$\pm$0.024 &   0.113$\pm$0.029\\
 $B$ &  99.346$\pm$0.010 &   0.624$\pm$0.016\\
 $V$ &  99.076$\pm$0.006 &   0.000$\pm$0.012\\
 $R$ &  99.099$\pm$0.010 &   $-$0.154$\pm$0.016\\
 $R'$ & 99.151$\pm$0.004 &   $-$0.267$\pm$0.013\\
 $I$ &  99.306$\pm$0.004 &   $-$0.293$\pm$0.015\\
&&\\
\hline
\end{tabular}\\[5pt]
\end{center}
\end{table}
\subsection{Polarization efficiency} \label {ss:poleff}
A total of 160 observations of several unpolarized stars were made with the
Glan-Taylor prism in the light path of the telescope beam to determine the
polarization efficiency of the instrument, which is the numerical
value obtained by the instrument for an input beam that is 100\% polarized.
In the top panel of Figure~\ref{f:poleff} we have plotted the individual
values of the polarization efficiencies obtained by us in $UBVRR'I$ spectral
bands against the corresponding mean wavelength.
The averages of the polarization efficiencies in different spectral bands
are given Table~\ref{t:poleff} along with their probable errors.
The continuous line in the top panel of the figure
connects these averages of polarization
efficiencies. It is clear from the table and figure that
the polarization efficiency
has a slight wavelength dependence, with lower values in the $V-R$ spectral
region. The polarization efficiency, which is halfway between the maximum
and the minimum, is 99.211\%, and it is indicated in the figure by the
dashed line.
The total amplitude of variation in the $U-I$ spectral region is only
0.271\%. The wavelength dependence of polarization seen in the figure
resembles closely the computed wavelength dependence of retardation
produced by a superachromatic halfwave plate given in Figure~\ref{f:hwpret}. 

The individual measurements of the
position angle of the polarization produced by the Glan-Taylor prism is plotted
in the bottom panel of Figure~\ref{f:poleff}. The average of the
position angles observed in the $V$ band was subtracted before plotting
in the figure. The continuous line shown in the figure
connects the averages of the position
angles, which  are given in Table~\ref{t:poleff}. The middle value of
the maximum and minimum of the position angles is 0.\degr 165, and it is
also shown in the figure. As in the case of the polarization efficiency,
the position angle also shows a slight wavelength dependence.
The wavelength dependence observed is almost an inverted and
scaled-down version of the variation of the position angle of
the effective optical axis theoretically computed for a super-achromatic
halfwave plate shown in Figure~\ref{f:hwpret},
indicating that the fixed super-achromatic
halfwave plate in the beam does not fully compensate for the variation in
the position angle of the effective optical axis of the rotating plate
because of the slight, but unavoidable errors in their fabrication.
The total amplitude of variation in the position angle is only 0.\degr 92,
and the wavelength-dependent offset in the position angle
can be incorporated in the data reduction procedure easily for
an accurate determination of the position angles.

It is advisable to check the constancy of the
polarization efficiency of the polarimeter
by observing unpolarized stars through the Glan-Taylor prism once in a while
during the observing runs.

The dead-time coefficient $\rho = $2.51$\times 10^{-8}$~s was incorporated in
the online reduction procedure on 14~March~2015. The calculations before this
date 
were made with $\rho = $ 0.0~s. The simple equation~\ref{e:dead1} to correct
the observed polarization for the dead-time effect is valid only
at low polarization values because it was derived on
the assumption that the count rates are similar at the maximum and minimum
of the light modulation.
When the polarization efficiency of the instrument is determined using the
Glan-Taylor prism, the counts registered at the lower regions of the
modulation-curve will not be affected by the dead-time effects.
From equation~\ref{e:ptrue} we find that the change in polarization efficiency
$\delta \eta_P$ due to a change $\delta N_{max}$ in $N_{max}$,
the counts at the maximum of the modulation, is given by
$$\delta \eta_P = \frac{2 N_{max} \eta\prime}{N_{max}(1 + \eta\prime)^2}
\,\delta N_{max} \,, $$
where $\eta \prime = (1 - \eta_P)/(1 + \eta_P)$. 
Making use of equation~\ref{e:ntrue}, the expression for $\delta \eta_P$ can be
written as
$$\delta \eta_P = \frac{2 \eta\prime \rho C}{(1 + \rho C)^2} \,, $$ where
$C$ is the observed count rate, taken as an approximation to the true
count rate. The observed polarization efficiency will be affected by
dead-time effects at large count rates only if the efficiency
is very different from unity.

\begin{figure}[htb]
\centerline{\includegraphics[width=12.0cm]{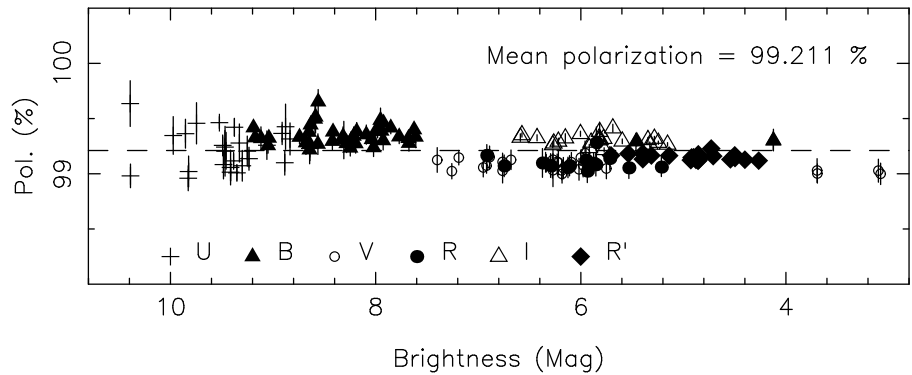}} 
\caption {Plot of the polarization efficiency against the corresponding
observed brightness.}
\label{f:poleffmag}
\end{figure}
The polarization efficiencies observed by us in $UBVRR'I$ bands
are plotted
against the corresponding brightness, expressed in the magnitude
scale in Figure~\ref{f:poleffmag}.
The observations plotted in the figure span a large range
in brightness and from the figure we find no perceptible dependence of
polarization efficiency on the brightness of the observed
object in any spectral band.
The count rates at the highest observed magnitude of 3.1 for
the $V$ band observations plotted in the figure is 
7.1$\times10^5$ and the error in the polarization efficiency
due to the dead-time effect, according to the above equation,
would be only 0.014\%, which is negligibly small when compared
to the errors of individual measurements in the $V$ band.
Therefore, the values for the polarization
efficiencies derived by us in all the spectral bands from the data
obtained before 14~March~2015 are free from any dead-time effects.

\begin{figure}[htb]
\centerline{\includegraphics[width=10cm]{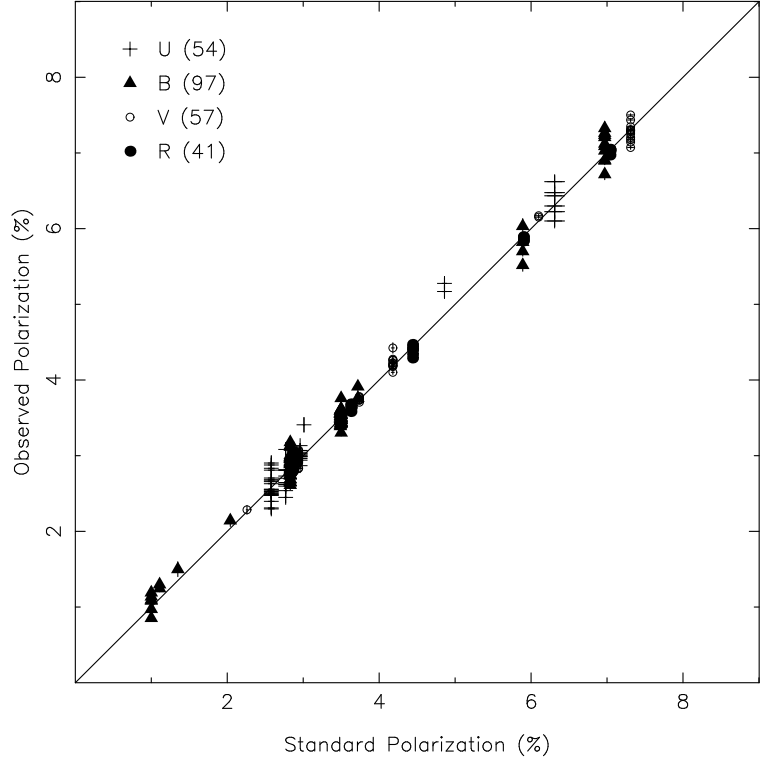}}
\caption {Plot of observed polarization against that available
in the literature.
The observed polarization was corrected for the wavelength-dependent
polarization efficiency and the instrumental polarization.
The straight line indicates an efficiency of 100 per cent for the polarimeter.
The number inside the brackets indicates the number of observations
obtained in the corresponding spectral band.}
\label{f:stanpol}
\end{figure}

\subsection {Observations of polarized stars}

During the observing runs we
observed the polarized stars, HD~21291, HD~23512, HD~43384, HD~147084,
HD~154445, HD~160529, HD~183143, HD~58439, HD~77581,
HD~94473, HD~127769 and HD~142863 in $UBVRR'I$ bands. Most of the above
objects were observed several times during the observing runs.
The first 7 objects are considered to be standard polarized stars, and
are normally used to determine the offset in the measured position angles
from the standard equatorial coordinate system.
Hsu \& Breger (1982) have reported polarizations and position angles for
these objects in $UBVR$ bands and in a band centred around
0.75~$\mu$. For the other 5 stars,
Mathewson \& Ford (1970) have given polarization
measurements in the $B$ spectral band.
The above 5 stars were included in the
present observations so as to have an extended range in the
polarization and brightness for the observed polarized stars.

\subsubsection{Polarization efficiency}
In Figure~\ref{f:stanpol}
we have plotted the polarization determined by us in $UBVR$
bands against the corresponding value available in the literature. The
observed polarizations were corrected for the wavelength-dependent
polarization efficiency of the instrument given in Table~\ref{t:poleff}
and also for the instrumental polarization given in Table~\ref{t:ipol}.
The dead-time corrections were applied only to the observations in $VRR'I$.
The polarization measurements made before 14~March~2015 were corrected using
the equation~\ref{e:dead1}, using the average count rates. In $UB$ bands
the corrections for the dead-time effects are expected to be negligibly small
in view of the low count rates in those bands.
The pulse pair resolution of the pulse amplifier-discriminator type AD6,
which is used in the $U$ and $B$ channels, is 20\,ns, while the type C9744, used
in the $VRI$ channel is 25\,ns. Therefore, the dead-time coefficient is expected
to be smaller than 2.5$\times 10^{-8}$~s, and hence, the correction even at 
the highest observed count rate of 38000 s$^{-1}$ for HD~147084 in the $B$
band would be less than 0.09\% of the observed polarization. 
It is clear from Figure~\ref{f:stanpol} that there is an
excellent agreement between the measured polarization
and the corresponding value available in the literature.
A linear least square fit to the data plotted in the figure
gave a value of 1.0044$\pm$0.0014 for the slope.

\begin{figure}[htb]
\centerline{\includegraphics[width=10cm]{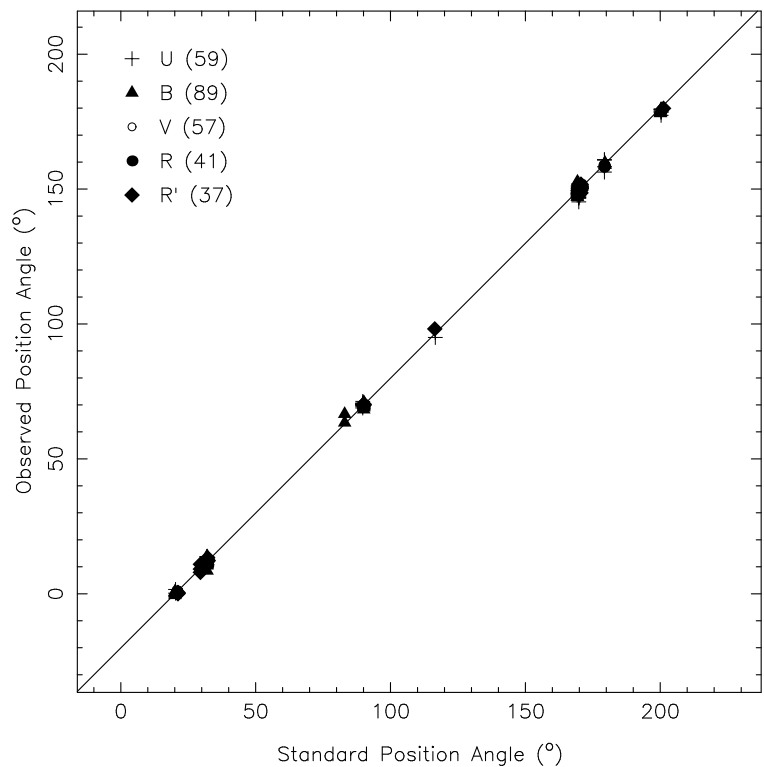}}
\caption {Plot of observed position angle against that available
in the literature. The observed position angle was corrected for the
wavelength-dependent offset. The straight line has a slope of unity.
The number inside the brackets indicates the number of observations
obtained in the corresponding spectral band.}
\label{f:stanang}
\end{figure}

\subsubsection{Offset in position angles} \label{ss:offset}
We have plotted in Figure~\ref{f:stanang}
the position angles observed by us in $UBVRR'$ against the
corresponding values available in the literature.
The position angles in $R'$ band were obtained by an interpolation of the
position angles in $R$ and 0.75~$\mu$ bands given in Hsu \& Breger (1982).
It is clear that the agreement between the measured values and those available
in the literature is very good.
The observed position angles were corrected for the wavelength-dependent
position angle of the effective optical axis of the
rotating halfwave plate given in Table~\ref{t:poleff}.
The offset in the position angles, in the
sense, standard $minus$ observed,
determined by a least square fit to the combined data plotted
in Figure~\ref{f:stanang} is 20.\degr 072$\pm$0.\degr 044; note that
this offset is with the Position Angle Device reading 300\degr.
The offset obtained using
such a procedure would be better than the
value determined using a single standard polarized star
since the position angles of some of the standards are found to vary by
more than 1\degr \, (Hsu \& Breger 1982).
The consistency in the observed
position angle throughout the observational period
indicates a good mechanical stability for the polarimeter, especially,
in the repeatability in the positioning of the halfwave plate during each cycle.

\begin{table}[htb]
\begin{center}
\caption{$UBVRI$ magnitudes and the effective wavelengths of the observation.}
\medskip
\label{t:effwls}
\begin{tabular}{lccccccccccc}
\hline
&&&&&&&&&&&\\
      & \multicolumn{5}{c}{Magnitudes} &\multicolumn{6}{c}{Effective wavelengths (nm)}\\
&&&&&&&&&&&\\
Object  &  $U$  &  $B$  & $V$ &  $R$  &  $I$ & $U$  &  $B$  & $V$ &  $R$  & $R'$ &  $I$ \\
&&&&&&&&&&&\\
\hline
&&&&&&&&&&&\\
HD 21291  & 4.40 &  4.63 &  4.22 &  3.84 &  3.46 & 365 & 440 & 561 & 663 & 726 & 800\\
HD 23512  & 8.72 &  8.44 &  8.09 &  7.79 &  7.49 & 372 & 441 & 559 & 647 & 701 & 799\\
HD 43384  & 6.31 &  6.70 &  6.25 &  5.78 &  5.38 & 363 & 440 & 561 & 651 & 708 & 800\\
HD 147084 & 5.99 &  5.40 &  4.57 &  3.69 &  2.93 & 375 & 442 & 566 & 675 & 745 & 804\\
HD 154445 & 5.10 &  5.73 &  5.61 &  5.39 &  5.34 & 362 & 439 & 558 & 647 & 700 & 797\\
HD 160529 & 8.17 &  7.87 &  6.66 &  5.50 &  4.51 & 367 & 443 & 569 & 665 & 736 & 806\\
HD 183143 & 8.25 &  8.08 &  6.86 &  5.74 &  4.79 & 366 & 443 & 568 & 655 & 735 & 806\\
&&&&&&&&&&&\\
\hline
\end{tabular}
\end{center}
\end{table}

\subsubsection{Effective wavelengths of observation}
In order to compare the results of the present
polarimetry of polarized stars with those
available in the literature
the effective wavelengths of observation are 
needed because the polarization is a function of
wavelength and differences in the effective wavelengths of observation
could give rise to substantial differences in the polarizations observed
in the same spectral band with different instruments.
Since the polarization varies slowly within a
spectral passband normally used, 
the effective wavelength is a fairly good approximation 
to the isophotal wavelength.
As these stars are reddened at different degrees,
resulting in large differences in their flux distributions, there could be
substantial differences in the
effective wavelengths of observation when there are significant
differences in the way the spectral passbands are isolated 
for the observations.
The measurements of Hsu \& Breger (1982), which
form the bulk of the data against
which we have compared the present observations, 
were obtained with the filter-detector combinations
that are completely different from what we used, and therefore, the
corresponding spectral passbands also would be significantly different.
They had used interference filters to separate the $BVR$ spectral bands,
while we used glass filters with much broader passbands.

\begin{figure}
\centerline{\includegraphics[width=8.5cm]{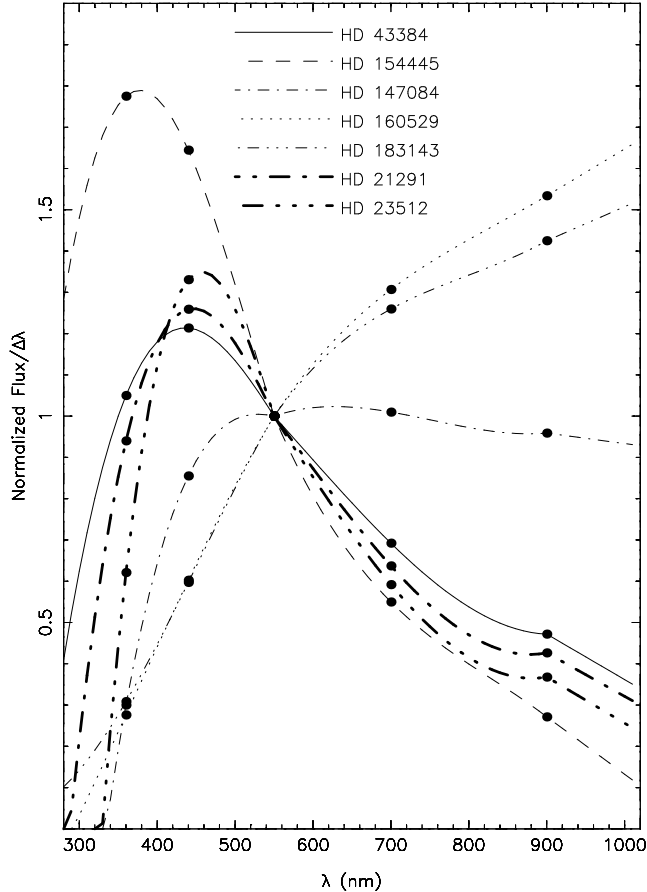}}
\caption {Plots of the normalized fluxes of the polarimetric standards,
which are used for the calculation of the effective wavelengths
of observation in $UBVRI$ bands.}\label{f:norflux}
\end{figure}

We computed the effective wavelengths of observation using the
$UBVRI$ magnitudes of these stars in the Johnson's system, which are listed
in Table~\ref{t:effwls},
and the response curves of the filter-detector combinations
given in Figure~\ref{f:filterdet}. 
The $UBVRI$ magnitudes were taken from Ducati (2002)
and were converted to fluxes normalized at the $V$ band,
using the mean wavelengths and fluxes for a 0.0 mag star 
given in Johnson (1966).
The error caused by using mean wavelengths for the effective wavelengths is
expected to be negligible. The effective wavelengths of
observation were calculated using
$$\lambda_{eff} = \frac{\int\displaylimits_{band} \lambda^2 f (\lambda) S (\lambda) \delta \lambda}
{\int\displaylimits_{band} \lambda
 f (\lambda) S (\lambda) \delta \lambda}\, , $$ since the number of
photons detected will be proportional to
$\int \lambda f (\lambda) S (\lambda) \delta \lambda $.
The interpolation of fluxes
at other wavelengths and extrapolation at wavelengths
shorter than $U$ band were done using a four-point Lagrange formula. The
fluxes at wavelengths longward of the $I$ band were obtained by linearly
extrapolating the fluxes at $R$ and $I$ bands.
The normalized fluxes used for the derivation of the effective wavelengths,
which are listed in Table~\ref{t:effwls}, are
plotted in Figure~\ref{f:norflux}.
The effect of atmospheric extinction was included in the calculation of the
effective wavelengths. We used the mean extinction coefficients 
at Kavalur, which are listed in Table~\ref{t:expcnts}, and the airmass of
observation was taken as 1.5. The extinction coefficients at
other wavelengths were computed again
using a four-point Lagrange interpolation formula.
The variation of reflectivity of the telescope mirrors with wavelength was
neglected because the effect is expected to be only marginal. 

\begin{table}[htb]
\begin{center}
\caption{Results of the 
$UBVRR'I$ polarimetry of standard stars}\label{t:stdstars}
\medskip
\begin{tabular}{cccccc}
\hline
&&&&&\\
Band & P (\%) & $\theta$ (\degr) &  & P (\%) & $\theta$ (\degr)\\
&&&&&\\
\hline
& \multicolumn{2}{c}{HD 21291}&\quad\quad\quad & \multicolumn{2}{c}{HD 23512} \\ 
 $U$ & 3.41$\pm$0.10 & 115.1$\pm$0.8 (1)&   &  --- & --- \\
 $B$ & 3.39$\pm$0.03 & 116.6$\pm$0.2 (1)&   &  2.14$\pm$0.08 &  29.9$\pm$1.0 (1) \\
 $V$ & 3.54$\pm$0.04 & 118.1$\pm$0.3 (1)&  &  2.28$\pm$0.03 &  29.5$\pm$0.5 (2) \\
 $R'$ & 3.20$\pm$0.02 & 118.2$\pm$0.2 (1)&  &  2.11$\pm$0.03 &  29.4$\pm$0.5 (3) \\
 $I$ & 2.81$\pm$0.03 & 117.1$\pm$0.3 (1)&  &  2.06$\pm$0.04 &  27.4$\pm$0.5 (2)  \\
& \multicolumn{2}{c}{HD 43384} &  & \multicolumn{2}{c}{HD 147084} \\
 $U$ & 2.62$\pm$0.03 & 168.7$\pm$0.3 (19)&  &  2.69$\pm$0.03 &  32.2$\pm$0.3 (11) \\
 $B$ & 2.86$\pm$0.02 & 169.9$\pm$0.2 (25)&  &  3.50$\pm$0.02 &  31.8$\pm$0.1 (17) \\
 $V$ & 2.97$\pm$0.01 & 170.4$\pm$0.1 (22)&  &  4.23$\pm$0.02 &  32.2$\pm$0.1 (11) \\
 $R$ & 2.86$\pm$0.01 & 170.8$\pm$0.1 (18)&  &  4.39$\pm$0.02 &  32.4$\pm$0.1 (6) \\
 $R'$ & 2.74$\pm$0.01 &170.6$\pm$0.1 (15)&  &  4.30$\pm$0.01 &  32.5$\pm$0.1 (8) \\
 $I$ & 2.56$\pm$0.01 & 170.9$\pm$0.1 (22)&  &  4.24$\pm$0.01 &  32.5$\pm$0.1 (11) \\
& \multicolumn{2}{c}{HD 154445} &  & \multicolumn{2}{c}{HD 160529} \\
 $U$ & 3.01$\pm$0.02 &  90.2$\pm$0.2 (13)&  &  6.33$\pm$0.05 &  19.4$\pm$0.2 (10) \\
 $B$ & 3.51$\pm$0.02 &  90.0$\pm$0.2 (13)&  &  7.08$\pm$0.05 &  19.7$\pm$0.2 (10) \\
 $V$ & 3.75$\pm$0.01 &  90.1$\pm$0.1 (8)&  &  7.28$\pm$0.02 &  20.3$\pm$0.1 (11) \\
 $R$ & 3.65$\pm$0.01 &  90.0$\pm$0.1 (8)&  &  7.02$\pm$0.01 &  20.9$\pm$0.1 (5) \\
 $R'$ & 3.47$\pm$0.01 & 90.2$\pm$0.1 (5)&  &  6.43$\pm$0.01 &  20.2$\pm$0.1 (5) \\
 $I$ & 3.23$\pm$0.01 &  89.8$\pm$0.1 (8)&  &  6.08$\pm$0.02 &  20.7$\pm$0.1 (10) \\
& \multicolumn{2}{c}{HD 183143}&&&\\
 $U$ & 5.04$\pm$0.09 & 179.1$\pm$0.7 (4) &&&\\
 $B$ & 5.77$\pm$0.07 & 179.4$\pm$0.1 (4) &&&\\
 $V$ & 6.16$\pm$0.01 & 178.8$\pm$0.3 (2) &&&\\
 $R$ & 5.87$\pm$0.01 & 178.6$\pm$0.1 (4) &&&\\
 $I$ & 5.18$\pm$0.02 & 178.6$\pm$0.2 (2) &&&\\
&&&&&\\
\hline
\end{tabular}
\end{center}
\end{table}

\subsubsection{Comparison with the literature values}
The averages of the polarization of HD~21291, HD~23512, HD~43384, HD~147084,
HD~154445, HD~160529 and HD~183143 observed by us
in $UBVRR'I$ spectral bands are given Table~\ref{t:stdstars}.
The numbers inside the brackets indicate the number of observations
in each case.
In this section we make a detailed comparison of the
present multiband polarimetric data of these stars,
which are considered to be polarization standards,
with those of Hsu \& Breger (1982) and Serkowski, Mathewson \& Ford (1975).

\begin{figure}
\centerline{\includegraphics[width=8.75cm]{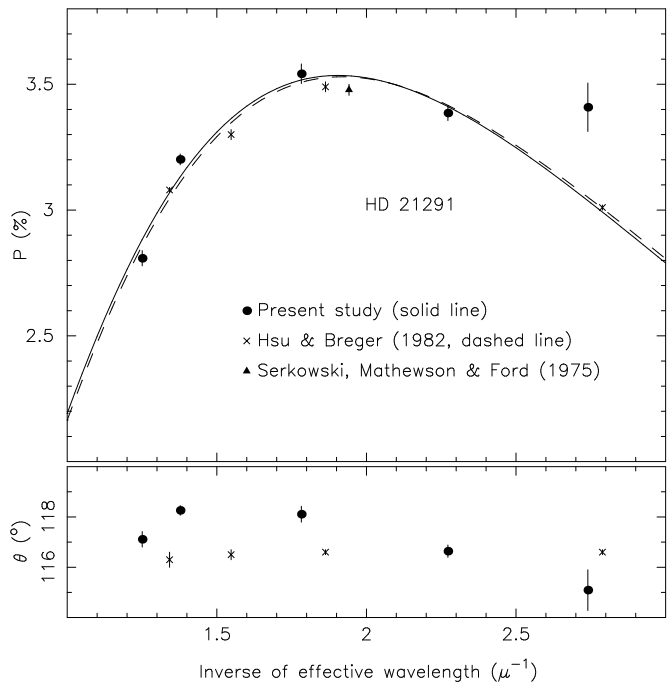}}
\caption {Plot of the polarization of HD~21291 observed in $UBVR'I$ bands 
against the inverse of the corresponding wavelength. The filled
circles indicate the data given in Table~\ref{t:stdstars}, crosses those
given in Hsu \& Breger (1982) and filled triangles those
given in Serkowski, Mathewson \& Ford (1975).
The dashed line shows the interstellar polarization curve computed using the
$P_{max}$ and $\lambda_{max}$ given by the former authors, and the solid line
that computed using the values derived from the present polarimetry.}
\label{f:hd21291}
\end{figure}
\begin{figure}
\centerline{\includegraphics[width=8.75cm]{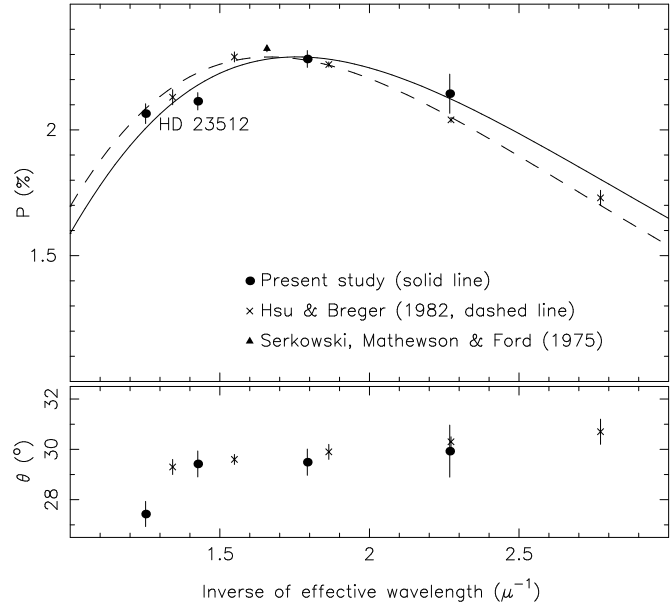}}
\caption {Plot of the polarization of HD~23512 observed in $BVR'I$ bands
 against the inverse of the corresponding wavelength. The filled
circles indicate the data given in Table~\ref{t:stdstars}, crosses those
given in Hsu \& Breger (1982) and filled triangles those
given in Serkowski, Mathewson \& Ford (1975).
The dashed line shows the interstellar polarization curves computed using the
$P_{max}$ and $\lambda_{max}$ given by the former authors, and the solid line
that computed using the values derived from the present polarimetry.}
\label{f:hd23512}
\end{figure}
\begin{figure}
\centerline{\includegraphics[width=8.75cm]{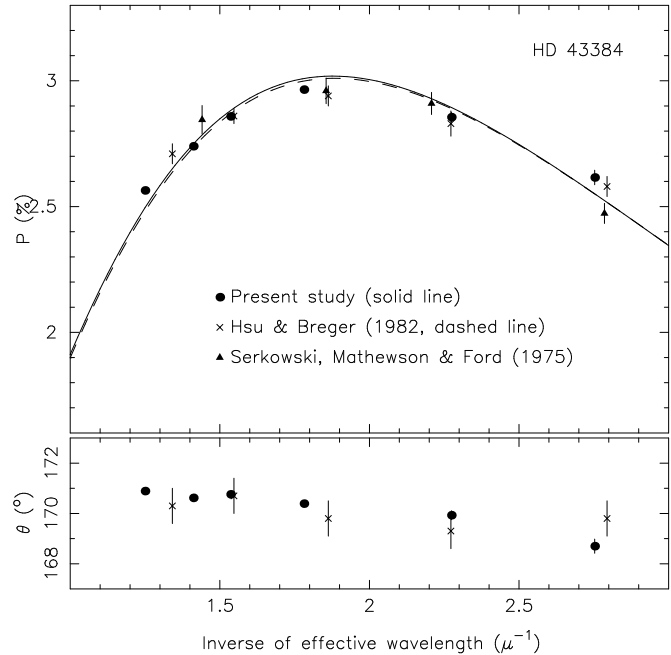}}
\caption {Plot of the polarization of HD~43384 observed in $UBVRR'I$ bands 
against the inverse of the corresponding wavelength. The filled
circles indicate the data given in Table~\ref{t:stdstars}, crosses those
given in Hsu \& Breger (1982) and filled triangles those
given in Serkowski, Mathewson \& Ford (1975).
The dashed line shows the interstellar polarization curves computed using the
$P_{max}$ and $\lambda_{max}$ given by the former authors, and the solid line
that computed using the values derived from the present polarimetry.}
\label{f:hd43384}
\end{figure}
\begin{figure}
\centerline{\includegraphics[width=8.75cm]{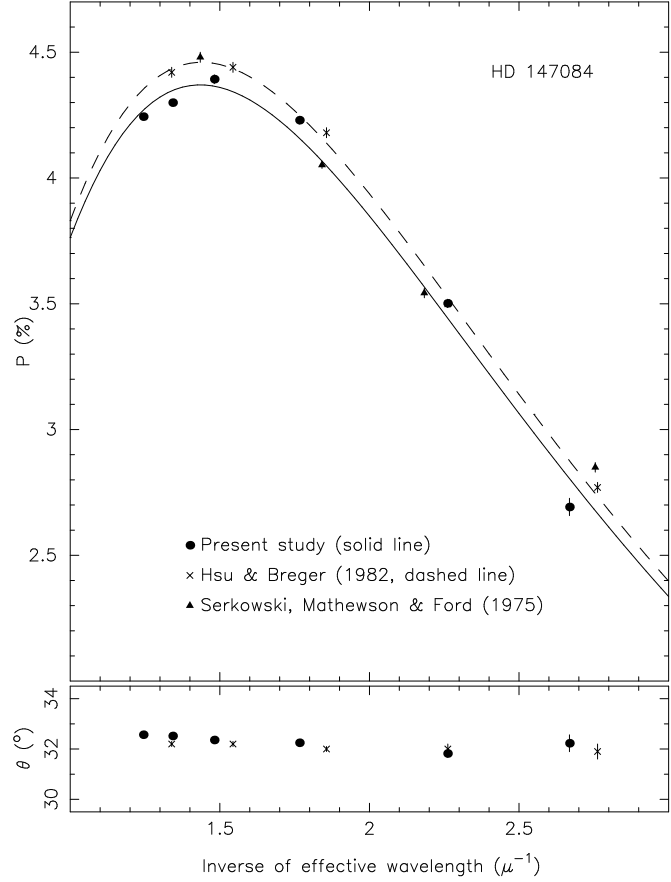}}
\caption {Plot of the polarization of HD~147084 observed in $UBVRR'I$ bands 
against the inverse of the corresponding wavelength. The filled
circles indicate the data given in Table~\ref{t:stdstars}, crosses those
given in Hsu \& Breger (1982) and filled triangles those
given in Serkowski, Mathewson \& Ford (1975).
The dashed line shows the interstellar polarization curves computed using the
$P_{max}$ and $\lambda_{max}$ given by the former authors, and the solid line
that computed using the values derived from the present polarimetry.}
\label{f:hd147084}
\end{figure}
\begin{figure}
\centerline{\includegraphics[width=8.75cm]{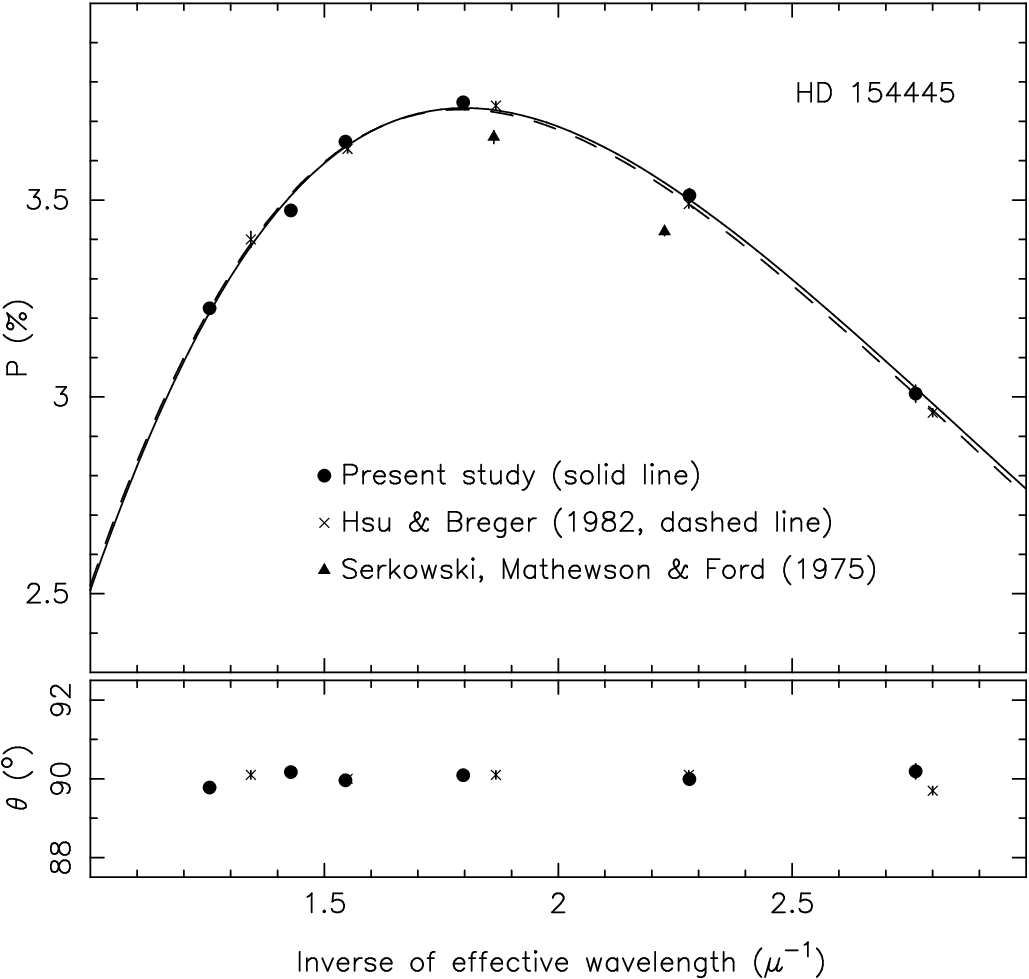}}
\caption {Plot of the polarization of HD~154445 observed in $UBVRR'I$ bands 
against the inverse of the corresponding wavelength. The filled
circles indicate the data given in Table~\ref{t:stdstars}, crosses those
given in Hsu \& Breger (1982) and filled triangles those
given in Serkowski, Mathewson \& Ford (1975).
The dashed line shows the interstellar polarization curves computed using the
$P_{max}$ and $\lambda_{max}$ given by the former authors, and the solid line
that computed using the values derived from the present polarimetry.}
\label{f:hd154445}
\end{figure}
\begin{figure}
\centerline{\includegraphics[width=8.75cm]{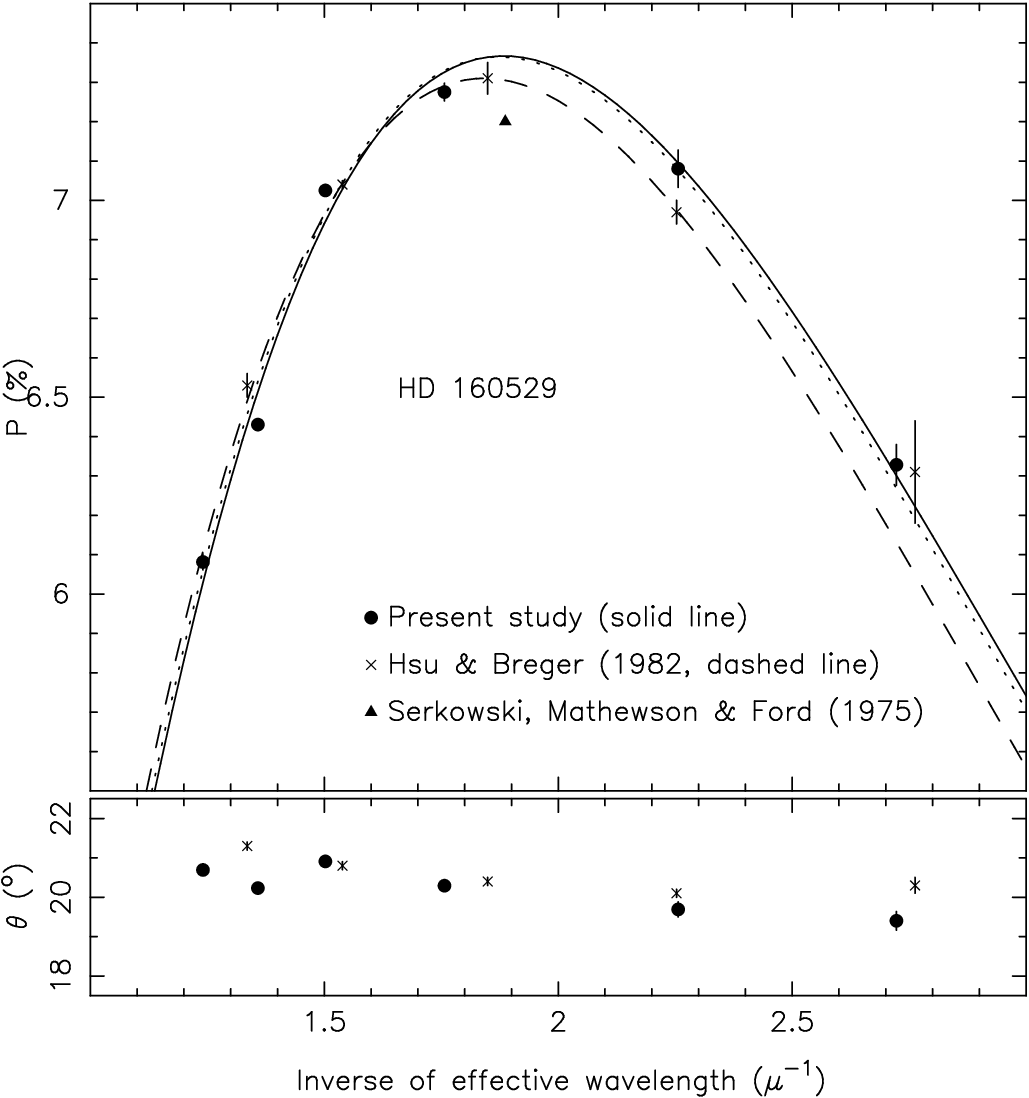}}
\caption {Plot of the polarization of HD~160529 observed in $UBVRR'I$ bands 
against the inverse of the corresponding wavelength. The filled
circles indicate the data given in Table~\ref{t:stdstars}, crosses those
given in Hsu \& Breger (1982) and filled triangles those
given in Serkowski, Mathewson \& Ford (1975).
The dashed line shows the interstellar polarization curves computed using the
$P_{max}$ and $\lambda_{max}$ given by the former authors, and the solid line
that computed using the values derived from the present polarimetry. The dotted
line shows the solution obtained by us using the data of Hsu \& Breger (1982).}
\label{f:hd160529}
\end{figure}

The polarizations in the $UBVRR'I$ spectral
bands given in Table~\ref{t:stdstars} are plotted
in Figures~\ref{f:hd21291}$-$\ref{f:hd183143}
against the corresponding effective wavelengths given in Table~\ref{t:effwls}.
The figures also show the observations of Hsu \& Breger (1982) and
Serkowski, Mathewson, \& Ford (1975) plotted against the respective
inverses of the effective wavelengths of observation, which were
computed using the empirical relations given by the authors.
The average airmass of observation was taken as
1.0 while computing the effective wavelengths of the observations of
Hsu \& Breger (1982). The errors in the measurements are also indicated
in the figures.

In Figures~\ref{f:hd21291}$-$\ref{f:hd183143} we have also
plotted the interstellar polarization curve, which is the same for
all stars when normalized to $\lambda_{max}$, the wavelength of maximum
polarization $P_{max}$, computed using the empirical relation,
$$ P = P_{max} \exp \{-1.15 \ln^2(\lambda_{max}/\lambda)\} ,$$ given in
Serkowski, Mathewson, \& Ford (1975).
We derived the values of $P_{max}$ and $\lambda_{max}$ for
all the standards which we observed from the present data using the non-linear
least square method,
and they are listed in Table~\ref{t:wlpmax} along with those given Hsu \&
Breger (1982). In the table we have also listed the differences in
the wavelength-averaged position angles obtained by them and us.

Hsu \& Breger (1982) had reported that several stars which are used as
standards definitely show polarimetric variability in the order of 0.3\%.
We observed two of the stars, namely, HD~43384 and HD~183143, for which
variabilities have been reported by them.
Hsu \& Breger (1982) found HD~43384 to show changes of 0.25\% and 2\degr \,
in $P$ and $\theta$ observed in the 
$V$ band.  A variability of 0.2--0.3\% in polarization
in the $V$ band has been reported by the above authors for
HD~183143 also. The averages values of polarization in $UBVRI$ bands obtained by us
for both these objects are in excellent agreement with those given
by Hsu \& Breger (1982), indicating that these objects exhibit no
long-term variations. The mean values of position angles observed by us are
essentially the same as those reported by the above authors, and so are the  
the values of $P_{max}$ and $\lambda_{max}$ .

Figure~\ref{f:hd154445} shows that the present observations of HD~154445
agree well with those of Hsu \& Breger (1982) and Serkowski, Mathewson \&
Ford (1975). The present values of $P_{max}$, $\lambda_{max}$ and
the average position angle of the
star are the same as those given by the former authors. The polarizations in
$B$ and $V$ reported by Serkowski, Mathewson \& Ford (1975) for this star
are about 0.1\% lower than those observed by Hsu \& Breger (1982) and us.

From Figure~\ref{f:hd147084} we find that in the case of
HD~147084 the polarizations in the $R-I$ region observed by us
are about 0.1\% below those observed by Hsu \& Breger (1982).
The $VRR'I$ observations of both HD~21291 and HD~147084 were done through
neutral density filters of optical density 1.0. The transmission of the
filter was taken into account while computing the effective wavelengths
in these bands. The vendor of this filter has provided the transmission values 
only up to 750~nm, above which the transmission was assumed to be constant
while computing the effective wavelengths. The effective wavelengths of
the $R'$ and $I$ bands
could be longer than what we derived. However, the close
agreement in the polarization  of HD~21291 in the $R-I$ obtained by us
with those obtained by Hsu \& Breger (1982) rule out any large deviations.
It may be noted from Figure~\ref{f:norflux} that the normalized flux
of HD~147084
in the $R-I$ spectral region is much higher than that of HD~21291.
The positions angles of HD~147084 in the $R-I$ spectral region are marginally
higher than those obtained by the above authors. It is possible that
the discrepancy in the polarization in the $R-I$ region is due to
a variability of the object.  The values of $\lambda_{max}$ are the
same in both the cases.

Figure~\ref{f:hd160529} shows that the polarizations of HD~160529 observed
by us agree well with those reported by Hsu \& Breger (1982). However, the
values of $\lambda _{max}$ differ by about 0.012~$\mu$, which is significant
when compared to the errors in their determinations. We did a least square
solution of their data, which yielded $P_{max} = $7.36$\pm$0.04 and
$\lambda _{max} = $ 0.534$\pm$0.004, which are almost the same as those
given by the present data. It may be noted that
Serkowski, Mathewson \& Ford (1975) had reported a value of 0.530~$\mu$ for
the $\lambda _{max}$ of this object.

The polarizations and position angles of HD~21291 and HD~23512 obtained by us
also agree with those obtained by  Hsu \& Breger (1982), even though
we have observed these objects just once or twice. In the case of
HD~23512, the difference in
the values of $\lambda _{max}$ obtained by us and the above authors is
because we do not have the $U$ band data and the polarization obtained by us
in $B$ band is slightly more than that obtained by them. 

\begin{table}[htb]
\begin{center}
\caption{$\lambda_{max}$ and $P_{max}$ of the observed polarized stars.}
\medskip
\label{t:wlpmax}
\begin{tabular}{lccccc}
\hline
&&&&&\\
       &\multicolumn{2}{c}{Present study} & \multicolumn{2}{c}{Hsu \& Breger 1982} & $\Delta \theta_{mean}$\\
Object & $P_{max}$ & $\lambda_{max}$ & $P_{max}$ & $\lambda_{max}$ &
 (HS $-$ PS)$^1$\\ 
&&&&&\\
\hline
&&&&&\\
HD 21291$^2$ & 3.54$\pm$0.03 & 0.525$\pm$0.008 & 3.53$\pm$0.02 & 0.521$\pm$0.003 & $-$1.03$\pm$0.27\\
HD 23512 & 2.29$\pm$0.02 & 0.569$\pm$0.011 & 2.29$\pm$0.01 & 0.600$\pm$0.006 & $+$0.89$\pm$0.41\\
HD 43384 & 3.02$\pm$0.02 & 0.533$\pm$0.005 & 3.01$\pm$0.04 & 0.531$\pm$0.011 & $+$0.24$\pm$0.28\\
HD 147084 & 4.37$\pm$0.03 & 0.697$\pm$0.007 & 4.46$\pm$0.03 & 0.695$\pm$0.006 & $-$0.25$\pm$0.08\\
HD 154445 & 3.73$\pm$0.01 & 0.556$\pm$0.002 & 3.73$\pm$0.01 & 0.558$\pm$0.002 & $-$0.03$\pm$0.07\\
HD 183143 & 6.12$\pm$0.01 & 0.552$\pm$0.002 & 6.08$\pm$0.05 & 0.551$\pm$0.006 & $-$0.35$\pm$0.11\\
HD 160529$^3$ & 7.37$\pm$0.02 & 0.531$\pm$0.002 & 7.31$\pm$0.02 & 0.543$\pm$0.003 & $-$0.38$\pm$0.21\\
&&&&&\\
\hline
\multicolumn{6}{l}{\footnotesize{(1)HS$-$Hsu \& Breger (1982); PS$-$ Present Study.}}\\
\multicolumn{6}{l}{\footnotesize{(2) $U$ band polarization was 
not used while deriving $P_{max}$ and $\lambda_{max}$.}}\\
\multicolumn{6}{l}{\footnotesize{(3) We get $P_{max} = $ 7.36$\pm$0.04 and $\lambda_{max} = $0.534
$\pm$0.004 from the data of Hsu \& Breger (1982).}}\\
\end{tabular}
\end{center}
\end{table}
\begin{figure}
\centerline{\includegraphics[width=8.75cm]{hd183143.ps}}
\caption {Plot of the polarization of HD~183143 observed in $UBVRI$ bands 
against the inverse of the corresponding wavelength. The filled
circles indicate the data given in Table~\ref{t:stdstars}, crosses those
given in Hsu \& Breger (1982) and filled triangles those
given in Serkowski, Mathewson \& Ford (1975).
The dashed line shows the interstellar polarization curves computed using the
$P_{max}$ and $\lambda_{max}$ given by the former authors, and the solid line
that computed using the values derived from the present polarimetry.}
\label{f:hd183143}
\end{figure}
\subsection{Gain-ratios}
\begin{figure}[htb]
\centerline {\psfig{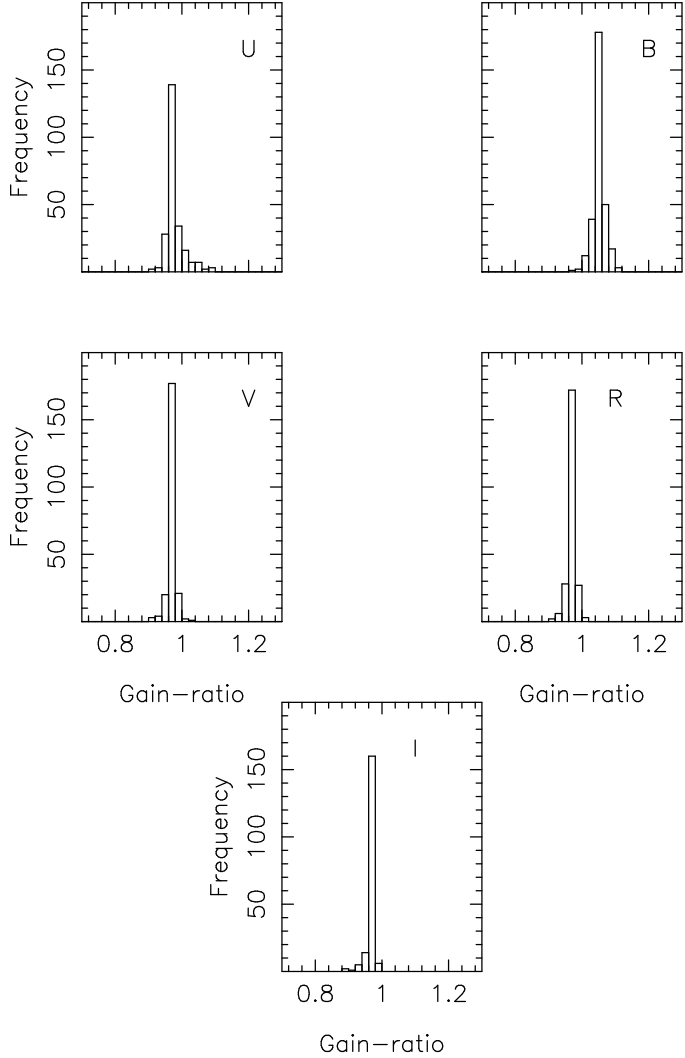}}
\caption {Gain ratios of the two beams in $UBVRI$ bands, and the total
number of observations are 241, 302, 228, 238 and 188, respectively.
The average values and their standard deviations are 0.977$\pm$0.028, 
1.051$\pm$0.018, 0.968$\pm$0.013, 0.966$\pm$0.013 and 0.967$\pm$0.012.}
\label{f:gratio}
\end{figure}
The histograms of the
gain-ratios of the two beams derived from the observations are plotted
in Figure~\ref{f:gratio}.  The gain-ratios observed in $R'$ were combined
with those observed in $R$ band in forming the histograms.
The average values and their standard deviations
in the $UBVRI$ spectral bands are 0.977$\pm$0.028,
1.051$\pm$0.018, 0.968$\pm$0.013, 0.966$\pm$0.013 and 0.967$\pm$0.012,
respectively. It may be noted that the mean values of the gain-ratios 
in the $VRI$ bands are almost the same.  The average values in
all the spectral bands are close to unity, as they are
expected to be, if the Fabry lenses are properly placed. 
The spread in the observed gain-ratios results mainly from
the slight differences in the image centring on different occasions.
Any significant deviations in the gain-ratios from the above
averages may be indicative of problems
with the counting electronics, or the presence of a faint component in one of
the beams.

\subsection{Overhead time}
The error in the measured polarization due to photon noise depends on the total
counts accumulated, which means that
for an efficient use of the telescope time
 the available time should be used entirely for counting.
 This is not possible because the two beams are observed
alternately, and the counting can be done only during the interval when
the corresponding slot in the rotating chopper does not obstruct the diaphragm
as seen by the photomultiplier. The slots in the chopper are
separated by a blind sector of 8 degree in size
during the passage of which below the diaphragm no counting can be done.
Similarly, counting should begin only after the halfwave plate stops completely
at each position, otherwise, there will be a certain
amount of depolarization.
Additionally, several delays, lasting a few milliseconds,
are introduced in the program
by trial and error for the effective
communication between the computer and the interface.

\begin{figure}[htb]
\centerline {\psfig{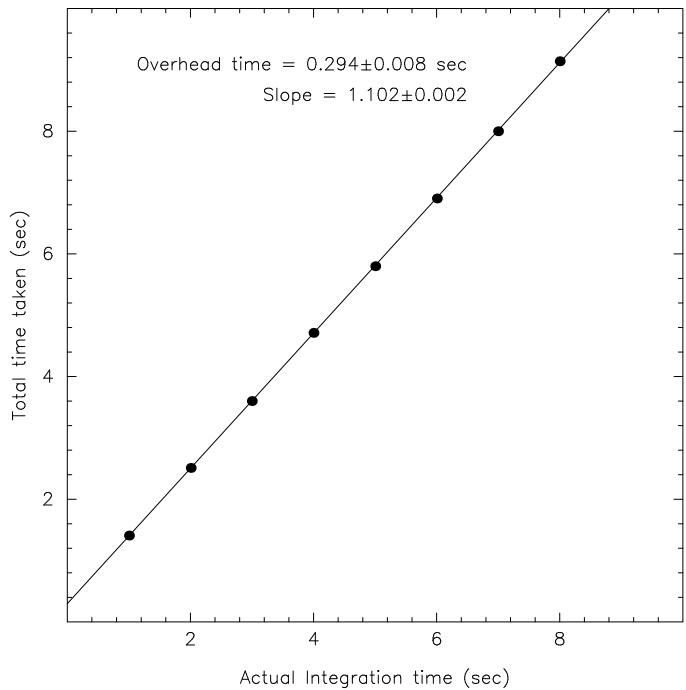}}
\caption {Plot of the total time taken per halfwave plate position
against the actual time over which counting is done.
The straight line represents a linear least square fit to the data.}
\label{f:inttime}
\end{figure}

In Figure~\ref{f:inttime}
 we have plotted the total time taken against the actual counting
time at each position of the halfwave plate. The total time was calculated
for each integration time from the time taken to complete one cycle of
rotation of the halfwave plate, and the actual time during which counting was
done was calculated from the chopper frequency and the sizes of the
blind sectors in the chopper disc. From the figure we 
find that there is an overhead of
about 300~ms spent at each position of the halfwave plate. The fractional time
lost can be reduced by giving a higher integration time and choosing
a lower number for the positions of the
halfwave plate at which integrations are made during its full rotation.

Most of the overhead time is spent while waiting for the halfwave plate to stop
completely
at each position. Probably, this can be reduced slightly by trial and error.

The factor of 1.10 is essentially caused by the blind sectors in the chopper,
and by re-designing the chopper this factor can be reduced slightly.
The latching pulses for the counters are derived from the sensors
which monitor the positions of the slots in the chopper. A perfect straight line
fit to the data shown in Figure~\ref{f:inttime} 
is a clear indication of the stability in
the rotational frequency of the chopper. If there were any systematic losses
in the detection of the latching pulses during the chopper
rotations, the line would have shown a larger slope.

\subsection{Limiting magnitudes}
From the observed data, we computed the count rates expected
from a 10.0 mag star in $UBVRI$
bands along with the probable errors in polarization
arising from the statistical fluctuations in the total counts in
10 minutes of integration in those bands, and the results are given
in Table~\ref{t:expcnts}.
We used the highest observed count rates in each band
for different stars to scale down
the count rates to that of a 10.0 mag star.
The average air mass of observation is also given in the table.
The overhead time and the time required for
background sky integration, which would be only a fraction of the
above timings even if the sky is moonlit and the object
brightness is comparable to the sky brightness,
were not included in the computation.
\begin{table}[htb]
\begin{center}
\caption{Wavelength-independent parameters used to compute the expected counts.}
\medskip
\label{t:windpar}
\begin{tabular}{ll}
\hline
&\\
Parameter & Value \\
&\\
\hline
&\\
Telescope aperture & 1~$m$ \\
Central obscuration & 0.40 \\
Reflectivity of the telescope & 0.70 \\
Transmittance of the calcite block & 0.90 \\
Transmittance of the half-wave plate & 0.95 \\
&\\
\hline
\end{tabular}
\end{center}
\end{table}

\begin{table}[htb]
\begin{center}
\caption{Wavelength dependent parameters used to compute the expected counts.}
\medskip
\label{t:wdeppar}
\begin{tabular}{lrrrrr}
\hline
&&&&&\\
Parameter & $U$ & $B$ & $V$ & $R$ & $I$ \\
&&&&&\\
\hline
&&&&&\\
Flux for 0 mag $F_\lambda$\,(W\,$m^{-2}\,nm^{-1}\times 10^{11})$& 4.35 &
   7.20 & 3.92 & 2.26 & 1.23 \\
Mean wavelength $\lambda$ ($nm$) & 357 & 437 & 561 & 652 & 801 \\
Bandwidth $\Delta \lambda$ ($nm$) & 69 & 54 &  94 & 129 & 149 \\
Filter-detector efficiency & 0.22 & 0.16 & 0.12 & 0.09 & 0.09 \\
&&&&&\\
\hline
\end{tabular}
\end{center}
\end{table}
\begin{table}[htb]
\begin{center}
\caption{Scaled down count rates and the  corresponding
probable errors in polarization (\%) for a 10.0 mag star in 10 minutes of
integration, the expected photon rates, and the deficiency factors in
the observed count rates ($\gamma$).
Overhead time was not included in computing the probable errors.}
\medskip
\label{t:expcnts}
\begin{tabular}{cccccccc}
\hline
&&&&&&&\\
 Filter & Air & Scaled & p. e. &  FWHM  &  & Expected & \\
  band &  mass & c s$^{-1}$ & (\%)   & (nm) &
$k_\lambda$  & p s$^{-1}$ & $\gamma$\\
&&&&&&&\\
\hline
&&&&&&&\\
 $U$  & 1.03 & 527   & 0.18  & 69 & 0.75 & 11482  & 22\\
 $B$  & 1.03 & 836   & 0.13  & 54 & 0.40 & 18461  & 22\\
 $V$  & 1.06 & 2289  & 0.08  & 94 & 0.25 &  19318  & 8\\
 $R$  & 1.08 & 2104  &  0.08  & 129 & 0.15 & 14623  & 7\\
 $I$  & 1.06 & 1501  & 0.10  & 149 & 0.10& 11904  & 8\\
&&&&&&&\\
\hline
\end{tabular}
\end{center}
\end{table}
The number of photoelectrons produced
per second can be calculated using
$$ n_{pe} =  \frac{10^{-0.4\,m}\,\pi\,F_\lambda\,\Delta
\lambda\,T_{atm}\,T_{total}
\,q\,\lambda\,D_{eff}^2}{8\times 10^6\,h\,c}, $$
where $m$ is the magnitude of the object, $F_\lambda$, the flux density for a
0 mag A0V star, and $h$ and $c$, the Planck constant and velocity of light,
$\lambda$ and $\Delta \lambda$ the effective wavelength and bandwidth
of observation, $T_{atm}$, the atmospheric transmittance, $T_{total}$,
the total effective transmittance and reflectivities of the various
optical elements in the
light path, $q$, the quantum efficiency of the detector, and $D_{eff}$, the
effective diameter of the telescope. On inserting the numerical
values of the constants, the above expression reduces to
$$ n_{pe} =  1.975\times 10^{15}\,F_\lambda\,10^{-0.4\,m}\,\Delta
\lambda\,T_{atm}\,T_{total}\,q\,\lambda\,D_{eff}^2, $$
where $F_\lambda$ is in $W\,m^{-2}\,nm^{-1}$, $\lambda$ and
$\Delta \lambda$ in $nm$, and D in $m$. The fact that only
50 per cent of the photons collected by the telescope are incident on the
photocathode has been taken into account in the above expression. 

Using the above expression and the
parameters given Tables~\ref{t:windpar} and \ref{t:wdeppar}
we computed the number of counts per second expected from
a 10.0 mag star, and the results are also given 
Table~\ref{t:expcnts}.
The absolute fluxes
for a 0 mag A0V star were taken from Johnson (1966) for the $UBV$ bands
and from Bessel (1979) for the $RI$ bands. The bandwidths and the
combined responses of the filter-detector combination were read from
Figure~\ref{f:filterdet}.  The table also gives the 
deficiency factor ($\gamma$) of observed counts, which is the ratio of the
the expected photon rates to the scaled down count rates for 10.0 mag star.
The FWHMs of the bands estimated from
the plots shown in Figure~\ref{f:filterdet},
and the mean extinction at VBO usually
observed during the January--March period, which were used
for the computations, are also given in the table.

\begin{figure}[htb]
\centerline {\psfig{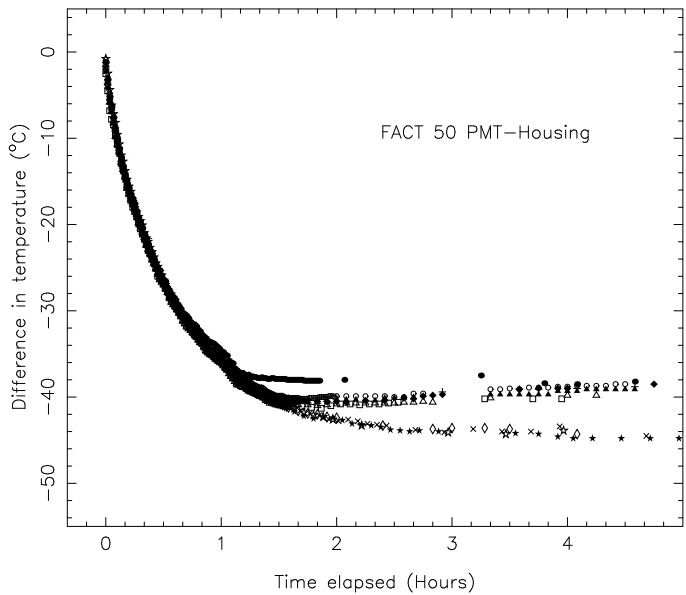}}
\caption {Plot of the difference in the PMT chamber- and ambient-temperatures
for different settings of cooling against the time elapsed
after switching on the cooling unit.}
\label{f:ccurve}
\end{figure}
\begin{figure}[htb]
\centerline {\psfig{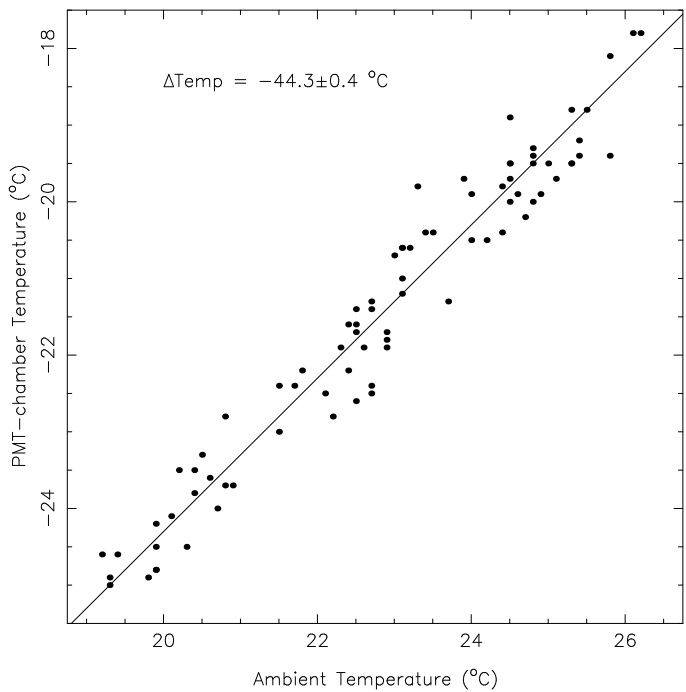}}
\caption {Plot of the temperature of the PMT chamber against the
ambient temperature. The straight line represents a constant
difference of 44.3\degr C with a standard deviation of
0.4\degr C.} \label{f:pmtamb}
\end{figure}

From Table~\ref{t:expcnts}
we see that the deficiency factors in the $UBVRI$ bands are
22, 22, 8, 7 and 8, respectively.
The reflectivity of the telescope was taken as 70\% for the computation of the
expected photon rates. The implied reflectivity of about 84\% for the mirrors is
likely to be on the higher side, considering the fact that the mirrors were
aluminized a few years back.
The dichroic filters used in the $U$ and $B$ channels were acquired
more than 12 years back. There is also a possibility that their reflectivities
and transmittances 
have deteriorated over time, even though they were kept inside their boxes.
These factors would have increased the deficiency factors in all the spectral
bands in a similar way.

 The identical deficiency factors in the
$U$ and $B$ are
almost three times those of $VRI$ bands, which are again identical within
observational uncertainties.
It may be noted that both the
$U$ and $B$ bands have identical photomultipliers and pulse 
amplifier-discriminators, and the photomultipliers were operated at
the same high voltages of 2300~V. The identical photon deficiency factors
for these bands indicate the definite possibility that
all the detected photons were not counted. The photomultipliers at
2300~V were already operated at very high gains (see Table~\ref{t:maxall}).
The discriminator level is factory set in AD6 model
pulse amplifier-discriminators,
and they may not be actually set at 
the valley of the pulse height distribution of
the photomultipliers used. The possibility of using a pre-amplifier in these
channels should be examined. 
It is likely that a fraction of the deficiency factor in the $VRI$ channel
also arises from not counting all the detected photons since
poor reflectivities of the telescope mirrors may not be able to account for
the rather high deficiency factor of 8 observed in that channel.
From Table~\ref{t:maxall} we see that the Hamamatsu tube also has
a very high gain when it is operated at a high voltage of 1900~V. The 
discriminator level of C9744 can be adjusted manually. The possibility
of lowering the discriminator level in the $VRI$ channel also should be
examined. If the output pulses from the amplifier-discriminators
show a large range,
the pulse counters may not be responding to the low amplitude pulses. In such
a case introducing a high bandwidth pulse amplifier before the counters
might help in reducing the deficiency factor.

\subsection{Cooling characteristics of FACT50}
As already mentioned, the Hamamatsu R943-02 tube is mounted in FACT50 model
chamber, which is of forced air-cooled type, for reducing the dark counts. In
Figure~\ref{f:ccurve} we have plotted the observed cooling curves of this model
 for different temperature settings. Along the vertical axis in the figure
we have plotted the difference in temperatures,
in the sense, chamber temperature
minus ambient temperature. We find from the figure that at the
 coolest setting, which is usually the case while
observing, it takes about two hours for the chamber to attain the maximum
difference in temperature with respect to the ambient, and hence, the
cooling unit should be switched on at least two hours before the beginning
of the observations.

In Figure~\ref{f:pmtamb}
 we have plotted the PMT chamber-temperatures observed after 
three hours of switching on of the cooling unit against the corresponding
ambient temperature. Probably, the difference in temperature attained
does not depend on the ambient temperature. The average difference in 
temperature observed is 44.3\degr C with a standard deviation of 0.4\degr C.
The PMT temperature will be
varying with the variation in the ambient temperature over the course night.
However, this will not have adverse effects since the ambient temperature
will not be varying significant during the period of observation of an object.

\section*{Acknowledgments}
We gratefully acknowledge the keen interests shown by
Professors H.~C.~Bhatt and P.~Sreekumar, and
Mr~A.~V.~Ananth in putting the instrument in
operation at the telescope, and Professors N. Kameswara Rao and R. Srinivasan
for their encouragements in the initial stages of the project.

We thank
Professors G. V. Coyne, F. Scaltriti, and A. M. Magalhaes
for their prompt responses to our queries, which helped us in
finalizing the optical layout, and
Professor F. Scaltriti for
sending copies of the drawings of the polarimeter at Torino
Astronomical Astronomical Observatory. 

Several of our colleagues helped us at various stages;
Mr~P.K. Mahesh, Mr~P.M.M. Kemkar, Mr~P.U.Kamath,
Dr~D. Suresh and Dr~G. Rajalakshmi helped us in the preparation of the
Auto CAD drawings of the mechanical parts of the polarimeter;
Professor T. P. Prabhu helped us in acquiring the dichroic mirrors and
glass filters;
Dr Baba Varghese helped us in creating the latex file of the manuscript;
Mr~N. Sivaraj helped us in checking the photomultiplier
tubes and the pulse-amplifier-discriminators;
Mr S.~Sriram, Mr~T.~Vishnu, Mr~R.~Selvendran and Mr~P.~Devendran helped
us in making the neutral density filter that was used for the
determination of the dead-time coefficient of the
pulse counting electronics; we thank all of them.

We also thank the staff at Vainu Bappu Observatory, especially,
Mr P.~Anbazhagan, for their extremely valuable help in carrying out the
observations with the polarimeter.
\nopagebreak

\vskip 1.5cm
\noindent
{\bf Publications}

\begin{enumerate}
\item{{\it A new three-band, two beam astronomical
photo-polarimeter}, G.~Srinivasulu, A.~V.~Raveendran,  S.~Muneer, M.~V.~Mekkaden
N.~Jayavel, M.~R.~Somashekar, K.~Sagayanathan, S.~Ramamoorthy, M.~J.~Rosario,
K.~Jayakumar, 2014, Bull. Astr. Soc. India, 42, 165}
\item{{\it Linear Polarization Measurements of Nova Sco 2015} \quad
({\it PNV} \quad J17032620-3504140), S.~Muneer, G.~C.~Anupama, A.~V.~Raveendan,
2015, \, The Astronomer's Telegram, \#7161}
\item{{\it UBVRI Polarimetry of Nova Sgr 2015 \#2},
S.~Muneer, G.~C.~Anupama, A.~V. Raveendran, A.~Muniyandi, R.~Baskar, 2015,
 The Astronomer's Telegram, \#7310}
\end{enumerate}
\end{document}